%% file: hlbl_pp.tex
\documentclass[aps,prd,12pt,nofootinbib,tightenlines,preprintnumbers,superscriptaddress,notitlepage]{revtex4-1}
\usepackage{graphicx}

\usepackage[utf8]{inputenc}

\usepackage{amssymb}
\usepackage{amsmath}
\usepackage{xcolor}
\usepackage{hyperref}
\usepackage{multirow}
\hypersetup{colorlinks, linkcolor = [rgb]{0,0.0,0.75}, citecolor = [rgb]{0,0.0,0.75}, urlcolor = [rgb]{0,0.0,0.75}}

\makeatletter
\g@addto@macro\bfseries{\boldmath}
\makeatother

\input{definitions}

\begin{document}

\preprint{MITP/21-019}
\preprint{CERN-TH-2021-047}

\title{Hadronic light-by-light contribution to $(g-2)_\mu$\\ from lattice QCD: a complete calculation}

\author{En-Hung~Chao}
\affiliation{PRISMA$^+$ Cluster of Excellence \& Institut f\"ur Kernphysik,
Johannes Gutenberg-Universit\"at Mainz,
D-55099 Mainz, Germany}

\author{Renwick~J.~Hudspith}
\affiliation{PRISMA$^+$ Cluster of Excellence \& Institut f\"ur Kernphysik,
Johannes Gutenberg-Universit\"at Mainz,
D-55099 Mainz, Germany}

\author{Antoine~G\'erardin}
\affiliation{Aix Marseille Univ, Universit\'{e} de Toulon, CNRS, CPT, Marseille, France}

\author{Jeremy~R.~Green}
\affiliation{Theoretical Physics Department, CERN, 1211 Geneva 23, Switzerland}

\author{Harvey~B.~Meyer}
\affiliation{PRISMA$^+$ Cluster of Excellence \& Institut f\"ur Kernphysik,
Johannes Gutenberg-Universit\"at Mainz,
D-55099 Mainz, Germany}
\affiliation{Helmholtz~Institut~Mainz,
Staudingerweg 18, D-55128 Mainz, Germany}
\affiliation{GSI Helmholtzzentrum f\"ur Schwerionenforschung, Darmstadt, Germany}

\author{Konstantin~Ottnad}
\affiliation{PRISMA$^+$ Cluster of Excellence \& Institut f\"ur Kernphysik,
Johannes Gutenberg-Universit\"at Mainz,
D-55099 Mainz, Germany}

\begin{abstract}
We compute the hadronic light-by-light scattering contribution to the muon $g-2$ from the up, down, and strange-quark sector directly using lattice QCD. Our calculation features evaluations of all possible Wick-contractions of the relevant hadronic four-point function and incorporates several different pion masses, volumes, and lattice-spacings. We obtain a value of $\ahlbl = 106.8(14.7) \times 10^{-11}$ (adding statistical and systematic errors in quadrature), which is consistent with current phenomenological estimates and a previous lattice determination. It now appears conclusive that the hadronic light-by-light contribution cannot explain the current tension between theory and experiment for the muon $g-2$.
\end{abstract}

\date{\today}

\maketitle

\input{intro}
\input{formalism}
\input{lattice_setup}

\input{integrand}
\input{light_contributions}
\input{strange_contributions}
\input{higher_order}
\input{combined_result}
\input{conclu}

\acknowledgments{This work is supported by the European Research Council (ERC) under the
  European Union's Horizon 2020 research and innovation programme
  through grant agreement 771971-SIMDAMA, as well as by  the Deutsche
  Forschungsgemeinschaft (DFG) through the Collaborative Research
  Centre 1044, through project HI 2048/1-2 (project No.\ 399400745) and through
 the Cluster of Excellence \emph{Precision Physics, Fundamental Interactions, and Structure of Matter} (PRISMA+ EXC 2118/1)  within the German Excellence Strategy (Project ID 39083149). The project leading to this publication has also received funding from the Excellence Initiative of Aix-Marseille University - A*MIDEX, a French “Investissements d’Avenir” programme, AMX-18-ACE-005.
Calculations for this project were partly performed on the HPC clusters ``Clover'' and ``HIMster II'' at the Helmholtz-Institut Mainz and ``Mogon II'' at JGU Mainz. The authors gratefully acknowledge the Gauss Centre for Supercomputing e.V. \href{(www.gauss-centre.eu)}{(www.gauss-centre.eu)} for funding this project by providing computing time on the GCS Supercomputer HAWK at Höchstleistungsrechenzentrum Stuttgart \href{(www.hlrs.de)}{(www.hlrs.de)} (Project ID GCS-HQCD).
Our programs use the deflated SAP+GCR solver from the openQCD package~\cite{Luscher:2012av}, as well as the QDP++ library
\cite{Edwards:2004sx}.
We are grateful to our colleagues in the CLS initiative for sharing ensembles.

\appendix
\input{matching_nf2p1}
\input{data_tables}

\bibliographystyle{apsrev4-1}
\nocite{bibtitles}
\bibliography{refs}

\end{document}

%% file: definitions.tex
\newcommand{\be}{\begin{equation}}
\newcommand{\ee}{\end{equation}}
\newcommand{\ba}{\begin{eqnarray}}
\newcommand{\ea}{\end{eqnarray}}
\newcommand{\la}{\label}

\newcommand{\Tr}{\,\hbox{\rm Tr}}

\newcommand{\<}{\langle}
\renewcommand{\>}{\rangle}

\newcommand{\tr}{{\rm Tr\,}}

\newcommand{\ahlbl}{a_\mu^{\text{Hlbl}}}

\newcommand{\kernel}{\mathcal{\bar L}}

%% file: intro.tex
\section{Introduction}

The anomalous magnetic moment of the muon, $a_\mu \equiv (g-2)_\mu/2$, is one of the most precisely measured quantities of the Standard Model (SM) of particle physics. Its value is of considerable interest to the physics community as, currently, there exists a $3.7\sigma$ tension between the experimental determination of Ref.~\cite{Bennett:2006fi} and the current theoretical evaluation (see Ref.~\cite{Aoyama:2020ynm} and references therein). Although the central value of the theoretical prediction is overwhelmingly dominated by QED effects, its uncertainty is dominated by low-energy QCD contributions. If the tension persists under more precise scrutiny, it is possible that a $5\sigma$ discrepancy could appear, heralding an indirect determination of Beyond the Standard Model (BSM) physics.

A new series of experimental results (E989 at Fermilab \cite{Grange:2015fou} and E34 at J-PARC \cite{Abe:2019thb}) intend to increase the precision of the experimental determination by a factor of about four; as it stands, the error on $a_\mu$ is at the level of $63\times10^{-11}$. Similarly, the theory community is striving to reduce the error of their determination to match the upcoming experimental precision. One of the contributions that is of specific interest is the hadronic vacuum polarisation (HVP), which enters at $O(\alpha_\text{QED}^2)$. Being a QCD quantity dominated by hadronic scales, this contribution can be directly obtained from first-priciples lattice QCD calculations, although currently its most precise estimate 
\cite{Aoyama:2020ynm,Davier:2017zfy,Keshavarzi:2018mgv,Colangelo:2018mtw,Hoferichter:2019gzf,Davier:2019can,Keshavarzi:2019abf,Kurz:2014wya}
comes from dispersive methods and is $6931(40)\times 10^{-11}$. Significant progress has been made in recent years within the lattice approach 
\cite{Chakraborty:2017tqp,Borsanyi:2017zdw,Blum:2018mom,Giusti:2019xct,Shintani:2019wai,Davies:2019efs,Gerardin:2019rua,Aubin:2019usy,Giusti:2019hkz,Lehner:2020crt,Borsanyi:2020mff}, and these determinations are quickly becoming competitive with the dispersive approach.

A much smaller contribution to the overall $(g-2)_\mu$ comes from hadronic light-by-light scattering (Hlbl), entering at $O(\alpha_\text{QED}^3)$. However, this quantity is currently only known at the 20\% level: the recent evaluation of Ref.~\cite{Aoyama:2020ynm}, omitting an estimate of the small charm-quark contribution, amounts to
\cite{Melnikov:2003xd,Masjuan:2017tvw,Colangelo:2017fiz,Hoferichter:2018kwz,Gerardin:2019vio,Bijnens:2019ghy,Colangelo:2019uex,Pauk:2014rta,Danilkin:2016hnh,Jegerlehner:2017gek,Knecht:2018sci,Eichmann:2019bqf,Roig:2019reh}
$89.0(19.0)\times 10^{-11}$. Thus the absolute uncertainty of the Hlbl contribution is only about half that of the recent average~\cite{Aoyama:2020ynm} for the HVP. To match the expected experimental precision, it is thought that the Hlbl contribution $\ahlbl$ needs to be known with a precision of around $10\%$. The task of directly computing this contribution using lattice QCD methods is quite daunting, as it requires the computation of connected and disconnected four-point functions. Few lattice groups have even performed measurements of the leading contributions, and none with the desired precision. The most-precise lattice determination to date \cite{Blum:2019ugy} uses the finite-volume $\text{QED}_L$ prescription and quotes a value of (adding their statistical and systematic errors in quadrature) $78.7(35.4)\times 10^{-11}$. In \cite{Chao:2020kwq}, we provided an estimate at the physical pion mass, starting from our $\text{SU}(3)_f$-symmetric point result and correcting for the neutral-pion exchange~\cite{Gerardin:2016cqj,Gerardin:2019vio}, of $104.0(20.8)\times 10^{-11}$.

We extend our previous determination of the Hlbl contribution to $(g-2)_\mu$ at the $\text{SU}(3)_f$-symmetric point \cite{Chao:2020kwq} by incorporating data from simulations at pion masses as low as $200$ MeV. We also provide estimates for the sub-leading $(3+1)$, $(2+1+1)$, and $(1+1+1+1)$ contributions, providing a full first-principles calculation using lattice QCD with a competitive overall error.

This work is organised as follows: first we introduce our approach and formalism  for measuring $\ahlbl$ using lattice QCD and infinite-volume perturbative QED in Sec.~\ref{sec:formalism}. In Sec.~\ref{sec:latt_setup}, we discuss the numerical techniques and effort for our determination. Section~\ref{sec:intgnd} contains a comparison of the integrand to the predictions of hadronic models.
We then present results for the leading fully-connected and $(2+2)$ diagram contributions with light (Sec.~\ref{sec:leading_contributions}) and strange (Sec.~\ref{sec:strange_leading}) quark content. In Sec.~\ref{sec:higher_order} we discuss the determination of the higher-order $(3+1)$, $(2+1+1)$, and $(1+1+1+1)$ contributions. We finally discuss the systematics of our largest contribution in Sec.~\ref{sec:final_res}, and combine all of our determinations and draw conclusions in Sec.~\ref{sec:concl}. 

%% file: formalism.tex
\section{Formalism}\label{sec:formalism}

\begin{figure}[t]
	\includegraphics*[height=0.18\linewidth]{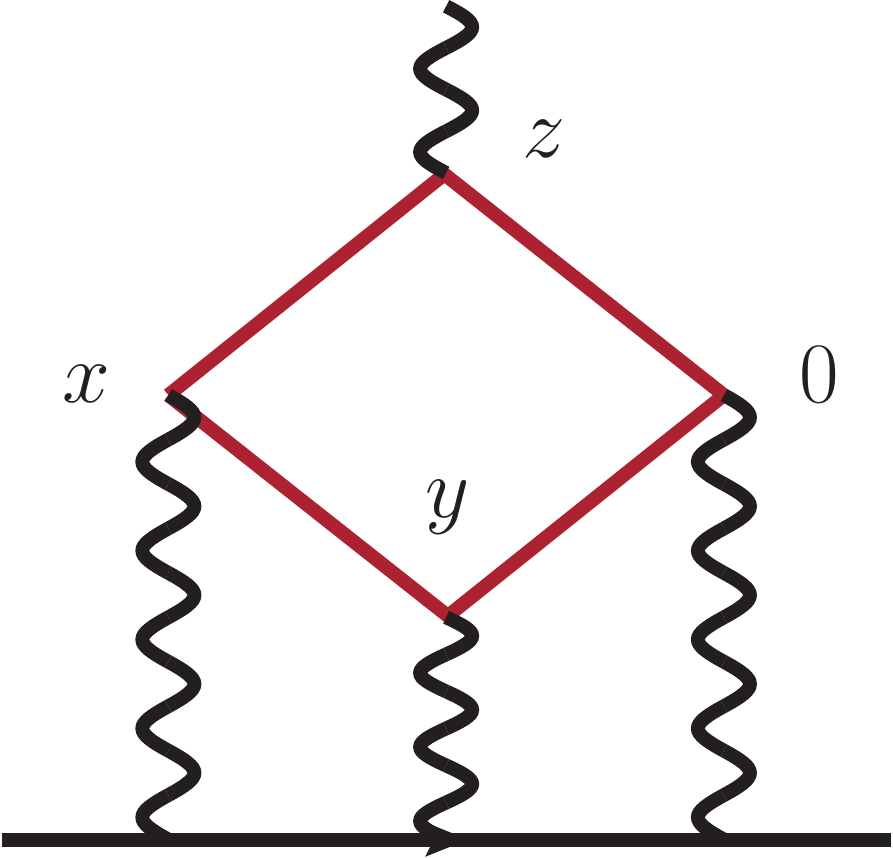} \hspace{2cm}
	\includegraphics*[height=0.18\linewidth]{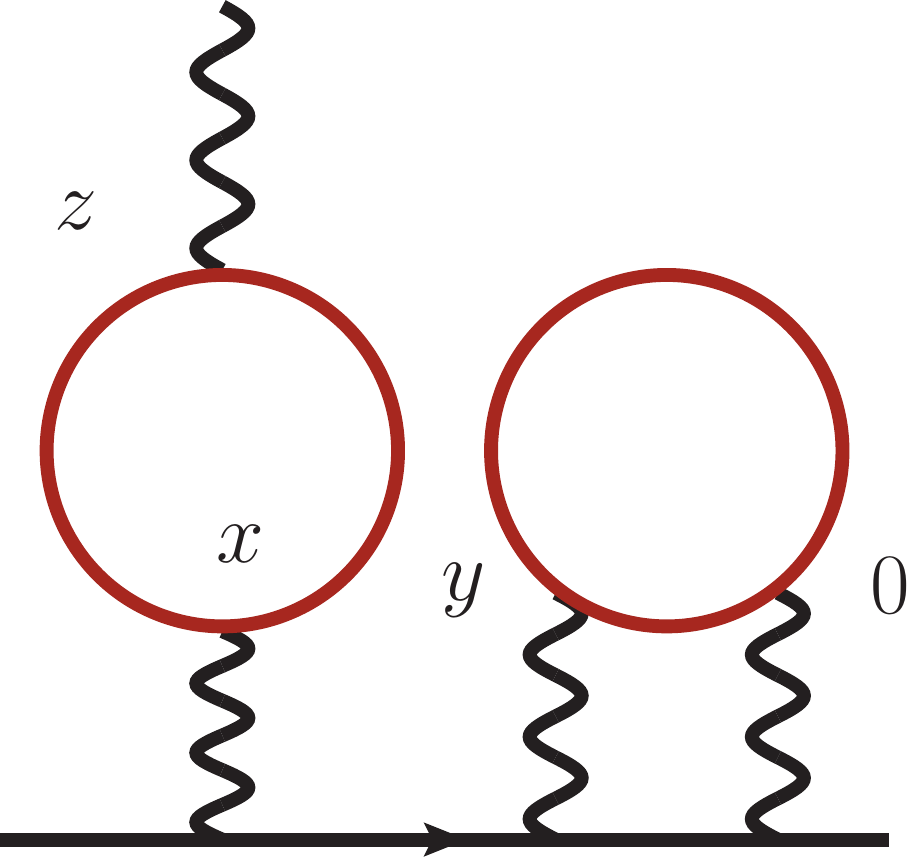} \\
	\includegraphics*[height=0.18\linewidth]{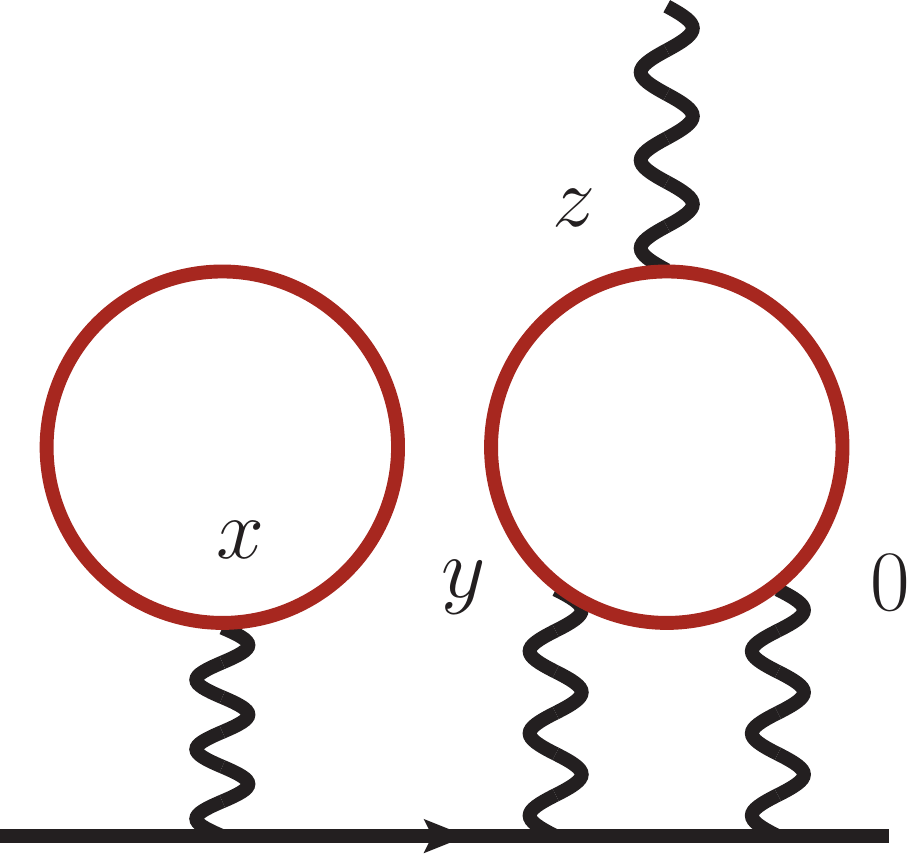} \hspace{1.5cm}
	\includegraphics*[height=0.18\linewidth]{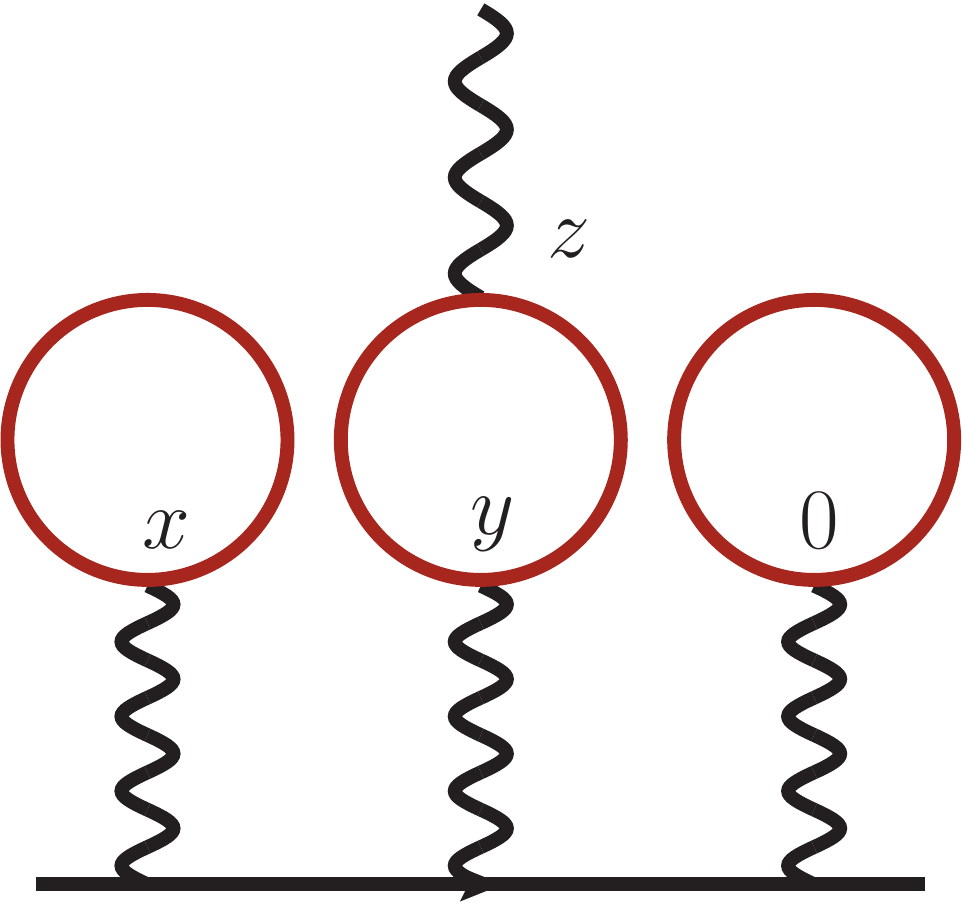} \hspace{1.5cm}
	\includegraphics*[height=0.18\linewidth]{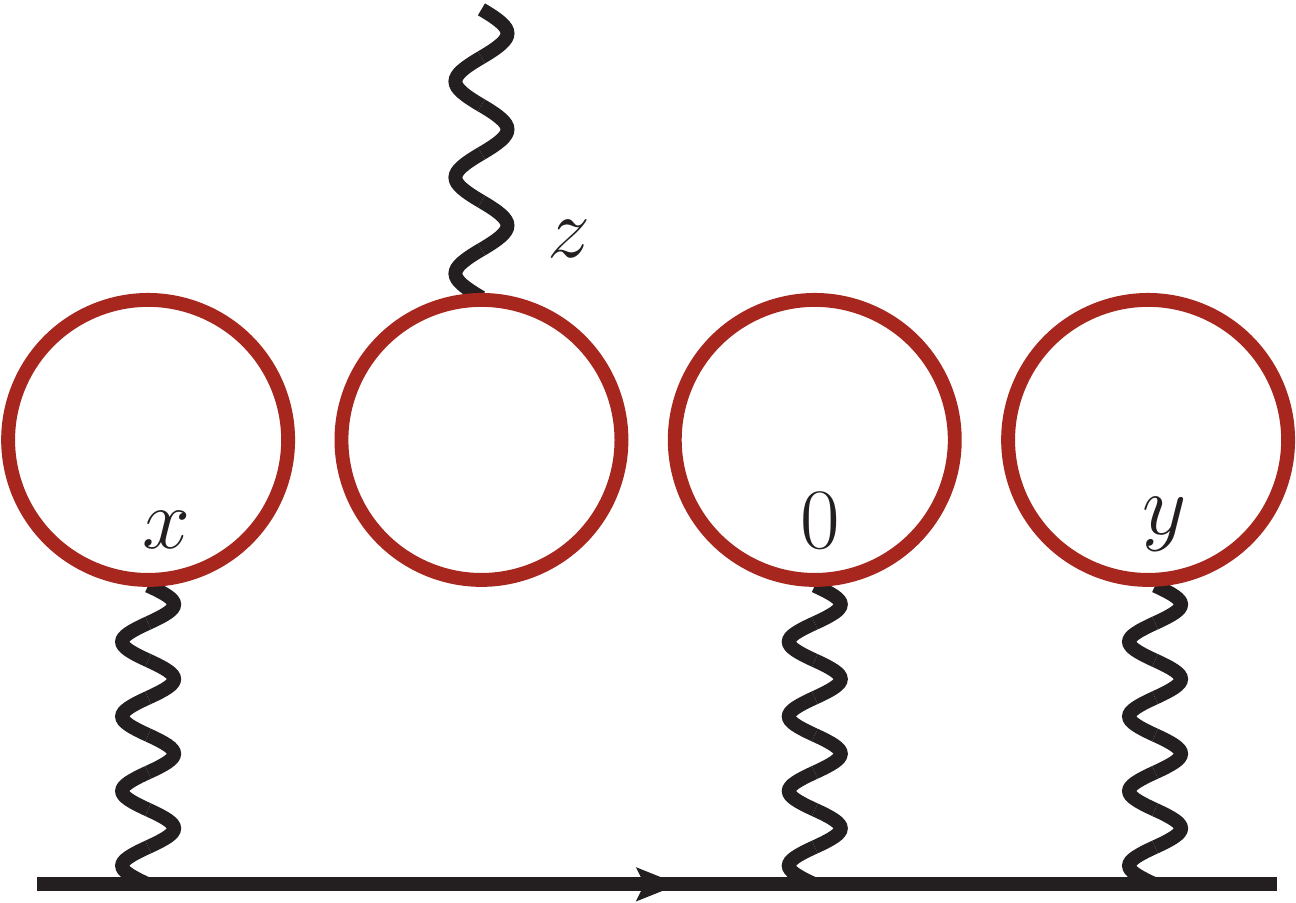} 
	\caption{Different quark Wick-contraction classes appearing in the computation of the QCD four-point correlation function. The straight horizontal lines represent muon propagators, wavy lines represent photon propagators. From left to right, top to bottom, they are the fully-connected, $(2+2)$, $(3+1)$, $(2+1+1)$ and $(1+1+1+1)$. Each class contains the digrams obtained from all the possible permutations of the four points attached to photons. }\label{formalism:fig:contractions}
\end{figure}

In order to have a better control over the long-distance QED effects, we use a position-space approach, which consists in treating the QED part perturbatively, in infinite-volume and in the continuum,
and the hadronic part non-perturbatively on the lattice~\cite{Asmussen:2016lse,Asmussen:2017bup,Asmussen:2019act}.
Due to the $O(4)$ symmetry in the Euclidean continuum, the hadronic light-by-light contribution to the anomalous magnetic moment of the muon, $\ahlbl$, admits the following integral representation 
\begin{equation}\label{eq:intmaster}
\ahlbl = \sum_{\text{Topology}}\int_0^\infty d|y|\, f^{(\text{Topology})}(|y|),
\end{equation}
where $f^{(\text{Topology})}(|y|)$, henceforth called the \textit{integrand} (for a fixed diagrammatic topology), is itself
obtained as an integral over spacetime (in our notation $\int_x = \int d^4x$),
\begin{equation}\la{eq:fy_master}
\sum_{\text{Topology}}f^{(\text{Topology})}(|y|) = \frac{m_\mu e^6}{3} 2\pi^2  |y|^3 \int_x \; \kernel_{[\rho,\sigma];\mu\nu\lambda}(x,y)\;i\widehat\Pi_{\rho;\mu\nu\lambda\sigma}(x,y).
\end{equation}
Here $e^2/(4\pi)=\alpha_{\rm QED}$ is the fine-structure constant and $m_\mu$ the muon mass.
The QED kernel $\kernel$ represents the contributions of the photon and muon propagators and vertices
(see Fig.~\ref{formalism:fig:contractions}),
and $i\widehat\Pi$ is the first moment of the connected, Euclidean, hadronic four-point function,
\begin{equation}
\begin{aligned}
i\widehat \Pi_{\rho;\mu\nu\lambda\sigma}( x, y)  &= -\int_z  z_\rho\, \widetilde\Pi_{\mu\nu\sigma\lambda}(x,y,z), \label{eq:pihat}
\\ 
\widetilde\Pi_{\mu\nu\sigma\lambda}(x,y,z)&\equiv\Big\<\,j_\mu(x)\,j_\nu(y)\,j_\sigma(z)\, j_\lambda(0)\Big\>_{\rm QCD}.
\end{aligned}
\end{equation}
The field $j_\mu(x)$ appearing above is the hadronic component of the electromagnetic current,
\begin{equation}
j_\mu(x) = \frac{2}{3} (\overline{u} \gamma_{\mu} u)(x) - \frac{1}{3} (\overline{d} \gamma_{\mu} d)(x) - \frac{1}{3} (\overline{s} \gamma_{\mu} s)(x).
\end{equation}
As for the QCD four-point function $\widetilde{\Pi}_{\mu\nu\sigma\lambda}$, it consists of five classes of Wick-contractions, illustrated in Fig.~\ref{formalism:fig:contractions}: the fully-connected, the $(2+2)$, the $(3+1)$, the $(2+1+1)$ and the $(1+1+1+1)$.
It can be shown that the contribution to $\ahlbl$ of each topology is itself a gauge-independent observable, therefore it is legitimate to focus on each independently. 

According to large-$N_c$ arguments and some numerical evidence provided by the RBC/UKQCD collaboration~\cite{Blum:2019ugy} on the $(3+1)$ topology, only the first two (the fully-connected and $(2+2)$) of the aforementioned classes are believed to be dominant, however no direct calculations of the subleading classes have been performed until now. In addition, the last three classes, which we refer to as \textit{higher-order} topologies, are suppressed by powers of the light-minus-strange quark-mass difference around the $\text{SU}(3)_f$-symmetric point,
and necessarily vanish exactly at that point.

As the integrand ($f(|y|)$) is a scalar function in $|y|$, our computational strategy consists in calculating the integrand, the inner integrals over $x$ and $z$ being replaced by sums, averaged over many equivalent instances of the origin and the $y$-vector for a given $|y|$ to enhance statistics, and then applying the trapezoidal rule to approximate the integral over $|y|$ of Eq.~\eqref{eq:intmaster}, in order to finally obtain $\ahlbl$ for each gauge ensemble. 
We then take the appropriate infinite-volume and continuum limits and extrapolate our result to physical quark masses.

In addition to showing the integrand, often we will find it useful to present the partially-integrated quantity,
\begin{equation}\label{eq:amuymax}
a_\mu(|y|_\text{Max.}) = \int_0^{|y|_\text{Max.}} d|y| f(|y|).
\end{equation}
This quantity is typically less sensitive to point-by-point fluctuations in $f(|y|)$ and adequately illustrates the salient features of the calculation. Our expectation is that the partially-integrated quantity admits plateau as $|y|_\text{Max.}$ is increased, indicating the integral has saturated within the uncertainties.
% , which would mean the integrand is statistically consistent with zero.

Exploiting the Ward identities associated with current conservation, the QED kernel can be modified by adding to it terms which do not contribute to $\ahlbl$ in the infinite volume limit~\cite{Blum:2016lnc,Asmussen:2019act}. 
To mitigate the signal-to-noise problem of vector-current lattice correlation functions at large seperations, one would like to choose a QED kernel which guarantees a rapid fall-off of the integrand $f(|y|)$ at large $|y|$, without picking up large discretisation effects by making it too-peaked at short distances.
Due to the gauge-invariance of each topology, one can even work with different choices of kernel for each topology individually.

In our previous work at the $\text{SU}(3)_f$-symmetric point~\cite{Chao:2020kwq}, we have shown the effectiveness of a certain one-parameter family of kernels, $\kernel^{(\Lambda)}_{[\rho,\sigma];\mu\nu\lambda}$, with positive, real $\Lambda$.
Our preferred choice for this parameter is $\Lambda=0.4$; this was motivated by several studies of the shape of the integrand: a continuum and infinite volume QED calculation of the lepton loop contribution, a study of the pion-pole contribution with a Vector Meson Dominance (VMD) parametrisation for the transition form factor in the continuum and finite volume, and our direct lattice calculations at the $\text{SU}(3)_f$-symmetric point.

While Eq.\ (\ref{eq:fy_master}) represents our general master formula for the integrand $f(|y|)$, 
for computational reasons it can be beneficial to exploit the translational invariance of the QCD correlation function to re-arrange the integrand in different ways, such that only the most favourable diagrams within each topology class have to be explicitly computed. In our previous work, we showed that for the fully-connected contribution such an approach reduced the computational cost significantly without introducing undesirable effects when used in conjunction with the kernel $\kernel^{(\Lambda)}_{[\rho,\sigma];\mu\nu\lambda}$~\cite{Chao:2020kwq}.
In the subsections below we present the specific integral representations that we use for each of the five topologies. 

For notational simplicity, we will find the following QED kernel combination useful,
\begin{equation}
\mathcal{L}^\prime_{[\rho,\sigma];\mu\nu\lambda}(x,y) = 
\bar{\mathcal{L}}^{(\Lambda)}_{[\rho,\sigma];\mu\nu\lambda}(x,y) 
+ \bar{\mathcal{L}}^{(\Lambda)}_{[\rho,\sigma];\nu\mu\lambda}(y,x) 
- \bar{\mathcal{L}}^{(\Lambda)}_{[\rho,\sigma];\lambda\nu\mu }(x, x-y).
\end{equation}
Also, in the equations below we anticipate our use of the local vector current on the lattice, which requires a multiplicative
renormalisation factor $\hat Z_{\rm V}$.

\subsection{The fully-connected contribution}

For the fully-connected calculation we use the following master equation for the integrand:
\begin{equation}
\begin{gathered}
f^{\text{(Conn.)}}(|y|) = -\sum_{j\in u,d,s}\hat{Z}_{\rm V}^4 Q_j^4 \frac{m_\mu e^6}{3}2\pi^2|y|^3 \times\\
 \int_x \bigg(
 \mathcal{L}^\prime_{[\rho,\sigma]\mu\nu\lambda}(x,y)
 \int_z z_\rho \widetilde\Pi^{(1),j}_{\mu\nu\sigma\lambda}(x,y,z)
+\kernel^{(\Lambda)}_{[\rho,\sigma];\lambda\nu\mu}(x,x-y)x_\rho \int_z \widetilde\Pi^{(1),j}_{\mu\nu\sigma\lambda}(x,y,z)\bigg)\,,
\end{gathered}
\end{equation}
with hadronic contribution
\begin{equation}
\widetilde\Pi_{\mu\nu\sigma\lambda}^{(1),j}(x,y,z) = -2\text{Re}\left\langle\Tr\left[ S^j(0,x) \gamma_{\mu} S^j(x,y) \gamma_{\nu} S^j(y,z) \gamma_{\sigma} S^j(z,0) \gamma_{\lambda} \right]\right\rangle_U.
\end{equation}
Here $S^{j}(x,y)$ is the flavour $j$-quark propagator from source $y$ to sink $x$, $Q_j$ is the charge factor ($Q_u = \frac{2}{3}$, $Q_d = -\frac{1}{3}$, $Q_s = -\frac{1}{3}$), and $\langle\cdot\rangle_U$ denotes the ensemble average.

\subsection{The \texorpdfstring{$(2+2)$}{(2+2)} contribution}

We start by defining the two-point function ``meson-field''
\begin{equation}\label{formalism:eq:pi2}
\Pi_{\mu\nu}^{j}(x,y)= -\text{Re}\left(\text{Tr}[S^j(y,x)\gamma_\mu S^j(x,y)\gamma_\nu]\right),
\end{equation}
which must have its vacuum expectation value (VEV) subtracted:
\begin{equation}
\hat{\Pi}^j_{\mu\nu}(x,y) = \Pi^j_{\mu\nu}(x,y) - \langle \Pi^j_{\mu\nu}(x,y) \rangle_U.
\end{equation}

We use the following integral representation for $a_\mu^{(2+2)}$
\begin{equation}
\begin{aligned}
f^{(2+2)}(|y|) =
&-\sum_{i,j\in u,d,s} Q_i^2 Q_j^2 \hat{Z}_{\rm V}^4\frac{m_\mu e^6}{3}2\pi^2 |y|^3 \times \\
\biggl\langle \int_x \bigg(
&(\kernel^{(\Lambda)}_{[\rho,\sigma];\mu\nu\lambda}(x,y)+\kernel^{(\Lambda)}_{[\rho,\sigma];\nu\mu\lambda}(y,x)) 
\hat{\Pi}^i_{\mu\lambda}(x,0)\int_z z_\rho
\hat{\Pi}^j_{\sigma\nu}(z,y)\\
&+\kernel^{(\Lambda)}_{[\rho,\sigma];\mu\nu\lambda}(x,y)\hat{\Pi}^i_{\mu\nu}(x,y)
\int_z z_\rho \hat{\Pi}^j_{\sigma\lambda}(z,0)\bigg)
\biggr\rangle_U.
\end{aligned}
\la{eq:2+2master}
\end{equation}
Note that the VEV subtraction is necessary to guarantee that the two quark loops are connected by gluons, in the perturbative picture. 
In this representation, the factorisation of the $x$- and $z$-integrations makes the lattice computation easier. 
Similar patterns can also be found in our choice of representation for the higher order topologies for the same reason.  

We call \textit{light-light} contribution the set of diagrams consisting exclusively of light quarks. Likewise, the \textit{strange-strange} contribution contains only strange quark loops. Finally, the \textit{light-strange} case covers all diagrams containing one light and one strange quark loop.
These sub-contributions can easily be constructed by combining different terms in Eq.\ (\ref{eq:2+2master}). As the integral is constructed as a post-processing step, the light-quark and strange-quark loops can easily be combined.

\subsection{The \texorpdfstring{$(3+1)$}{(3+1)} contribution}

As we work with $N_f=2+1$ lattice ensembles, we assume the mass-degeneracy between the $u$- and $d$-quark from here on to simplify our expressions. We begin by defining the two hadronic building blocks (here $l$ and $s$ refer to light and strange quarks respectively),
\begin{equation}\label{formalism:eq:disc_loop}
T_\mu(x) = \text{Im} \Big( \tr[\gamma_\mu S^{l}(x,x)] - \tr[\gamma_\mu S^{s}(x,x)] \Big),
\end{equation}
and
\begin{equation}\label{formalism:eq:triangle}
R^i_{\mu\nu\lambda}(x,y,z) = \text{Im}\Big(\tr[\gamma_\mu S^{i}(x,y)\gamma_\nu S^{i}(y,z)\gamma_\lambda S^{i}(z,x)] \Big).
\end{equation}
The quantity $R^{i}$ will be referred to as a \textit{triangle} with quark species $i$, and $T$ will be called
\textit{disconnected loop}.

Our expression for the integrand for this contribution reads
\begin{equation}\label{formalism:eq:f3p1}
\begin{aligned}
f^{(3+1)}(|y|) = \frac{2m_\mu e^6}{9}&\sum_{j\in{u,d,s}} \hat{Z}^4_{\rm V} Q_j^3 2\pi^2 |y|^3\times \\
&\Big\langle\int_{x} \mathcal{L}^\prime_{[\rho,\sigma]\mu\nu\lambda}(x,y) T_\mu (x)\int_z z_\rho R^j_{\lambda\nu\sigma}(0,y,z) \\ 
&+ \int_x\bar{\mathcal{L}}^{(\Lambda)}_{[\rho,\sigma]\lambda\nu\mu}(x,x-y) x_\rho T_{\mu}(x)\int_z R^j_{\lambda\nu\sigma}(0,y,z)\\
&+ \int_{x}\bar{\mathcal{L}}^{(\Lambda)}_{[\rho,\sigma]\mu\nu\lambda}(x,y) R^j_{\mu\nu\lambda}(x,y,0)\int_{z}z_\rho T_\sigma(z)\Big\rangle_U.
\end{aligned}
\end{equation}
It is worth noting that unlike in the $(2+2)$ case, no VEV-subtraction is needed for the $(3+1)$ contribution, because the VEV of the three-point function and the one-point function vanish due to the charge conjugation symmetry of the QCD action. In later sections, we will call $(3+1)_{\rm{light}}$ and  $(3+1)_{\rm{strange}}$ the sub-contribution with light and strange quark triangle respectively. 

\subsection{The \texorpdfstring{$(2+1+1)$}{(2+1+1)} contribution}

We can derive a representation for  $f^{(2+1+1)}$  from the expression for the $(3+1)$ topology; the idea is to split the triangles appearing in the expression of the $(3+1)$ integrand into a sum of products of two- and one-point functions, and then correct the diagram double-counting. In doing so, the terms involving the disconnected quark loop $T$ in Eq.~\eqref{formalism:eq:f3p1} can be reused for the $(3+1)$ calculation, as we perform more self-averages for this noisy, more-disconnected quantity (see Sect.~\ref{sec:latt_setup}). Moreover, we apply a change of variables to avoid the case where a disconnected loop is located at the origin to increase the number of available samples per $|y|$.

More explicitly, we define the two quantities
\begin{equation}
\begin{split}
& h_{\mu\nu\lambda}^{i}(x,y) = \hat{\Pi}^{i}_{\mu\lambda}(x,0)T_\nu(y),
\\
& g_{\mu\nu\lambda}^{i}(x,y) = 
h^{i}_{\mu\nu\lambda}(x,y) 
+ 2h^{i}_{\nu\mu\lambda}(y,x),
\end{split}
\end{equation}
and we write
\begin{equation}\label{formalism:eq:f2p1p1}
\begin{aligned}
f^{(2+1+1)}(|y|) = \frac{m_\mu e^6}{54}&\hat{Z}_{\rm V}^4 \sum_{i\in u,d,s}Q_i^2  2\pi^2 |y|^3 \times \\
\Big\langle
&-\int_{x}\mathcal{L}^{\prime}_{[\rho,\sigma]\mu\nu\lambda}(y-x,y)T_\mu(x)\int_z (z_\rho - y_\rho) h^{i}_{\sigma\lambda\nu}(z,y)\\
& + \int_{x} (x-y)_\rho \bar{\mathcal{L}}^{(\Lambda)}_{[\rho,\sigma]\lambda\nu\mu}(x-y,x) T_\mu(x)\int_{z} h^{i}_{\sigma\lambda\nu}(z,y)\\ 
& + \int_{x}
\mathcal{L}^\prime_{[\rho,\sigma]\mu\nu\lambda}(x,y)T_\mu(x)
\int_z z_\rho g^i_{\sigma\nu\lambda}(z,y)
\\
& + 
\int_x \bar{\mathcal{L}}^{(\Lambda)}_{[\rho,\sigma]\lambda\nu\mu}(x,x-y)x_\rho T_\mu(x)\int_z g^i_{\sigma\nu\lambda}(z,y)
\Big\rangle_U.
\end{aligned}
\end{equation}

\subsection{The \texorpdfstring{$(1+1+1+1)$}{(1+1+1+1)} contribution}

Here we finally give our parametrisation of the fully-disconnected $(1+1+1+1)$ contribution. This again takes advantage of the quantities computed in the previous cases. Here, one needs to carefully subtract the non-vanishing VEVs appearing in different pieces in this contribution. We define the following quantity:
\begin{equation}
\begin{aligned}
\langle T_\mu(x)T_\nu(y)T_\sigma(z)T_\lambda(0) \rangle_U^c =
&+\langle T_\mu(x)T_\nu(y)T_\sigma(z)T_\lambda(0) \rangle_U\\
&-\langle T_\mu(x)T_\nu(y) \rangle_U \langle T_\sigma(z)T_\lambda(0) \rangle_U \\
&-\langle T_\mu(x)T_\sigma(z) \rangle_U \langle T_\nu(y) T_\lambda(0) \rangle_U\\
&-\langle T_\mu(x) T_\lambda(0) \rangle_U \langle T_\nu(y)T_\sigma(z) \rangle_U.
\end{aligned}
\end{equation}
With this definition in place, we can write down the expression we used for the integrand for this topology, after correcting the triple-counting of the diagrams,
\begin{equation}\label{formalism:eq:f1p1p1p1}
\begin{aligned}
f^{(1+1+1+1)}(|y|) &= - \frac{m_\mu e^6}{729} \hat{Z}_{\rm V}^4 2\pi^2 |y|^3 \times \\
\Big\langle &\int_x \bar{\mathcal{L}}^{(\Lambda)}_{[\rho,\sigma]\lambda\nu\mu}(x,x-y)x_\rho T_\mu(x)\int_z T_{\nu}(y) T_{\lambda}(0) T_{\sigma}(z)
\\
& +	
\int_{x}
\mathcal{L}^\prime_{[\rho,\sigma]\mu\nu\lambda}(x,y) 
T_\mu(x)
\int_z z_\rho T_{\nu}(y) T_{\lambda}(0) T_{\sigma}(z)\Big\rangle_U^c.
\end{aligned}
\end{equation}

As a concluding remark for this section, in some of the provided expressions, terms with a $z$-integral without a $z$-dependent weight factor appear. These could be reduced and in some cases vanish in the infinite-volume limit due to the Ward-identity associated with current conservation. Such a modification would in general change the shape of the integrand, as well as its statistical variance. For definiteness, our lattice calculations are done precisely with the expressions given in this section.

%% file: lattice_setup.tex
\section{Numerical setup}\label{sec:latt_setup}

This section presents the gauge ensembles used in our calculation of $\ahlbl$, as well as the
different strategies we applied to compute the contributions of the different topology classes.

\subsection{Ensemble details}

\begin{figure}
  \centering
  \includegraphics{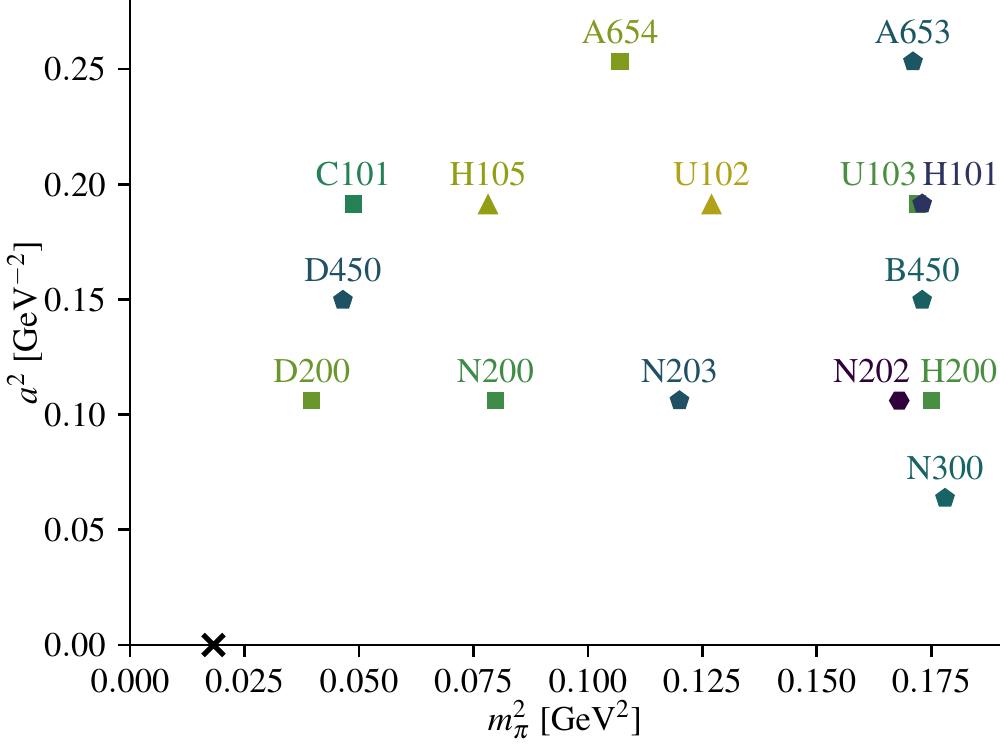}
  \caption{Ensembles used in this work, shown on the $(m_\pi^2,a^2)$ plane. The colour depends smoothly on $m_\pi L$, with darker colours corresponding to larger values, and the symbols have $\lfloor m_\pi L \rfloor$ sides. The cross indicates the physical point.}
  \label{fig:landscape}
\end{figure}

In this work we use $N_f=2+1$ O($a$)-improved Wilson fermion ensembles generated by the CLS initiative \cite{Bruno:2014jqa}, for which the improvement coefficient $c_{SW}$ was determined non-perturbatively in \cite{Bulava:2013cta}. We extend our previous work at the $\text{SU}(3)_f$-symmetric point to ensembles with $m_l< m_s$ down to pion masses of $200\text{ MeV}$, while maintaining $\Tr[M]=\text{Constant}$, with $M=\textrm{diag}(m_u, m_d, m_s)$ the quark mass matrix. In addition, we make use of two ensembles at a further, coarser lattice spacing.
We combine the symmetric-point results of our previous determination~\cite{Chao:2020kwq} with measurements taken from nine other ensembles to create a large data set from which all sources of systematic error can be estimated. 
Table~\ref{tab:ensembletab} summarises the gauge ensembles used, their pion and kaon masses, the lattice spacings, as well as the quark-mass dependent renormalisation factors, $\hat{Z}_V$. The latter is either measured directly as part of this work or taken from~\cite{Gerardin:2018kpy}, here we only use un-improved local vector currents in this work\footnote{With this setup, O$(a)$-discretisation effects might arise, but this turns out to be irrelevant to our target precision. Also, the difference between the relevant mass-dependent renormalisation factors is only at around the one-percent level.}.
The coverage of the lattice spacing and pion mass variables by the gauge ensembles used in this work is illustrated in Fig.~\ref{fig:landscape}.

\begin{table}
\begin{tabular}{cccccc|c|ccc|cc}
\toprule
Ensemble & (4) & (22) & (31) & (211) & (1111) & $\beta$ & $a^2\text{ [GeV]}^{-2}$ & $m_\pi^2 \text{ [GeV]}^2$ & $m_K^2 \text{ [GeV]}^2$ & $m_\pi L $ & $\hat{Z}_{\rm V}$ \\
\hline
A653 & $l,s$ & $l,s$ & 0 & 0 & 0 & \multirow{2}{*}{3.34} & \textbf{0.2532} & \textbf{0.171} & \textbf{0.171} & \textbf{5.31} & \textbf{0.70351} \\
A654 & $l,s$ & $l,s$ & $l$ & & & & \textbf{0.2532} & \textbf{0.107} & \textbf{0.204} & \textbf{4.03} & \textbf{0.69789} \\
\hline
U103 & $l,s$ & $l,s$ & 0 & 0 & 0 & \multirow{5}{*}{3.40} &0.1915 & 0.172 & 0.172 & 4.35 & 0.71562 \\
H101 & $l,s$ & $l,s$ & 0 & 0 & 0 & &0.1915 & 0.173 & 0.173 & 5.82 & 0.71562 \\
U102 & $l$ & $l$ & $l$ & & & & 0.1915 & 0.127 & \textbf{0.194} & 3.74 & 0.71226 \\
H105 & $l,s$ & $l,s$ & $l,s$ & & & & 0.1915 & 0.0782 & 0.213 & 3.92 & 0.70908 \\
C101 & $l,s$ & $l,s$ & $l,s$ & $l$ & $l,s$ & & 0.1915 & 0.0488 & 0.237 & 4.64 & 0.70717 \\
\hline
B450 & $l,s$ & $l,s$ & 0 & 0 & 0 & \multirow{2}{*}{3.46} & 0.1497 & 0.173 & 0.173 & 5.15 & 0.72647 \\
D450 & $l$ & $l$ & $l$ &   &   & & 0.1497 & \textbf{0.0465} & \underline{0.226} & \textbf{5.38} & \textbf{0.71921} \\
\hline
H200 & $l,s$ & $l,s$ & 0 & 0 & 0 & \multirow{5}{*}{3.55} & 0.1061 & 0.175 & 0.175 & 4.36 & 0.74028 \\
N202 & $l,s$ & $l,s$ & 0 & 0 & 0 & & 0.1061 & 0.168 & 0.168 & 6.41 & 0.74028 \\
N203 & & & $l$ & $l$ & & & 0.1061 & 0.120 & 0.194 & 5.40 & 0.73792 \\
N200 & $l$ & $l$ & $l$ & & & & 0.1061 & 0.0798 & 0.214 & 4.42 & 0.73614 \\
D200 & $l$ & $l$ & $l$ & & & & 0.1061 & 0.0397 & 0.230 & 4.15 & 0.73429 \\
\hline
N300 & $l,s$ & $l,s$ & 0 & 0 & 0 & \multirow{1}{*}{3.70} & 0.06372 & 0.178 & 0.178 & 5.11 & 0.75909 \\
\botrule
\end{tabular}
\caption{Details of the ensembles used to compute the various contributions to $\ahlbl$. Lattice spacings were determined in \cite{Bruno:2016plf}, apart from the ``A'' ensembles, where the lattice spacing was estimated from ratios of the Wilson flow parameter $t_0$ at the flavour-symmetric point. Pion and kaon masses primarily come from \cite{Gerardin:2019rua} unless directly measured as part of this work (indicated in \textbf{bold}) or in a recent project~\cite{sin2thetaw} (underlined). Likewise, values of $\hat{Z}_{\rm V}$ can be obtained from \cite{Gerardin:2018kpy} unless also measured as part of this project, using the same approach. Columns two through six indicate the flavour content computed for each class of diagrams: fully connected $(4)$, leading disconnected $(2+2)$, and subleading $(3+1)$, $(2+1+1)$, and $(1+1+1+1)$, where ``$+$'' has been omitted for space reasons. Zeros indicate diagrams that vanish by SU(3) flavour symmetry.}\label{tab:ensembletab}
\end{table}

\subsection{Computational strategies and numerical cost} 

Tab.~\ref{tab:cost} illustrates the number of gauge configurations used for our study and the number of point-source propagator inversions per configuration performed for both the fully-connected and $(2+2)$ disconnected. For the disconnected, over an order of magnitude higher statistics was used in comparison to the connected. Typically we favor having larger multiplicities of $|y|$ per configuration, as opposed to a larger number of configurations for the disconnected piece, since it is more effective at reducing the noise. Ideally the number of self-averages per-$|y|$ range in the thousands per configuration, overall the total statistics (configurations$\times$sources$\times$self-averages) lies in the low millions per point. We follow the same setup as in \cite{Chao:2020kwq}, building a grid of point sources in such a way as to maximise the number of self-averages available per $|y|$.

Although the total number of propagator solves per ensemble are comparable, the computational cost per solve as the pion mass is reduced grows significantly, as the lattice volume increases such that $m_\pi L \geq 4$, and generically the cost of a solve grows like $\frac{V^n}{m_\pi^m}$ with $m$ and $n$ both being greater than unity. Although we used a particularly sophisticated propagator-solving routine \cite{Luscher:2007se}, this prohibitive growth in cost is presently unavoidable.

To partially ameliorate the overhead from propagator solves, a truncated solver/AMA technique \cite{Bali:2009hu,Blum:2012uh} was used for all of the $(2+2)$ contributions on ensembles away from the $\text{SU}(3)_f$-symmetric point, with a sloppy stopping criteria of $10^{-3}$ on the norm of the residual. As the propagator solve cost was dominant in the $(2+2)$ calculation, a sloppy solve on one of our most expensive ensembles (D200) was approximately $6\times$ faster than a high-precision solve to $10^{-10}$.

\begin{table}
\begin{tabular}{c|cc|cc|c|cc|cc}
\toprule
\multirow{2}{*}{Ensemble} & \multicolumn{2}{|c|}{Fully-connected} & \multicolumn{2}{|c|}{$(2+2)$} & \multirow{2}{*}{Ensemble} & \multicolumn{2}{|c|}{Fully-connected} & \multicolumn{2}{|c}{$(2+2)$} \\
\cline{2-5}
\cline{7-10}
 & $N_\text{conf}$ & $N_\text{solve}$ & $N_\text{conf}$ & $N_\text{solve}$ & & $N_\text{conf}$ & $N_\text{solve}$ & $N_\text{conf}$ & $N_\text{solve}$ \\
\hline
A653 & $1258$ & $24$ & $628$ & $576$ & B450 & $402\times 4$ & $16$ & $1611$ & $128$ \\
A654 & $5007$ & $24$ & $629$ & $576$ & D450 & $500\times 4$ & $64$ & $500$ & $2048$ \\
\hline
U103 & $529\times 4$ & $12$ & $1030$ & $360$ & H200 & $250\times 4$ & $16$ & $500$ & $272$ \\
H101 & $250\times 4$ & $16$ & $1008$ & $272$ & N202 & $225\times 4$ & $24$ & $450$ & $816$ \\
U102 & $890\times 6$ & $12$ & $750$ & $720$ & N200 & $856\times 3$ & $24$ & $856$ & $816$ \\
H105 & $1027\times 3$ & $16$ & $1027$ & $448$ & D200 & $250\times 11$ & $32$ & $500$ & $1536$ \\
\cline{6-10}
C101 & $2000$ & $24$ & $500$ & $841$ & N300 & $384\times 4$ & $17$ & $384$ & $600$ \\
\botrule
\end{tabular}
\caption{Statistics gathered for the fully-connected and $(2+2)$ disconnected contributions. For the fully connected different hypercubically-equivalent orientations were used (hence the $\times$). $N_\text{conf}$ indicates the number of gauge configurations used and $N_\text{solve}$ indicates the number of propagator inversions performed per $N_\text{conf}$. The $\text{SU}(3)_f$-symmetric point ensembles' data was already used in \cite{Chao:2020kwq}, although an update for the $(2+2)$ on N202 has been performed here and the coarse ensemble A653 has been added.}\label{tab:cost}
\end{table}

\subsection{Higher-order contributions}

For the quark loops containing a single (local) vector current insertion,
we make use of an extensive general-purpose data set generated as part of a different project~\cite{sin2thetaw}.
Therefore we restrict our description of the computational aspects related to these loops to those directly relevant
to the Hlbl calculation. Since we are dealing exclusively with the electromagnetic current, it is always the difference
of a light and a strange quark loop that is needed. To compute this difference, the ``one-end trick'',
which has been applied extensively in twisted-mass fermion calculations~\cite{Jansen:2008wv,McNeile:2006bz}, is used 
as proposed in Ref.~\cite{Giusti:2019kff}.
The one-end trick yields an efficient estimator for the required difference of Wilson-quark loops based on the identity
\begin{equation}
 \mathrm{tr}\left[\gamma_\mu (S^{l}(x,x) - S^{s}(x,x))\right] = (m_{s}-m_l) \sum_y\mathrm{tr} \left[\gamma_\mu S^{l}(x,y) S^{s}(y,x)\right] \,.
 \label{eq:oet}
\end{equation}
The right-hand side of this equation is evaluated using
stochastic volume sources, inserted between the two propagators, without spin or color dilution.
In this way, gauge noise is reached after a few hundred sources at most.
The stochastic estimate of the quantity (\ref{eq:oet}) is averaged over blocks of individual volume
sources, leading to four ``effective'' sources that are stored separately as entire fields. Having access to four
effective sources is sufficient to compute all higher-order disconnected
diagrams for $\ahlbl$ without introducing any bias into the
final result.  For further technical details of the general computational setup we refer to
the description in Ref.~\cite{sin2thetaw}. \par

The parametrisations of Eq.~\eqref{formalism:eq:f3p1}, Eq.~\eqref{formalism:eq:f2p1p1}, and Eq.~\eqref{formalism:eq:f1p1p1p1} share certain $x$- or $z$-integrals, which allows us to precompute and recycle these terms for the different contributions.
For the $(3+1)$ contribution, the \textit{triangle} term defined in Eq.~\eqref{formalism:eq:triangle} and the terms derived from it can be conveniently obtained from the intermediate quantities in the calculation of the fully-connected contribution.
Based on this observation, we choose for the $(3+1)$ topology the same set of points for our origin and $y$-vector as for the fully-connected. 
Once the factorised terms in Eq.~\eqref{formalism:eq:f3p1} are computed, the Lorentz contraction with the terms which contain a disconnected loop can be performed off-line as a post-processing step. 

For the $(2+1+1)$ contribution, we first compute and save the lattice-wide two-point functions, Eq.~\eqref{formalism:eq:pi2}, for each source position, and then do the VEV subtraction and construct Eq.~\eqref{formalism:eq:f2p1p1} again off-line.
The sources are chosen to be the same set of points as for the fully-connected case.
Nonetheless, after setting the origins at these source points, our parametrisation Eq.~\eqref{formalism:eq:f2p1p1} still allows us to have many choices for the $y$-vector for a given $|y|$, because we have at our disposal the two-point function Eq.~\eqref{formalism:eq:pi2} and the disconnected loop Eq.~\eqref{formalism:eq:disc_loop} as entire lattice fields. 

A good choice for the $y$-vectors is hence to pick from the elements on the same orbit under the cubic group.
As an example, to obtain $(|y|/a)^2 = 12$, one can choose the 4-vector $y$ to be $(a,b,c, d)$ with $a,b,c\in\{-1,1\}$ and $d\in \{-3, 3\}$, if all these points fit in a range where boundary effects can be neglected. 
A summary of the choices of the $y$-vector for the ensembles used for the $(2+1+1)$ computation is given in Table~\ref{lattice_setup:tab:2p1p1}. Likewise, the $(1+1+1+1)$ calculation also benefits from this strategy because of the reuse of the data generated for the $(2+1+1)$ integrand.

\begin{table}
  \begin{tabular}{c|c|c}
  \toprule
  id & $N_{\rm{conf}}$ & $y=(a,b,c,d)$ \\
  \hline
  C101 & 1000 & $a,b,c\in \{-n,n\}$, $d\in\{-3n,3n\}$  \\
  N203 & 376 $\times$ 4 & $a,b,c\in\{-2n, 2n\}$, $d=0$ \\
  \botrule
  \end{tabular}
  \caption{Choice of the $y$-vectors and statistics for the (2+1+1). Here, $n$ is an integer.}\label{lattice_setup:tab:2p1p1}
\end{table}

%% file: integrand.tex
\section{The integrand of the two dominant contributions \label{sec:intgnd}}

In this section, we describe the integrands of the light connected and
light (2+2) disconnected contributions obtained in our lattice QCD calculations. Our
goal is on the one hand to present some of the available data at small pion masses, and on the
other to compare it to the predictions of hadronic models, such as the
$\pi^0$ exchange contribution.  Finally, an observation on the
approximate analytic form of the integrand for the latter
contribution motivates the analysis of the lattice data 
presented in the next section.

\begin{figure}[h!]
  \includegraphics[scale=0.64]{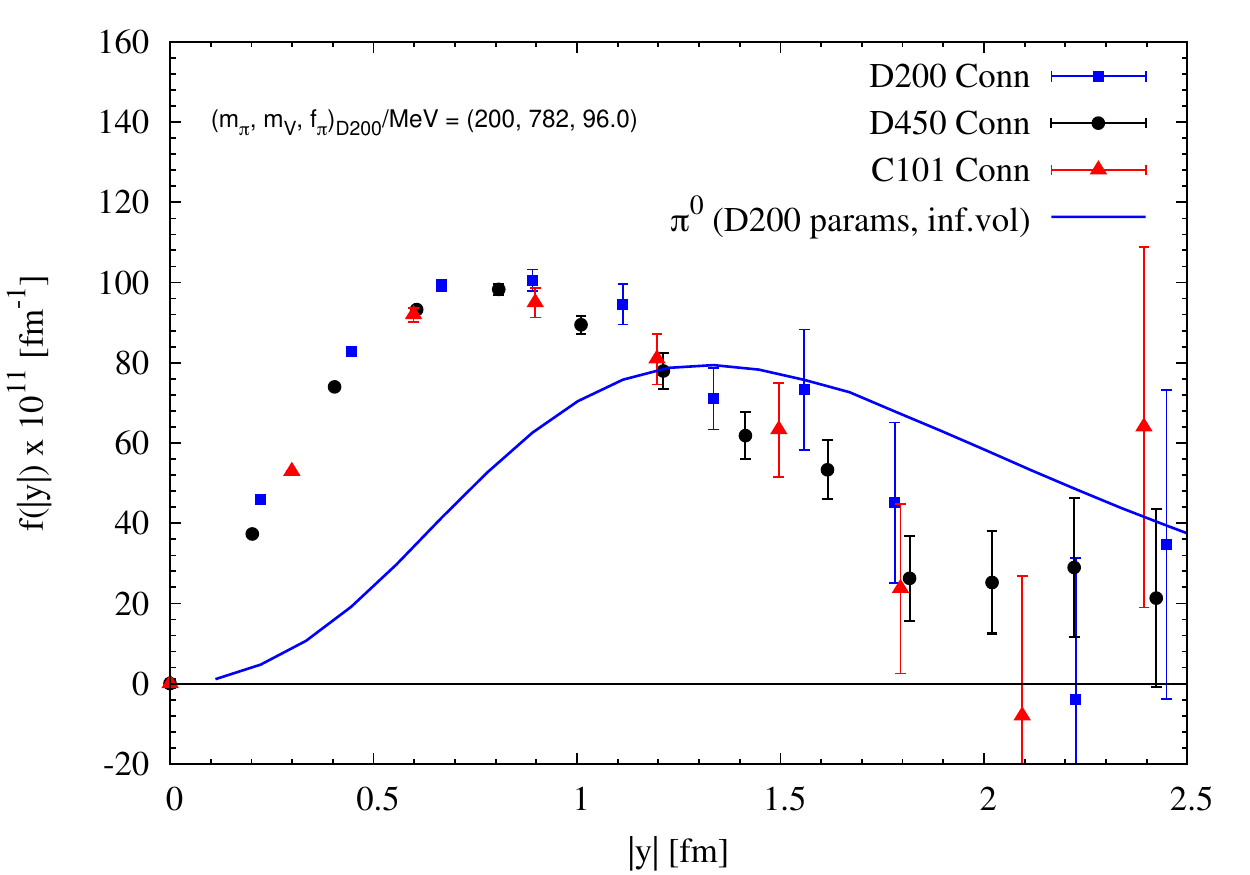}
  \includegraphics[scale=0.64]{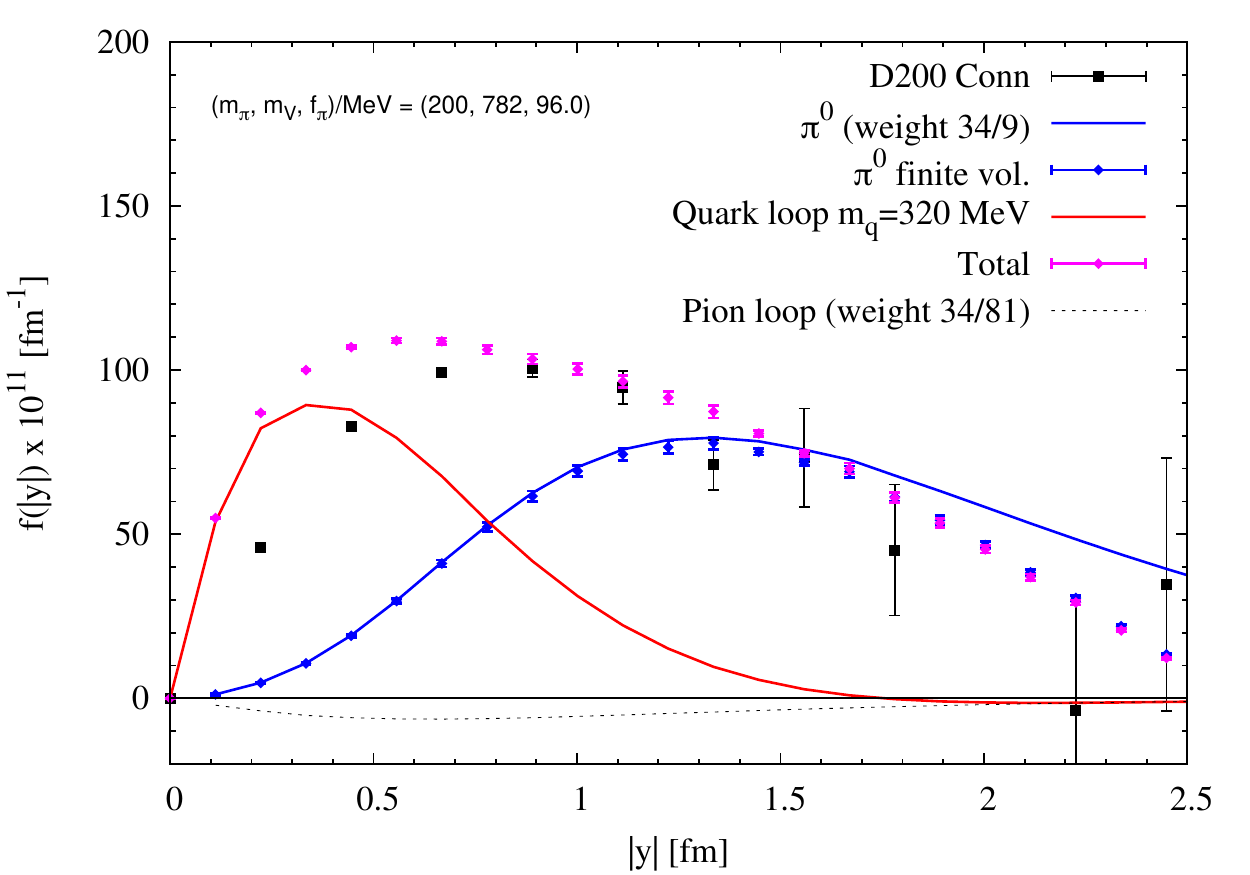}
  \caption{Left: The light connected contribution on the three most chiral ensembles.
    %As a first comparison to a hadronic model prediction,
    The solid curve represents the $\pi^0$ exchange in infinite volume, computed with the
     parameters directly determined on ensemble D200~\cite{Gerardin:2019vio}.
    Right: The light connected contribution on ensemble D200, compared to the predictions of the $\pi^0$ exchange
    (with a VMD transition form factor), the constituent quark loop, as well as the charged pion loop.
    The latter two contributions are computed  within spinor and scalar QED, respectively.
  %  The dashed curve  shows an approximate representation of the latter by an elementary function.
  }\label{fig:conn_chiral_ens}
\end{figure}

We begin with the left panel of Figure~\ref{fig:conn_chiral_ens}, showing an overview of the
integrand of the light connected contribution for our three most chiral ensembles (C101, D450, D200), for
which the pion mass lies in the interval 200 to 220\,MeV. These three
ensembles have different lattice spacings and different volumes,
nevertheless the corresponding data points fall within one recognisable band.
The maxima of these integrands, which lie between 0.7 and 0.9\;fm,
are followed by a slow fall-off. Beyond $|y|=2$\,fm, the integrand vanishes
within the uncertainties. The height of the maximum is about 20\% higher than
at the SU(3)-flavour symmetric point~\cite{Chao:2020kwq}, $m_\pi=m_K\approx 420\,$MeV.

\begin{figure}[t!]
  \includegraphics[scale=0.62]{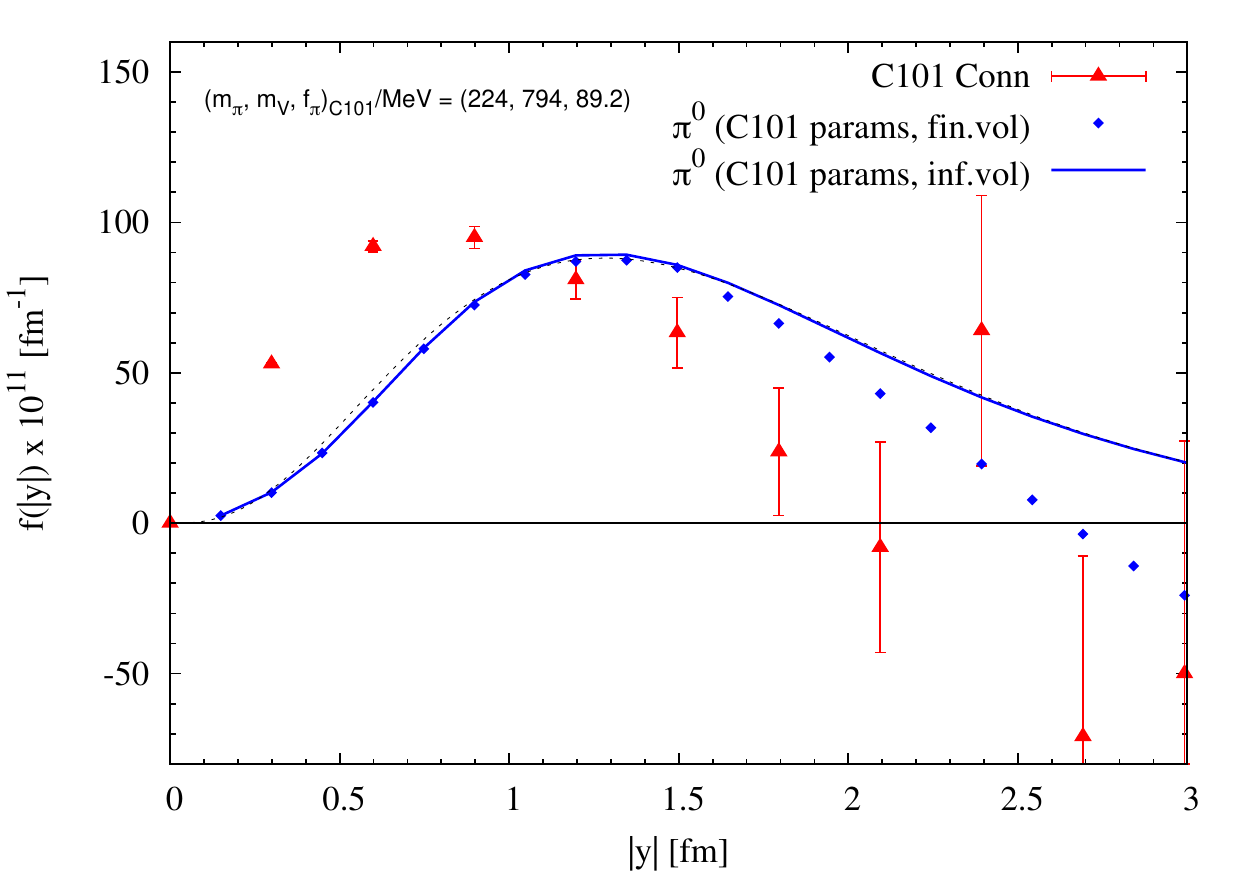}
  \hspace{-8pt}
  \includegraphics[scale=0.62]{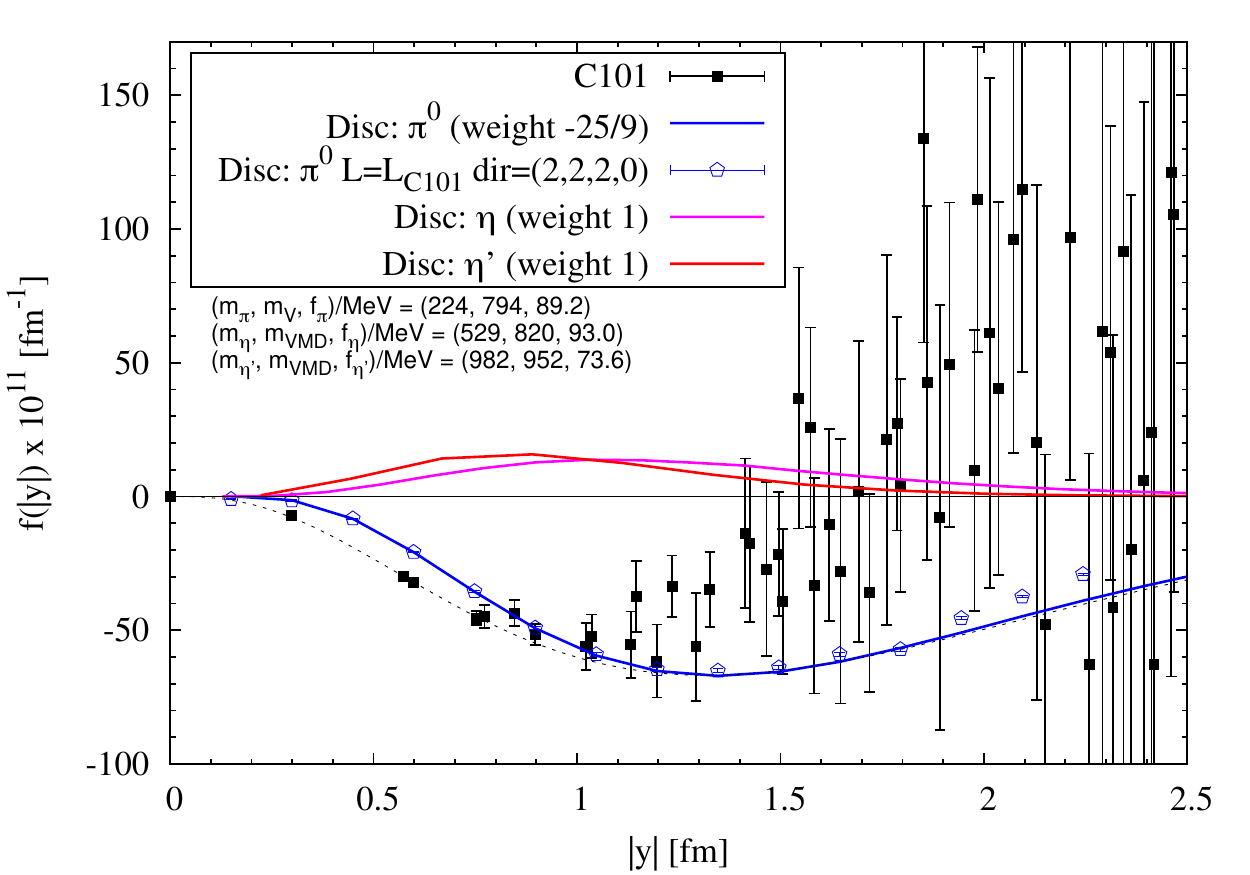}
           \caption{The connected (left) and the (2+2) disconnected (right) contributions on ensemble C101, compared
             to pseudoscalar-meson exchange contributions in infinite volume (continuous curves), as well
             as the $\pi^0$ exchange contribution in finite volume (blue points). 
             The dashed curve shows an approximate representation of the infinite-volume $\pi^0$ exchange integrand
             by the function $A|y|^3\exp(-B|y|)$, with $(A[{\rm fm}^{-4}],B[{\rm fm}^{-1}])=(840,\;2.34)$
             in the connected and $(-582,\;2.27)$ in the disconnected case.}
           \label{fig:C101intgnds}
\end{figure}

Figure~\ref{fig:C101intgnds} focuses on the data of ensemble C101.
The connected and (2+2) disconnected data are displayed separately in
the two panels.  The disconnected integrand is negative and admits a
minimum at $|y|\approx 1.2\,$fm.  The signal degrades sooner than in
the connected case, and is lost around 1.5\,fm.  The ordinate of the
minimum is about twice as large as the one found on ensemble H101 at
the SU(3)-flavour symmetric point~\cite{Chao:2020kwq}, despite the fact
that the latter case includes the strange quark, so that this contribution is weighted with the
electric-charge factor $36/81$ rather than $25/81$.
Thus we
anticipate a very strong chiral dependence of the (2+2) disconnected
contribution to $\ahlbl$.

Figure~\ref{fig:C101intgnds} also compares the integrand to
pseudoscalar-exchange predictions based on the vector-meson dominance
(VMD) parametrisation of the corresponding transition form factor.
As weight factors with which $(\pi^0,\eta,\eta')$ contribute
to the connected diagrams, we have used $(34/9,0,0)$; the weight factors
we have applied for the disconnected diagrams are $(-25/9,1,1)$.
While these weight factors are certainly the expected ones for the $\pi^0$,
the issue of which weight factors are appropriate for the isoscalar mesons is more complicated and
depends in particular on their mixing; see Tab.~\ref{tab:simple_results}
and the corresponding analysis presented in appendix, as well as Refs.\ \cite{Bijnens:2016hgx,Gerardin:2017ryf}.
For the $\pi^0$
exchange, the contribution has also been computed in finite volume.
As can be seen on the left panel, the finite-volume connected
integrand is predicted to dive towards negative values at long
distances.  Whether the lattice data does the same is uncertain due to
the growing statistical errors. The lattice data points lie below the
$\pi^0$-exchange prediction. A very similar observation was made at
the SU(3)-flavour symmetric point~\cite{Chao:2020kwq}. We do not have a clear explanation
for the difference, but note that for ensemble
D200, we observe a somewhat better agreement (see the right panel of figure
\ref{fig:conn_chiral_ens} discussed below).  For the disconnected
contribution, the finite-size effect on the integrand are predicted to
become significant only around $|y|=2\,$fm, which is beyond the useful
range of our lattice data. The $\eta$ and $\eta'$ contributions have been
estimated very roughly by using the parameters indicated in the figure.
The $\eta$ mass estimate comes from using the Gell-Mann--Okubo formula,
knowing the pion and kaon masses, while the same $\eta'$ parameters are used
as in~\cite{Chao:2020kwq}. The two isoscalar mesons contribute comparably to $\ahlbl$.
In the region between 0.8 and 1.2\,fm, the $\pi^0$-exchange prediction is consistent with the lattice data.

While the disconnected contribution does not have a strong
short-distance contribution, the connected contribution
does. Following~\cite{Chao:2020kwq}, we attempt to explain the
integrand semi-quantitatively by combining a constituent quark loop
with the long-distance contributions, i.e.\ the $\pi^0$ exchange and
the charged pion loop.  The right panel of figure~\ref{fig:conn_chiral_ens} illustrates the
comparison of this rough hadronic model with the lattice data. The
quark loop as well as the pion loop are calculated in the spinor and
scalar QED frameworks respectively, i.e.\ without the inclusion of
form factors.  Including the quark loop leads to a satisfactory
description of the shape of the integrand, even though the total
prediction overshoots the data at distances $|y|\lesssim0.6\,$fm. This
difference can partly be explained by the cutoff effects present in
the data, which tend to reduce the lattice integrand, and it is likely
that including a form factor for the constituent quarks would improve
the agreement. At distances $|y|\gtrsim0.9$\,fm, the model prediction
is consistent with the lattice data.

In summary, both in the connected and the disconnected case, the
prediction for the $\pi^0$ exchange alone gives a good first estimate
of the magnitude of the integrand.  It also predicts the approximate
shape of the integrand in the disconnected case.  Hence it is worth
asking whether the integrand for the $\pi^0$ exchange can be
approximated by a simple analytic function. Figure
\ref{fig:C101intgnds} shows that the infinite-volume $\pi^0$-exchange
integrand can be approximated very well at its extremum and beyond
with the form $f(|y|) = A|y|^3 \exp(-B|y|)$, displayed as a dashed
line.  In the connected case, the approximation holds to an excellent
degree even at short distances. These observations,
which apply to our specific choice of kernel $\kernel^{(\Lambda)}$,
form part of our motivation
to use this functional form in the next section to extend the
integrand obtained in lattice QCD to long distances.

%% file: light_contributions.tex
\section{Light-quark fully-connected and \texorpdfstring{$(2+2)$}{(2+2)} contributions}\label{sec:leading_contributions}

In this section, we describe the extraction of the dominant contributions, namely the light-quark fully-connected and (2+2) contributions.
In the previous section, the integrands are illustrated and compared semi-quantitatively to the main known hadronic contributions.
The rapid increase of the relative error on the integrand with growing $|y|$ leads us to employ a method to extend
the useful range of the data. 
In our previous calculation~\cite{Chao:2020kwq}, the long-distance tail was assumed to come from $\pi^0$, $\eta$, and $\eta'$-exchange contributions, with the dominant $\pi^0+\eta$ part determined from a lattice calculation of the $\pi^0\gamma\gamma$ transition form factor~\cite{Gerardin:2019vio}.
The fact that the integrand of the $\pi^0$ exchange itself is well described by a simple functional form has led us to adopt a more
self-contained and data-driven approach, which relies on extending the tail via a fit to the data.
In both the connected and $(2+2)$ contributions, we perform a fully-correlated fit to the data from each ensemble using the ansatz
\begin{equation}\label{eq:y3exp}
f(|y|) = |y|^3 Ae^{-B|y|},
\end{equation}
where $A$ and $B$ are free parameters. In the intermediate $|y|$ regime, this fit form describes all of our data well, with $\chi^2/\text{dof}$ close to 1. As our data become noisy at large $|y|$, the fit significantly reduces the error for the integral of the long-distance tail. In our approach, we will choose a cutoff distance: below it, we will numerically integrate the lattice data using the trapezoid rule; above it, we will switch to integrating Eq.~\eqref{eq:y3exp}. The cutoff is chosen to be a point where the integrated $a_\mu$ exhibits stability. The values of $a_\mu$ from all ensembles will then be used in a combined chiral, infinite-volume, and continuum extrapolation.
In particular, while in~\cite{Chao:2020kwq} the volume effects were corrected for on each ensemble
using the prediction for the $\pi^0+\eta$ exchange, here we rely on a global fit to all ensembles to eliminate these effects,
with an ansatz for the $L$-dependence motivated by the same meson exchange.

\begin{figure}
\includegraphics[scale=0.3]{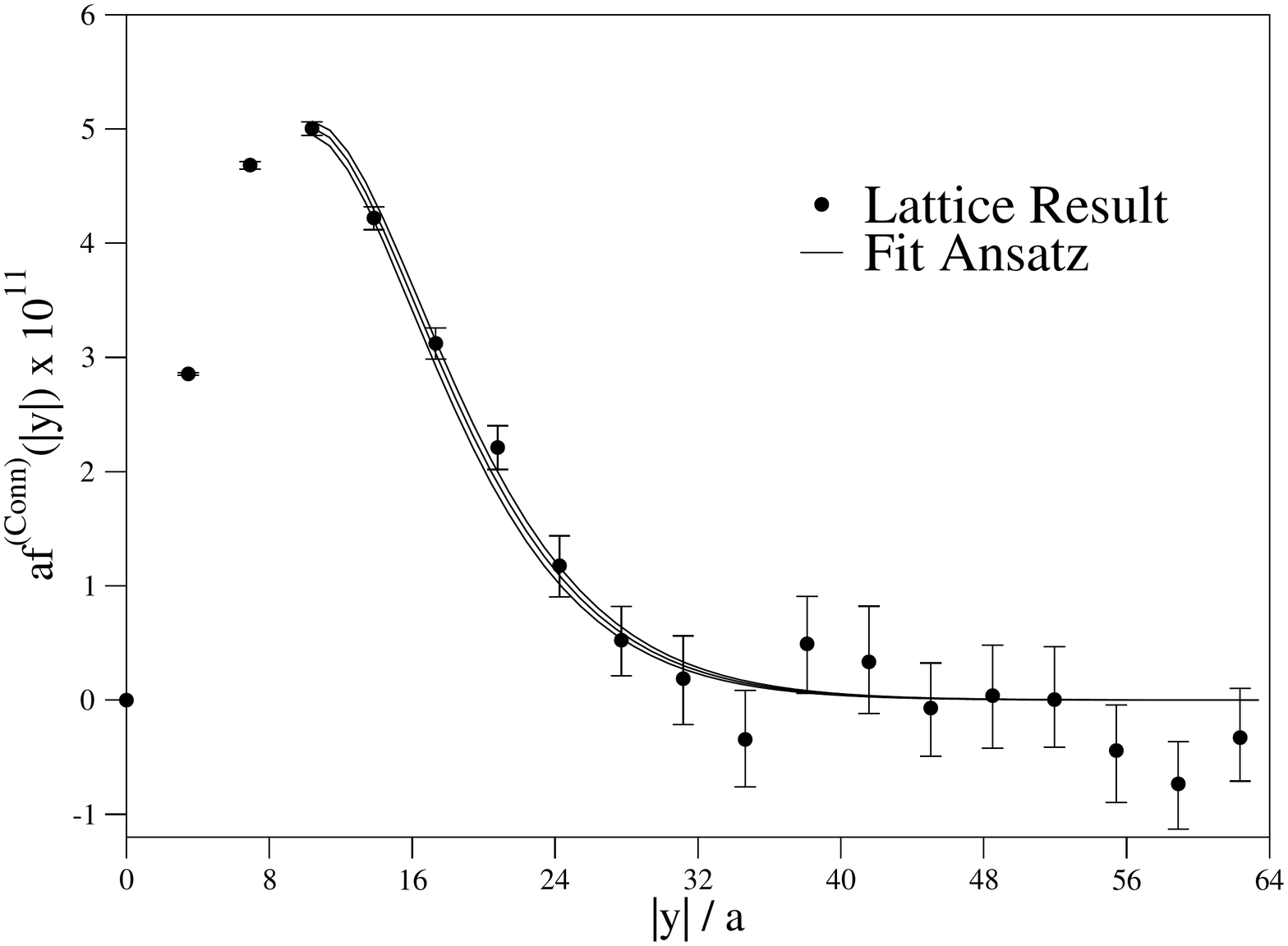}
\caption{An example of our fit ansatz Eq.~\eqref{eq:y3exp} for the fully-connected contribution from ensemble N202.}\label{fig:N202fit}
\end{figure}

Fig.~\ref{fig:N202fit} shows an example of our ability to describe the lattice data with Eq.~\eqref{eq:y3exp} for the fully-connected contribution. The displayed data~\cite{Chao:2020kwq}, corresponding to ensemble N202, are among the most precise at the $\text{SU}(3)_f$-symmetric point, and the correlated fit has a $\chi^2/{\rm dof}$ of 1.1. The figure illustrates that the data is very well described by our ansatz all the way to the point where the data fluctuates around zero and the signal is likely lost.

\begin{table}
\begin{tabular}{c|cc|c}
\toprule
Ensemble & Connected$\times 10^{11}$ & $(2+2)\times 10^{11} $ & Sum$\times 10^{11}$\\
\hline
A653 & 64.5(1.0) & $-31.8(2.8)$ & 32.7(3.1) \\
A654 & 79.4(1.8) & $-36.7(4.8)$ & 42.6(5.3) \\
\hline
U103 & 59.3(0.9) & $-22.1(4.3)$ & 37.2(4.3)\\
H101 & 70.2(1.8) & $-30.3(4.4)$ & 39.9(4.8)\\
U102 & 66.5(1.2) & $-23.6(2.5)$ & 42.9(2.9) \\
H105 & 92.9(2.8) & $-46.8(5.5)$ & 46.1(6.0) \\
C101 & 127.7(5.6) & $-62.2(6.6)$ & 65.5(9.0) \\
\hline
B450 & 70.4(1.3) & $-27.8(8.2)$ & 42.6(8.3) \\
D450 & 144.5(11.9) & $-84.8(14.5)$ & 59.7(20.0)\\
\hline
H200 & 65.3(1.3) & $-19.5(3.4)$ & 45.8(3.5) \\
N202 & 83.2(2.0) & $-28.5(3.0)$ & 54.8(3.6) \\
N200 & 116.9(4.9) & $-54.2(6.3)$ & 62.7(7.9) \\
D200 & 151.7(9.6) & $-80.4(13.4)$ & 71.4(16.6) \\
\hline
N300 & 75.7(1.3) & $-15.8(2.6)$ & 59.9(2.9) \\
\botrule
\end{tabular}
\caption{The two leading light-quark contributions to $\ahlbl$ for each gauge ensemble.}\label{tab:light_results}
\end{table}

Table~\ref{tab:light_results} summarises our results for the two leading sets of diagrams containing only light quarks.
For the extrapolation to the infinite-volume, physical pion mass, and continuum limit for both the fully-connected and $(2+2)$ disconnected contributions we use the following ansatz,
\begin{equation}\label{eq:fvol3}
a_\mu(m_\pi^2,m_\pi L,a^2) = A\, e^{-m_\pi L/2} + B\, a^2 + C\; S(m_\pi^2) + D + E\, m_\pi^2 \;,
\end{equation}
where we have identified several candidates for the non-analytic function $S(m_\pi^2)$,
\begin{equation}\label{eq:singterm}
  \begin{aligned}
    \text{Pole :: } &\frac{1}{m_\pi^2} \\
    \text{Log :: }  &\log{m_\pi^2}  \\
    \text{Log2 :: } &\log^2{\left(m_\pi^2\right)}  \\
    \text{m2Log :: } &m_\pi^2 \log{\left(m_\pi^2\right)}\;.  \\
  \end{aligned}
\end{equation}
These functions are inspired by the divergent chiral behaviors at the large-$N_c$ limit of the pion-pole exchange and the charged-pion loop contribution~\cite{Prades:2009tw}.

\begin{figure}[ht!]
  \includegraphics[scale=0.27]{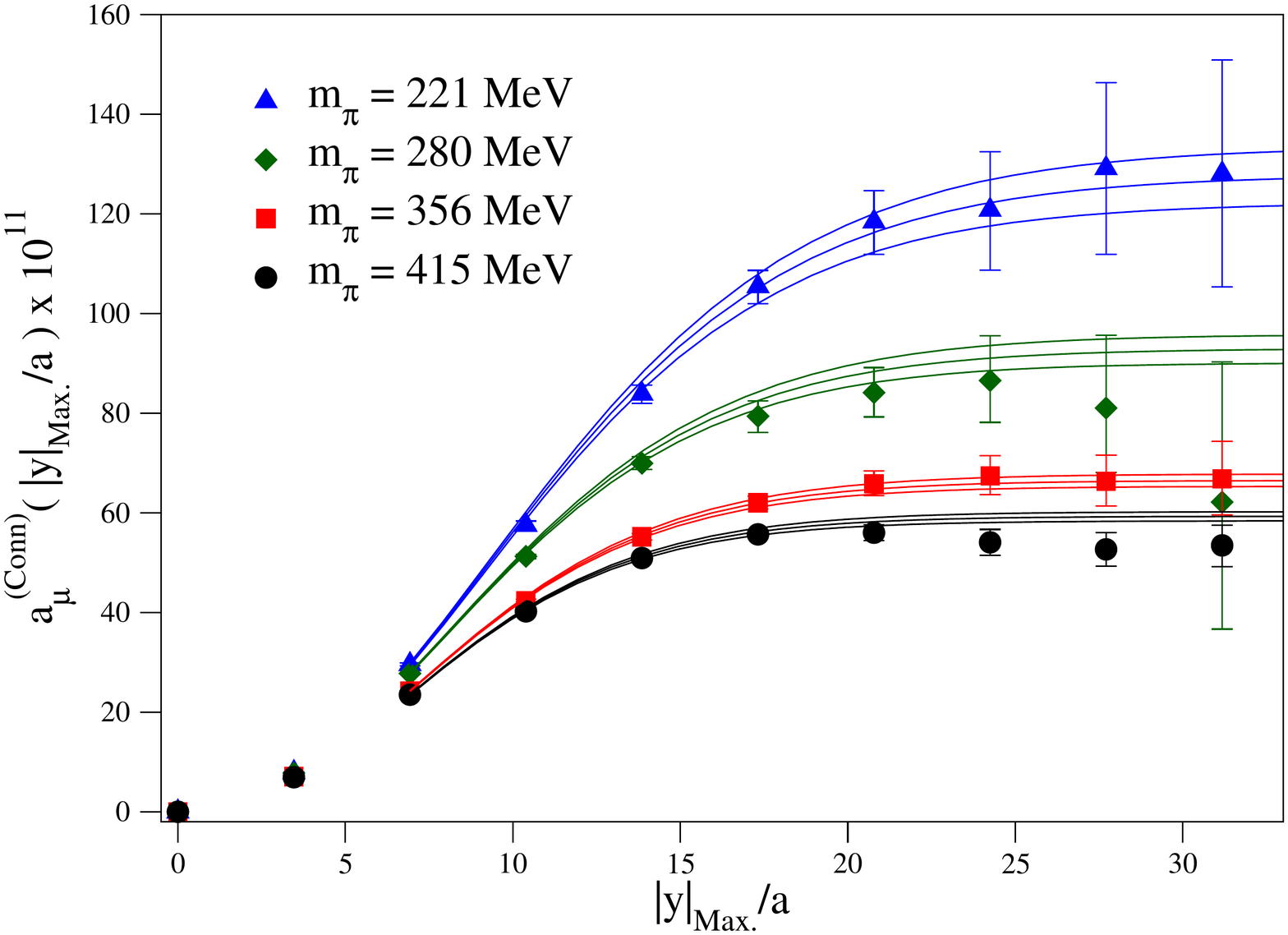}
  \hspace{-8pt}
  \includegraphics[scale=0.27]{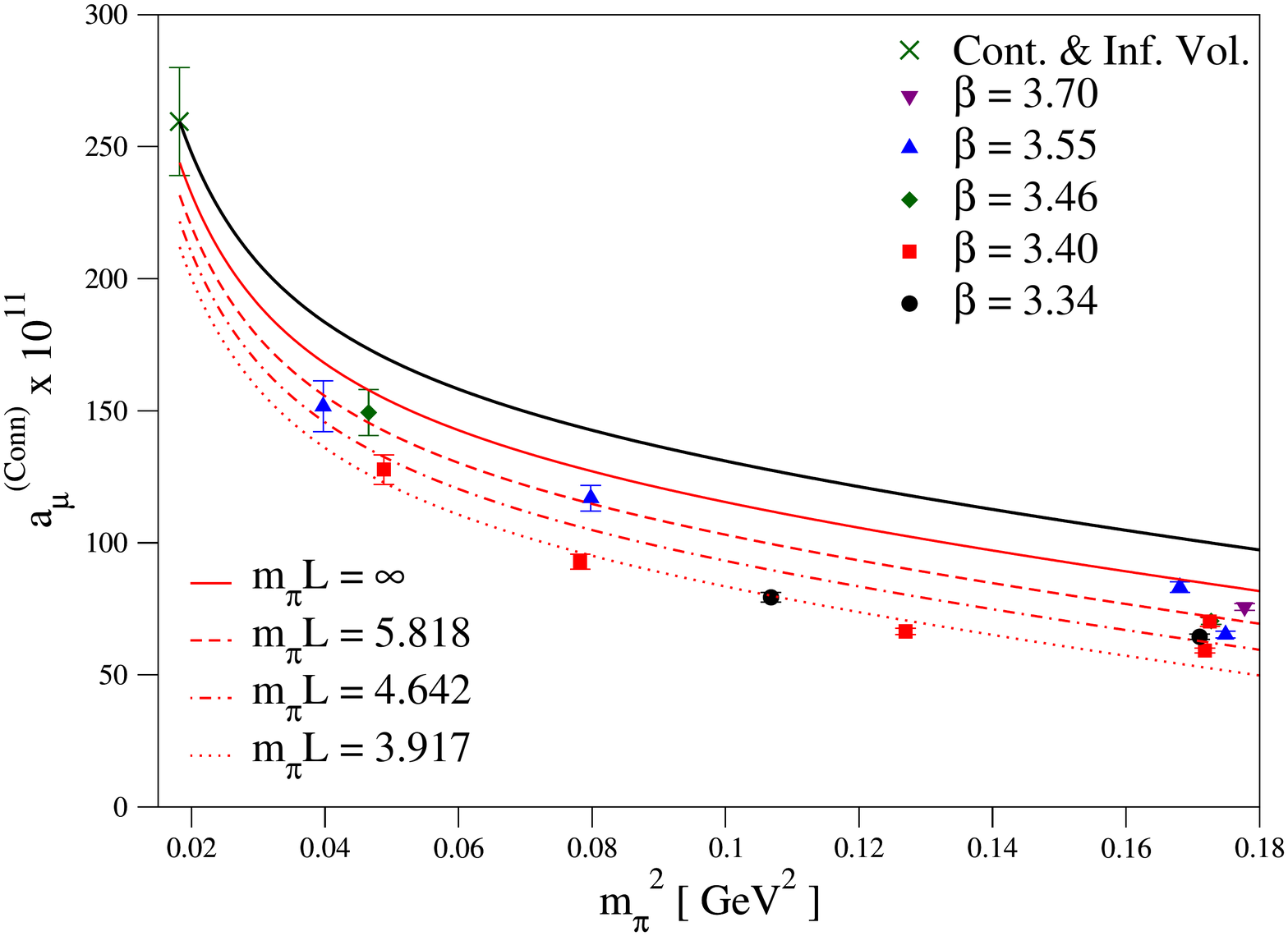}
  \caption{Left: Partially-integrated light-quark connected contribution to $\ahlbl$ versus $|y|_\text{Max.}/a$ for ensembles C101, H105, U102, and U103, which have a broad range of pion masses but the same lattice spacing and similar $m_\pi L$. The points are the numerically integrated lattice data and the curves result from switching the integrand to the fit of Eq.~\eqref{eq:y3exp} above a cutoff. Right: Chiral, continuum, and infinite-volume extrapolation of the light-quark connected contribution using the Pole ansatz, shown versus $m_\pi^2$. The points are lattice data at finite $L$ and nonzero $a$, and the cross indicates the extrapolated result at physical pion mass. Curves show the dependence on $m_\pi^2$ for fixed $a$ and $m_\pi L$, with the black curve corresponding to the continuum and infinite volume. The four red curves have different values of $m_\pi L$ but the same lattice spacing corresponding to $\beta=3.40$; three of them correspond to $(a,m_\pi L)$ of ensembles H101, C101, and H105. }\label{fig:connplot} 
\end{figure}

\begin{figure}[h!]
  \includegraphics[scale=0.27]{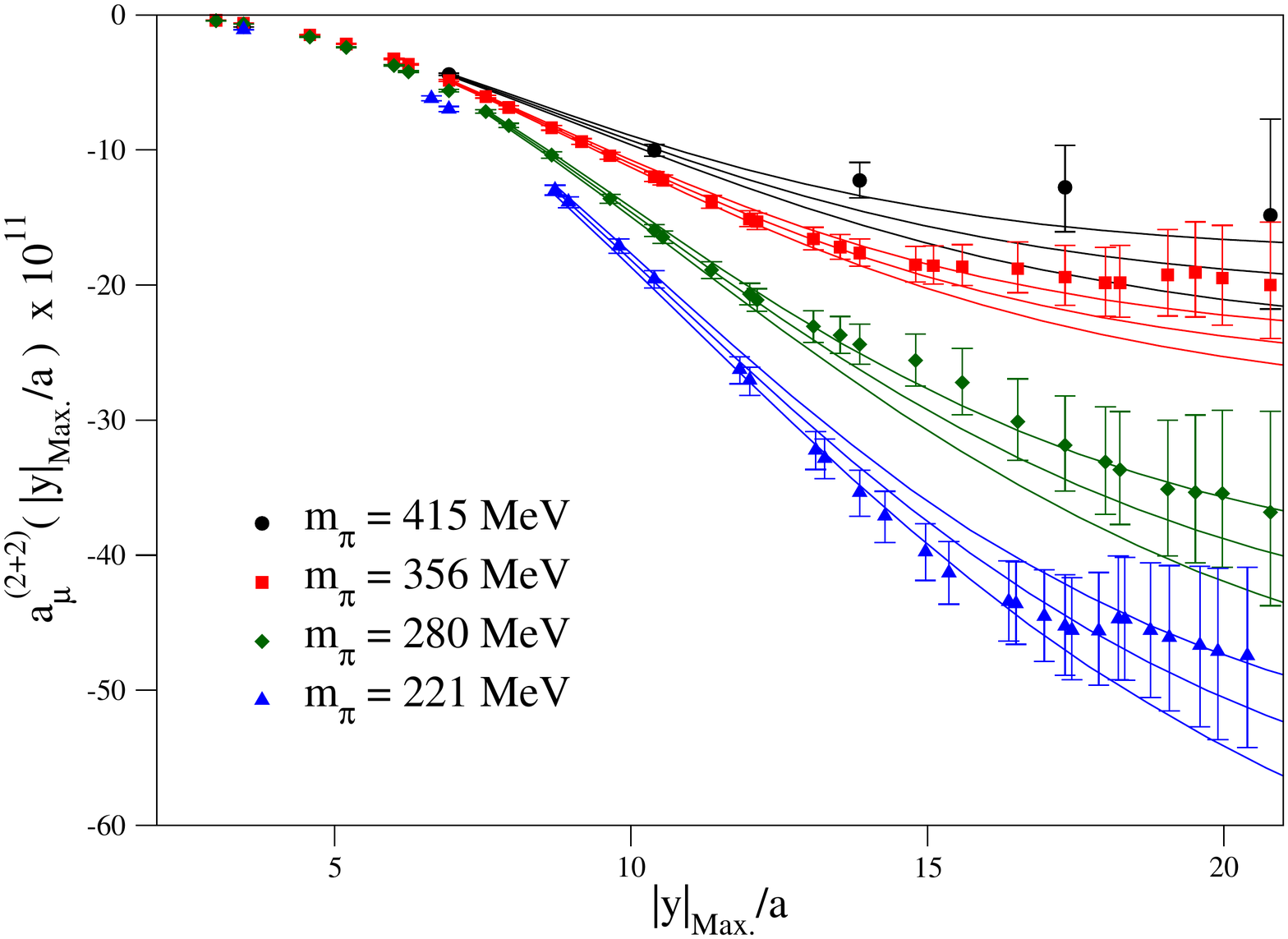}
  \hspace{-8pt}
  \includegraphics[scale=0.27]{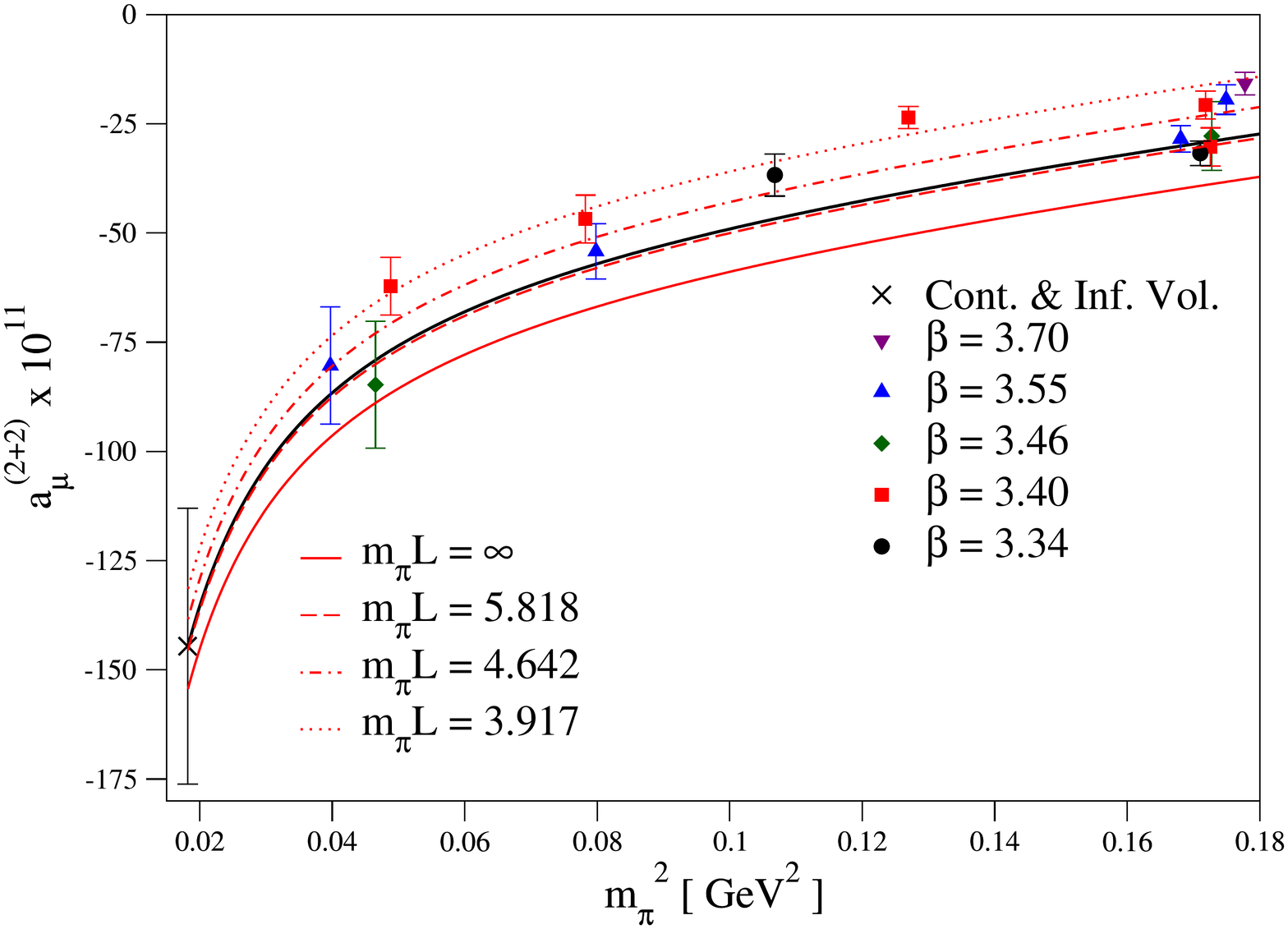}
  \caption{Light-quark $(2+2)$ disconnected contribution to $\ahlbl$. See the caption of Fig.~\ref{fig:connplot}.}\label{fig:discplot}
\end{figure}

\subsection{Light-quark fully-connected results}

The left plot of Fig.~\ref{fig:connplot} uses the partially-integrated $a_\mu(|y|_\text{Max.})$ defined in Eq.~\eqref{eq:amuymax} to illustrate the growth in the size of the connected contribution with decreasing pion mass. Here we consider a constant lattice spacing ($a=0.0864$\,fm) and include data from similar $m_\pi L$ to help isolate the chiral behaviour. The curves begin at the cutoff where we switch to integrating the fitted Eq.~\eqref{eq:y3exp}, with the trapezoidal-rule integrals of the lattice data up to the cutoff added to them. The fit adequately reproduces the lattice data above the cutoff, and saturates where the trapezoidal-rule integral does, within the uncertainties. At large $|y|_\text{Max.}$, some of the lattice points begin to drop below where the fit asymptotes; this is probably a mixture of finite volume effects and loss of signal in the integrand.

The right plot of Fig.~\ref{fig:connplot} shows a combined chiral, infinite-volume, and continuum limit extrapolation based on the ``Pole'' ansatz. The horizontal axis is $m_\pi^2$, and we illustrate the dependence of the global fit on $a$ and $m_\pi L$ via curves that show the fit at various fixed $(a,m_\pi L)$. The result increases along all three dimensions of the extrapolation (larger volumes, finer lattice spacings, and lighter pion masses), which produces a large combined effect.

\subsection{Leading light-quark disconnected results}

The $(2+2)$ disconnected analogue of Fig.~\ref{fig:connplot} can be found in Fig.~\ref{fig:discplot}. It is clear that much like the connected data, the size of the contribution increases with decreasing pion mass and so a very significant cancellation will occur at the physical pion mass between these two contributions with opposite signs. This cancellation was predicted in Ref.\ \cite{Bijnens:2016hgx} on the basis of the $\pi^0$ exchange contribution and is illustrated in Fig.\ \ref{fig:C101intgnds}. It is also worth noting that the statistical precision of the disconnected data is significantly worse than the connected, even though almost an order of magnitude more measurements were performed.

On the right-hand side of Fig.~\ref{fig:discplot}, we show the chiral-continuum-infinite-volume extrapolation with different $m_\pi L$ at fixed lattice spacing. Much like in our previous work \cite{Chao:2020kwq}, we find the lattice-spacing dependence to have a slope of the same sign as the connected contribution. It is also evident that an accidental partial cancellation occurs between the finite-volume and lattice-spacing terms, with the approach to the infinite volume making the $(2+2)$ contribution more negative and the approach to the continuum limit making it less negative.

\subsection{Combined light-quark results}

Due to the significant cancellation between the connected and the (2+2) contribution, we find it useful to take the ensemble-by-ensemble sum of the contributions and then perform the extrapolation. For this sum, our data cannot resolve any of the terms non-analytic in $m_\pi^2$ of Eq.\ (\ref{eq:singterm}), and it appears that these contributions largely cancel. We find that the fit ansatz
\begin{equation}\label{eq:fvol2}
a_\mu(m_\pi^2,m_\pi L,a^2) = a_\mu(0,\infty,0)(1+Am_\pi^2 + Be^{-m_\pi L/2} + Ca^2),
\end{equation}
describes our data very well. This ansatz assumes that any potential singular terms in our data cancel to a large extent, an assumption that we address along with the discussion of systematics in Section~\ref{sec:final_res}. Here we simply note that, in addition to the cancellation between the connected and the (2+2) contribution, the chirally singular behaviour expected in $a_\mu^{{\rm Conn}+ (2+2)}$ from the $\pi^0$ exchange and the charged pion loop is numerically suppressed over the pion-mass interval 135 to 200\,MeV, due to a partial cancellation between these two long-distance contributions.

\begin{figure}[h!]
  \includegraphics[scale=0.35]{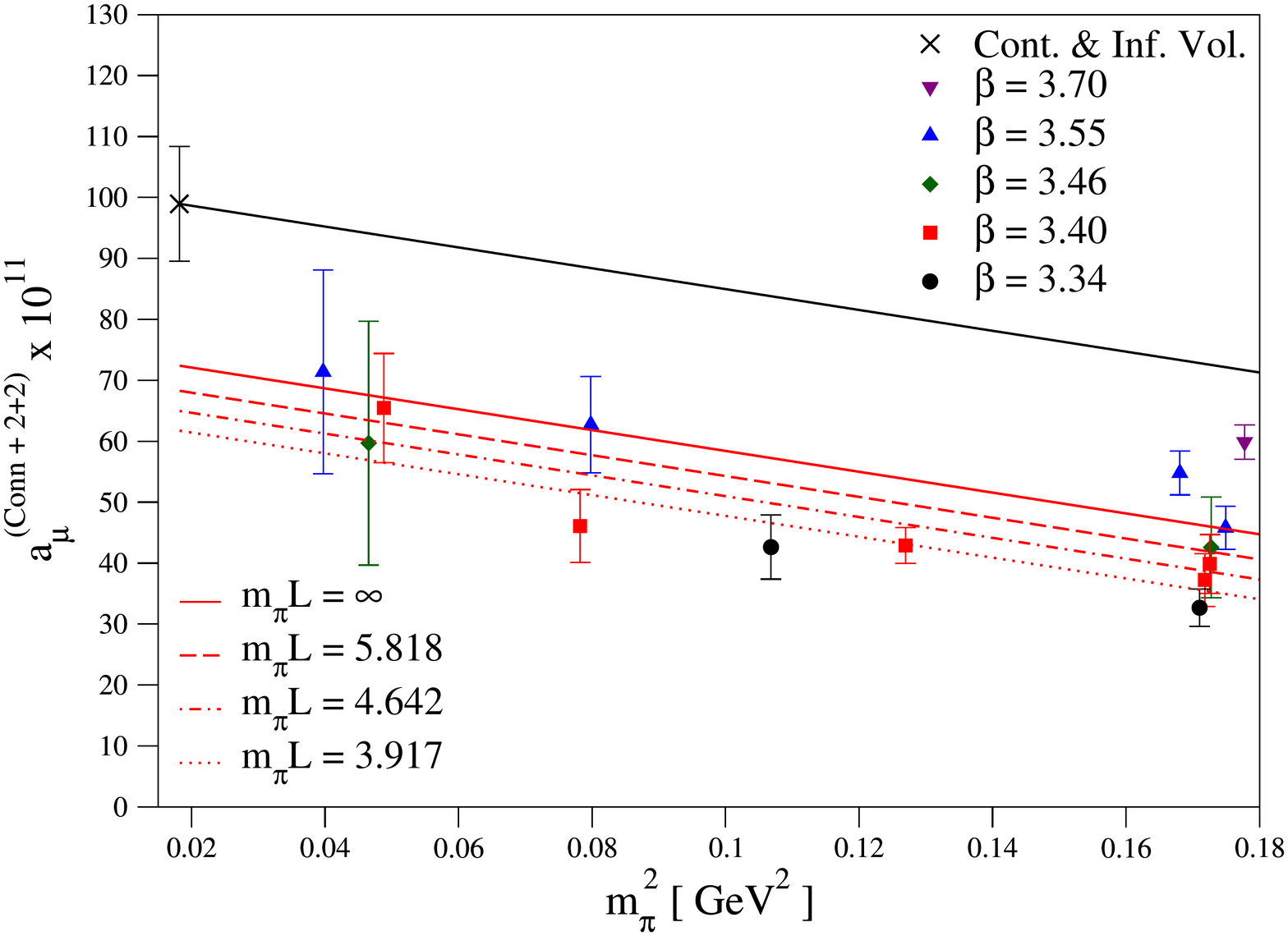}
  \caption{Chiral-continuum-infinite-volume extrapolation of the sum of the light-quark, fully-connected and $(2+2)$ contributions to $\ahlbl$. See the caption of Fig.~\ref{fig:connplot} (right panel).}\label{fig:sumplot}
\end{figure}

Fig.~\ref{fig:sumplot} shows an extrapolation for the sum of the light-quark, fully-connected and $(2+2)$ disconnected contributions. It appears in the plot that no chiral curvature is present in this combination and the error grows at lighter pion masses; this is due to the large cancellation between the connected and disconnected contributions. Considering the final column in Tab.~\ref{tab:light_results}, we do not appear to benefit from a cancellation of statistical errors due to correlations between the two measurements. It is also clear that the approach to the infinite volume is less severe in the combination of these two quantities compared to fitting them individually; this is likely due to large cancellations in the long-distance contributions such as the pion pole. We still see significant discretisation effects in this fit, but fortunately we have several lattice spacings to constrain this behavior; nevertheless this will form our largest systematic as is discussed later on in Sec.~\ref{sec:final_res}.

\subsection{Consistency checks}

It is useful to compare the present analysis to our previous work at the $\text{SU}(3)_f$-symmetric point \cite{Chao:2020kwq}. In that work, we combined light and strange contributions and obtained $98.9(2.5)\times 10^{-11}$ for the connected contribution in the infinite-volume and continuum limit, and $-33.5(4.2)\times 10^{-11}$ for the disconnected. Combining these with charge factors adjusted to isolate the $(u,d)$ quark contribution yields $70.1(3.8)\times 10^{-11}$. The extrapolation in Fig.~\ref{fig:sumplot} at $m_\pi^2 \approx 0.173 \text{ GeV}^2$ is $72.5(4.3)\times 10^{-11}$, so these results are in good agreement, even though the underlying methodology is considerably different.

\begin{figure}[h!]
  \includegraphics[scale=0.35]{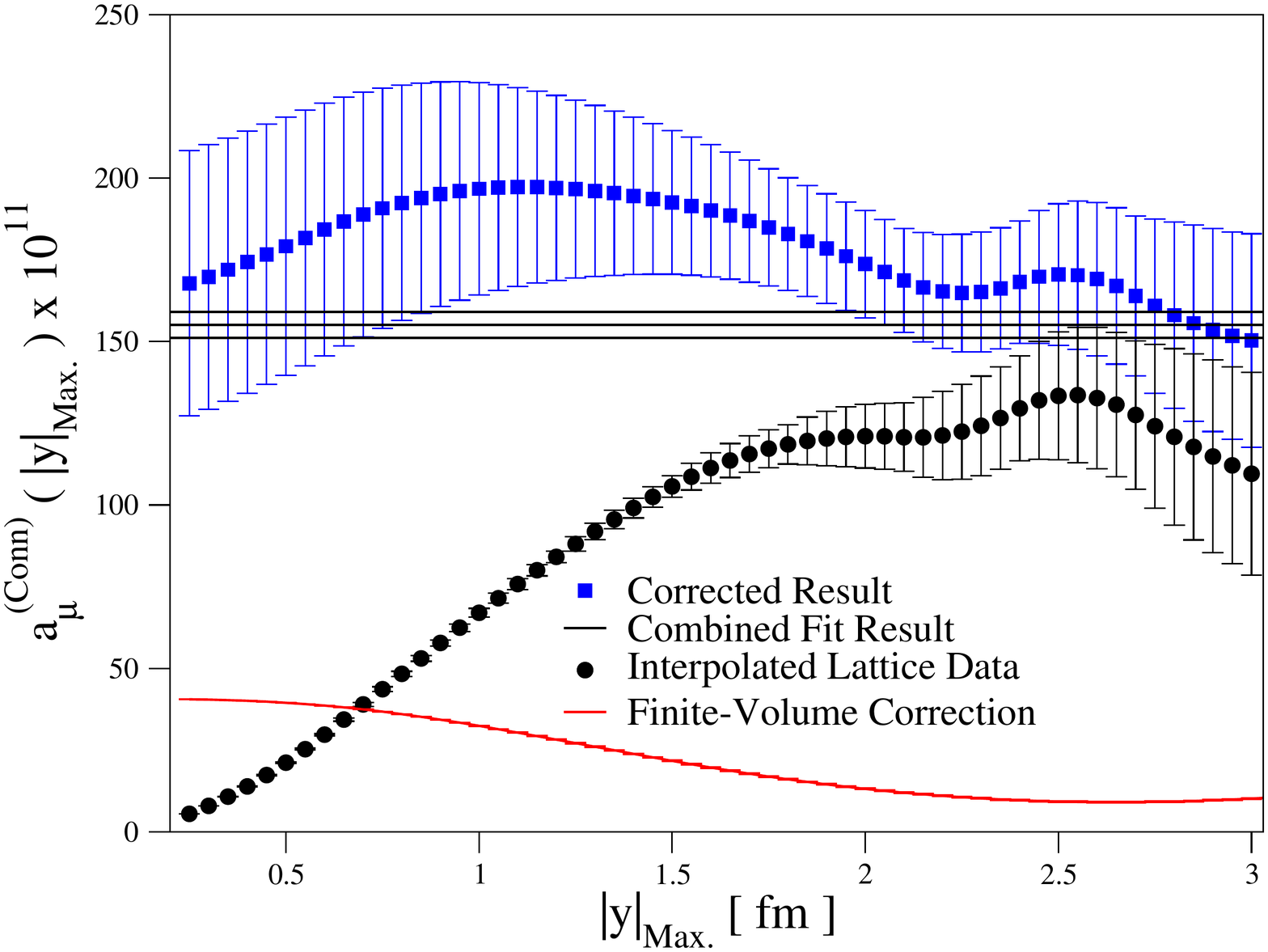}
  \caption{Consistency of the infinite-volume estimate for the light connected contribution on ensemble C101 between the analysis performed here (horizontal band) and the analysis method of Ref.~\cite{Chao:2020kwq} (blue points), which corrects the lattice data (black points) using the $\pi^0$ exchange prediction. The horizontal band is obtained
    by adding the finite-size correction from the global fit displayed in Fig.~\ref{fig:connplot} (right panel) to the $a_\mu^{\rm (Conn)}$ value obtained on ensemble C101 using the tail extension parametrisation Eq.~\eqref{eq:y3exp}. Note that the band does not include
  the systematic uncertainty of varying the fit ansatz in the global fit, which is addressed in section \ref{sec:final_res}.}\label{fig:C101consistency}
\end{figure}

Fig.~\ref{fig:C101consistency} illustrates the consistency between our previous `tail and finite-size correction'  methodology (blue and black points have been interpolated) and the result of our global fit with the Pole ansatz. The two are consistent within the combined statistical and systematic error of the blue data points, although the black line does lie a bit lower than the central value. In the C101 data, there is a strong upward fluctuation (visible in Fig.\ \ref{fig:C101intgnds}) that pushes both the black and the blue points up and likely hides a stable plateau region.

%% file: strange_contributions.tex
\section{Strange contributions}\label{sec:strange_leading}

For the fully-connected strange, the $(2+2)$-light-strange ($ls$) and the $(2+2)$-strange-strange ($ss$) contributions, we use results from a subset of the ensembles (listed in Tab.~\ref{tab:strange_results}) to cut down on cost for what turns out to be a very small contribution to the overall $\ahlbl$. Here we can reuse the results from the symmetric point with the appropriate charge factors.

\begin{table}[h!]
\begin{tabular}{c|ccc}
\toprule
Ensemble & Connected$\times 10^{11}$ & $(2+2)\times 10^{11} $ & Sum$\times 10^{11}$\\
\hline
A653 & 3.79(0.06) & $-14.0(1.2)$ & $-10.2(1.2)$ \\
A654 & 2.65(0.02) & $-9.8(1.1)$ & $-7.1(1.1)$ \\
\hline
U103 & 3.49(0.05) & $-9.7(1.9)$ & $-6.2(1.9)$ \\
H101 & 4.13(0.10) & $-13.3(1.9)$ & $-9.2(1.9)$\\
H105 & 2.50(0.06) & $-8.2(1.0)$ & $-5.7(1.0)$ \\
C101 & 2.25(0.02) & $-8.0(1.0)$ & $-5.8(1.0)$ \\
\hline
B450 & 4.14(0.07) & $-12.2(3.6)$ & $-8.1(3.6)$\\
\hline
H200 & 3.84(0.08) & $-8.6(1.5)$ & $-4.7(1.5)$ \\
N202 & 4.90(0.12) & $-12.5(1.3)$ & $-7.6(1.3)$ \\
\hline
N300 & 4.45(0.07) & $-7.0(1.1)$ & $-2.5(1.1)$ \\
\botrule
\end{tabular}
\caption{Fully-connected and the combined $(2+2)$ $ls$ and $ss$ contributions to $\ahlbl$.}\label{tab:strange_results}
\end{table}

\begin{figure}
\includegraphics[scale=0.25]{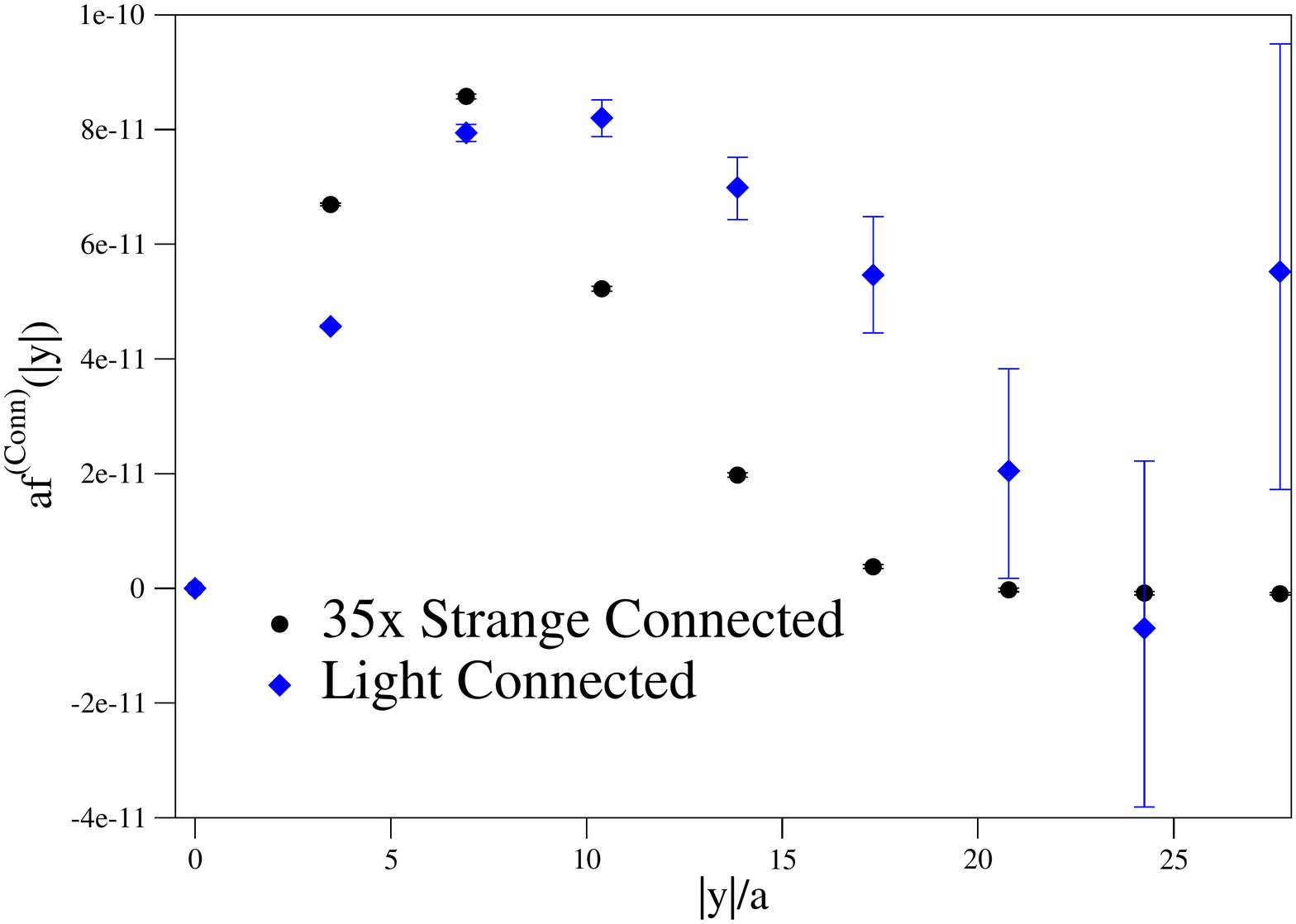}
\includegraphics[scale=0.25]{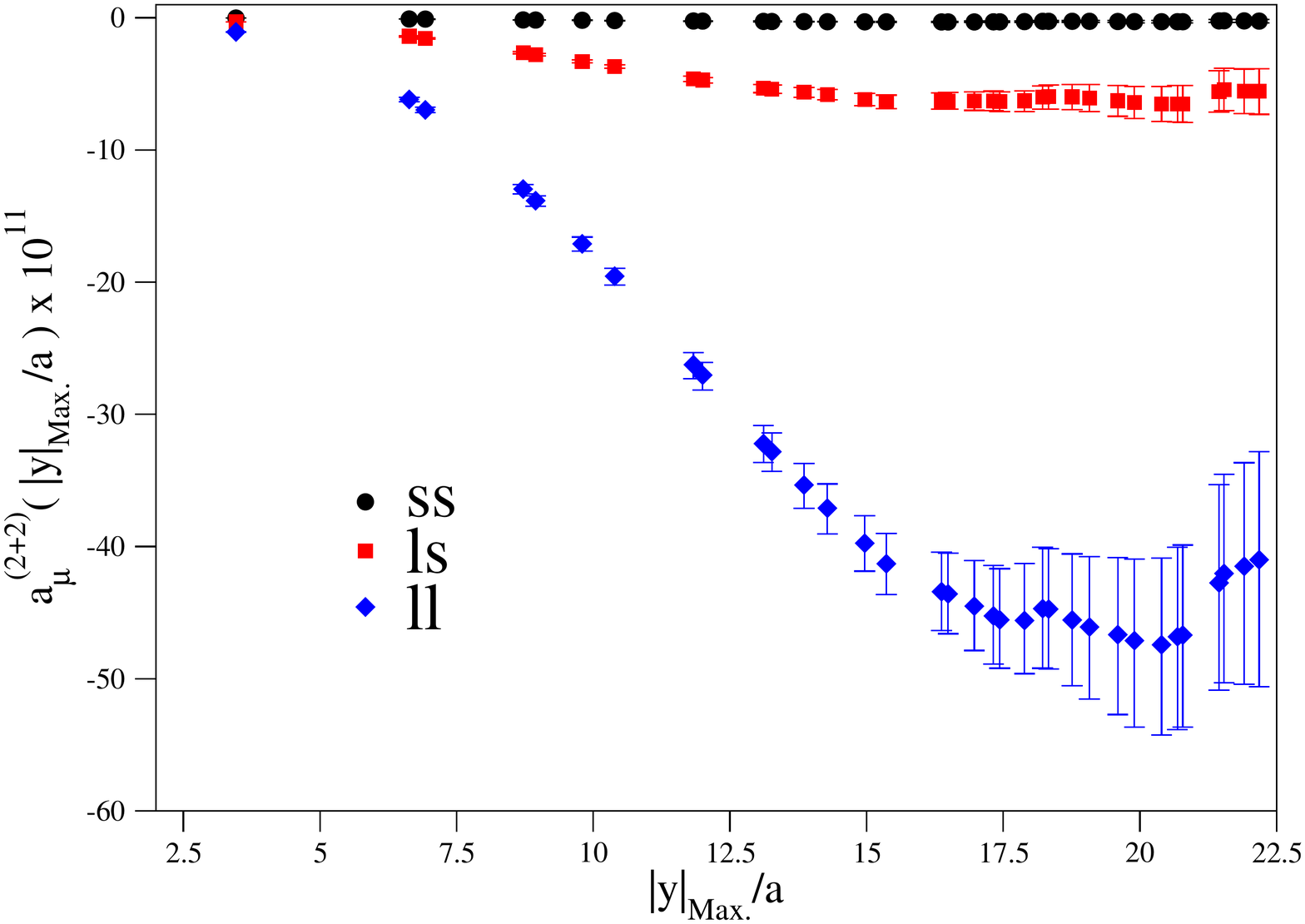}
\caption{Strange contributions for ensemble C101. Left: Strange and light connected integrands, in lattice units. The strange integrand has been multiplied by $35$ for visibility. Right: Partially integrated $(2+2)$ contributions with three different flavour combinations.}\label{fig:strange_contribs}
\end{figure}

The left plot of Fig.~\ref{fig:strange_contribs} illustrates the magnitude of light and strange quark contributions to the fully-connected diagrams for ensemble C101, which has a light pion mass. The light-quark contribution is much longer ranged than the strange and statistically far noisier. The peak of the strange integrand for this ensemble is about 35 times smaller than the light one, and the overall integrated contribution is calculated to be about 55 times smaller; as one approaches the physical pion mass this difference will only grow.

The right plot of Fig.~\ref{fig:strange_contribs} illustrates the size of the contributions from different flavour combinations within the $(2+2)$ calculation: light-light ($ll$), light-strange ($ls$), and strange-strange ($ss$). We again show ensemble C101 and use the same statistics for all flavours. For this ensemble the $ls$ contribution is a bit larger than $10\%$ of the $ll$, and the $ss$ contribution is about $0.6\%$. It can be seen that the integrated $a_\mu$ plateaus earlier for heavier quark content and the statistical precision is better too. As the majority of the data in this analysis comes from the previous $\text{SU}(3)_f$-symmetric work, the same conclusions apply; finite-volume effects and cut-off effects are still sizeable even for the contributions including strange quarks.

We choose to extrapolate the sum of all the strange and light-strange contributions to the infinite-volume, physical quark mass, continuum limit using the Ansatz
\begin{equation}\label{eq:fvol2_strange}
a_\mu(m_K^2,m_\pi L,a^2) = a_\mu(0,\infty,0)(1+Am_K^2 + Be^{-m_\pi L} + Ca^2).
\end{equation}
It is worth noting that the exponential volume factor here is $m_\pi L$ instead of $m_\pi L/2$ for the light-quark contribution as there is no $\pi^0$-exchange.

\begin{figure}[h!]
\includegraphics[scale=0.35]{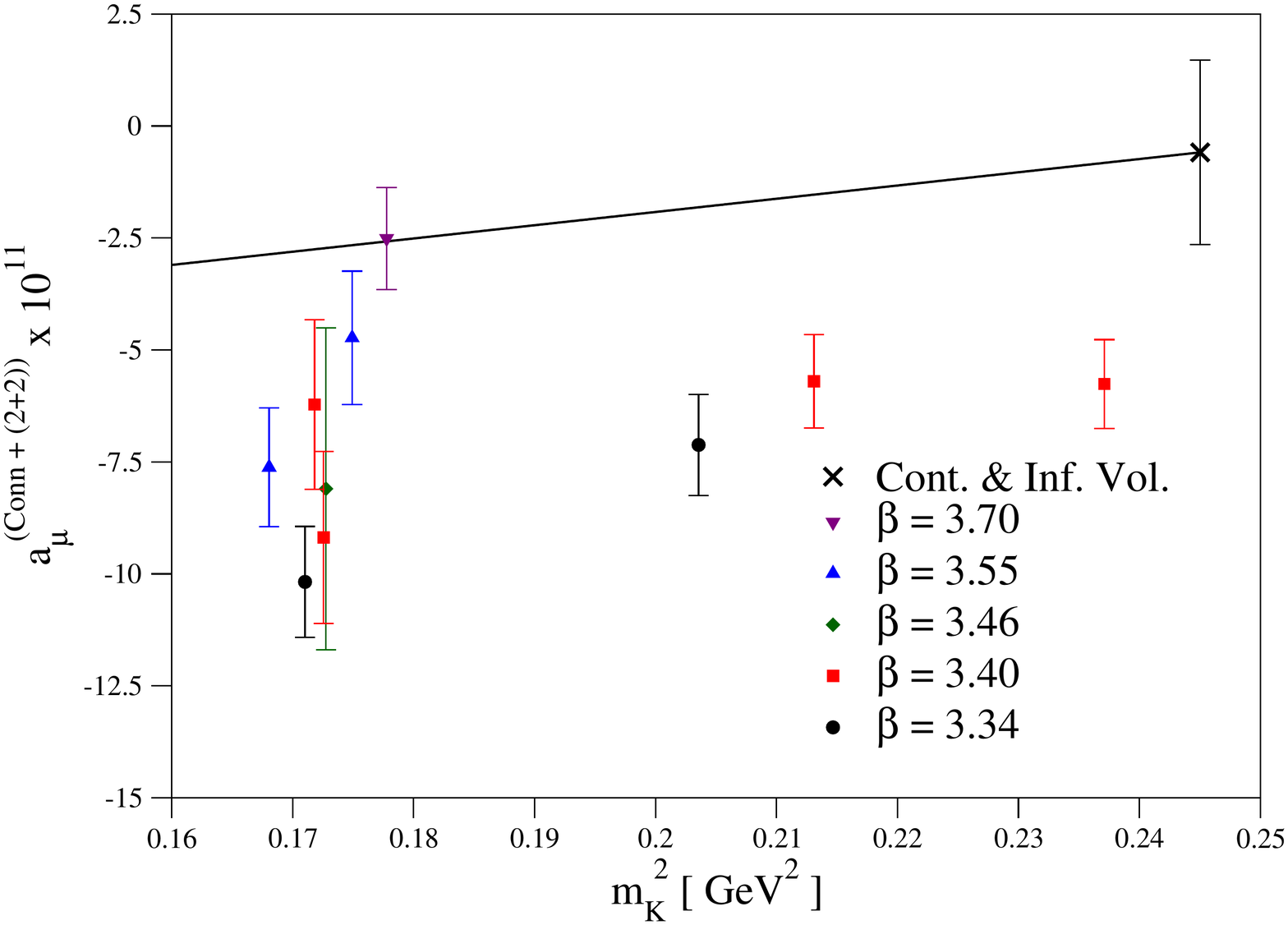}
\caption{An extrapolation of the sum of the fully-connected strange and $(2+2)$ $ls$ and $ss$ contributions to $\ahlbl$.}\label{fig:strange_extrap}
\end{figure}

A plot of this extrapolation can be found in Fig.~\ref{fig:strange_extrap}. The fit of Eq.~\eqref{eq:fvol2_strange} gives a $\chi^2/\text{dof} = 0.6$. Again, we see the $(2+2)$ contribution approaching the continuum limit with the same sign as the fully-connected contribution; in the continuum, these two contributions effectively cancel. Our final result at the physical point is
\begin{equation}
a_\mu^{\text{(Conn. + (2+2))-}s} = -0.6(2.0)\times 10^{-11}.
\end{equation}

%% file: higher_order.tex
\newcommand{\dm}{\Delta M^2}

\section{Higher-order contributions}\label{sec:higher_order}

The remaining three topologies, $(3+1)$, $(2+1+1)$, and $(1+1+1+1)$,
contain one, two, and four disconnected loops,
respectively. Empirically, diagrams containing a disconnected loop
with a vector current have been found to be suppressed in QCD
correlation functions~\cite{Green:2015wqa, Blum:2015you,
  DellaMorte:2017dyu, Gerardin:2019vio}. Furthermore, these loops
vanish at the SU$(3)_f$-symmetric point where light and strange quarks are
degenerate. Finally, diagrams containing more loops are suppressed at
large $N_c$. These considerations lead to the expectation that the
$(2+1+1)$ and $(1+1+1+1)$ topologies are suppressed relative to
$(3+1)$, which is itself suppressed relative to the two leading
topologies.

Our goal is thus to compute the $(3+1)$ class of diagrams as well as we can, and provide evidence that the $(2+1+1)$ and $(1+1+1+1)$ are small enough to be neglected from our targeted error budget. 
In particular, we will give details about how we treat the $|y|$-integration of our $(3+1)$ data.

\subsection{The \texorpdfstring{$(3+1)$}{(3+1)} contribution}

The charge factor of $(3+1)_{\rm{strange}}$ is $-1/7$ of that of $(3+1)_{\rm{light}}$.
(Recall that here the flavour label corresponds to the triangle and that for the disconnected loop, we always use the combined light and strange contributions.)
Furthermore, because of the larger mass of the strange quark, the $(3+1)_{\rm{strange}}$ contribution is expected to be much smaller compared to $(3+1)_{\rm{light}}$.
We will thus put our main effort on the $(3+1)_{\rm{light}}$ contribution.
Numerical evidence of the smallness of the $(3+1)_{\rm{strange}}$ contribution is given in Section~\ref{ho:sect:strange3p1};
it turns out that this quantity is at least ten times smaller than the contribution with light triangle. 

\subsubsection{Treatment of the tail of the \texorpdfstring{$(3+1)_{\rm{light}}$}{(3+1)-light} integrand}

\begin{figure}[h!]
  \includegraphics[scale=0.35]{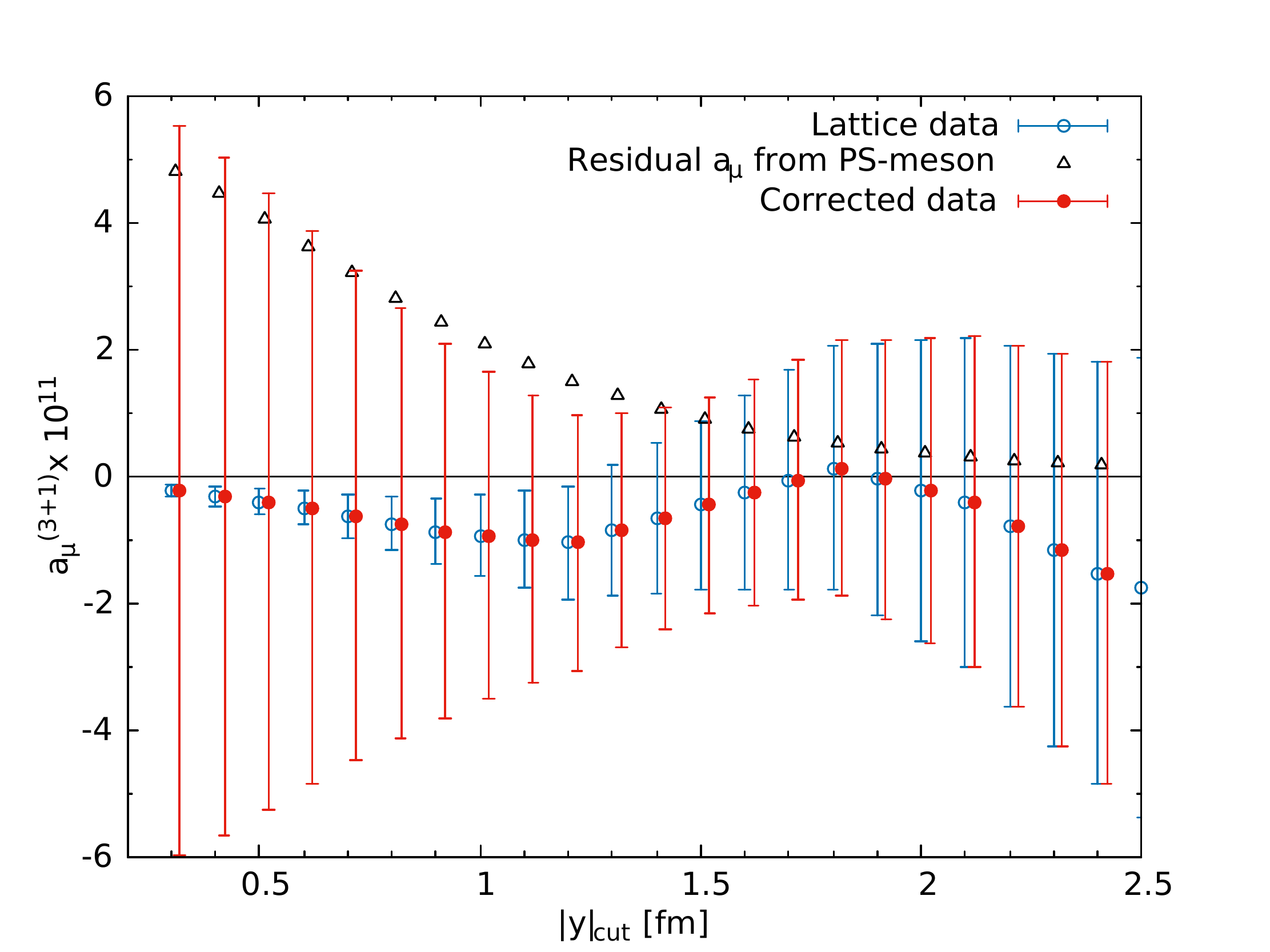}
  \caption{Determination of the $(3+1)_{\rm{light}}$ contribution to $\ahlbl$ on ensemble C101 using the tail treatment procedure. Horizontal offsets are applied for visibility. One can see that the total error is minimised for $|y|_{\rm{cut}}$ between 1.2 and 1.7 fm.}\label{ho:fig:ycut_C101}
\end{figure}

At large distances, the physics is dictacted by the lightest particles, \textit{i.e.}, the pseudoscalar mesons.
The most relevant contributions are neutral pseudoscalar-meson poles and charged pseudoscalar-meson loops (\cite{Aoyama:2020ynm} and the references therein).
The computation based on Partially-Quenched Chiral Perturbation Theory (PQChPT) in Appendix~\ref{matching_nf2p1:sect} shows that there is no contribution at leading order coming from pseudoscalar-meson poles.
Nevertheless, the $(3+1)_{\rm{light}}$ receives contributions from pseudoscalar-meson loops (cf. Fig.~\ref{matching_nf2p1:fig:loop} and Table~\ref{matching_nf2p1:tab:loop_coef}).
As the signal of the integrand degrades rapidly with increasing $|y|$, we decide to use the knowledge from light pseudoscalar contributions in infinite volume as a guideline for cutting the integral at some $|y|=|y|_{\rm{cut}}$.

The procedure is as follows.
We split the $|y|$-integral for $a_\mu^{(3+1)}$ into two intervals, below and above $|y|_{\rm{cut}}$, so that $a_\mu^{(3+1)}=a_\mu^<+a_\mu^>$.
Below $|y|_{\rm{cut}}$, we numerically integrate the lattice data to obtain $a_\mu^<$.
Above $|y|_{\rm{cut}}$, we take the central value of $a_\mu^>$ to be zero and assign an uncertainty to this omitted tail contribution based on a calculation of the charged pseudoscalar-meson loop contribution in scalar QED. Final we add the two uncertainties (statistical for $a_\mu^<$ and systematic for $a_\mu^>$) in quadrature.

For the tail uncertainty, we compute the integrand in infinite-volume scalar QED and integrate from $|y|_{\rm{cut}}$ to infinity, which gives an order-of-magnitude estimate of the possible missing mesonic contributions in the tail. We then assign a very conservative systematic error, namely $w_{\rm{sys.}}=120\%$ of this contribution. (This choice of $w_{\rm{sys.}}$ will be justified in the next subsection.) As scalar QED corresponds to pointlike photon-pseudoscalar-pseudoscalar vertices, which tend to overestimate the pseudoscalar-meson loop contributions to $a_\mu$, we assume that the assigned systematic error also covers the possible finite-size effects in the $|y|<|y|_{\rm{cut}}$ region.
Finally, we determine $|y|_{\rm{cut}}$ for each lattice ensemble by finding the value that minimises the total error. 
An example of this procedure is shown in Fig.~\ref{ho:fig:ycut_C101}.

\subsubsection{Numerical results for the \texorpdfstring{$(3+1)_{\rm{light}}$}{(3+1)-light} contribution}

\begin{table}
  \centering
  \begin{tabular}{c|ccc|c}
  \toprule
  Ensemble & $|y|_{\rm{cut}}$[fm] & $a_\mu^{<}\times 10^{11}$ & $a_\mu^{>}\times 10^{11}$ & $a_\mu^{(3+1)}\times 10^{11}$ \\
  \hline  
  A654 & 1.47 & $-0.23(0.16)$ & 0.00(0.11) & $-0.23(0.20)$ \\
  \hline
  U102 & 1.17 & $-0.23(0.28)$ & 0.00(0.16) & $-0.23(0.34)$\\
  H105 & 1.28 &  0.61(0.58) & 0.00(0.45) &  0.61(0.79)\\
  C101 & 1.43 & $-0.60(1.38)$ & 0.00(1.01) &  $-0.60(1.83)$ \\
  \hline
  D450 & 1.37 &  0.64(1.61) & 0.00(1.18) &  0.64(2.14)\\
  \hline
  N203 & 1.01 &  0.19(0.48) & 0.00(0.29) &  0.19(0.59) \\
  N200 & 1.32 &  0.01(0.65) & 0.00(0.42) &  0.01(0.82) \\
  D200 & 1.50 & $-0.57(1.53)$ & 0.00(1.32) & $-0.57(2.21)$ \\
  \botrule
  \end{tabular}  
  \caption{Results for $(3+1)_{\rm{light}}$ on each ensemble (using $w_{\rm{sys.}}=120\%$), along with the choice of $|y|_{\rm{cut}}$ and the contributions to $a_\mu$ below and above the cut.}\label{ho:tab:3p1_ycut}
\end{table}

\begin{figure}[h!]
\includegraphics[scale=0.35]{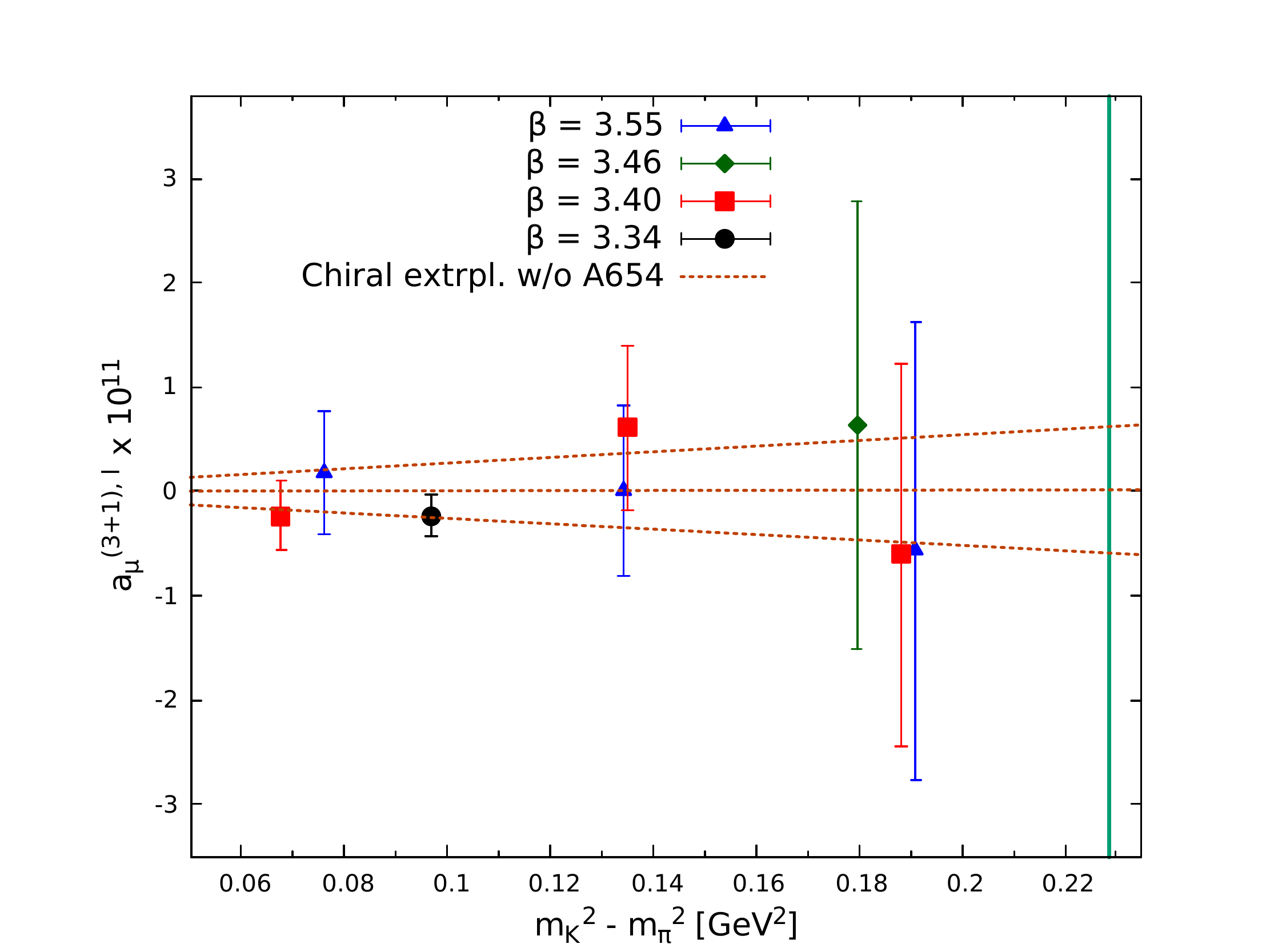}
\hspace{-8pt}
\includegraphics[scale=0.35]{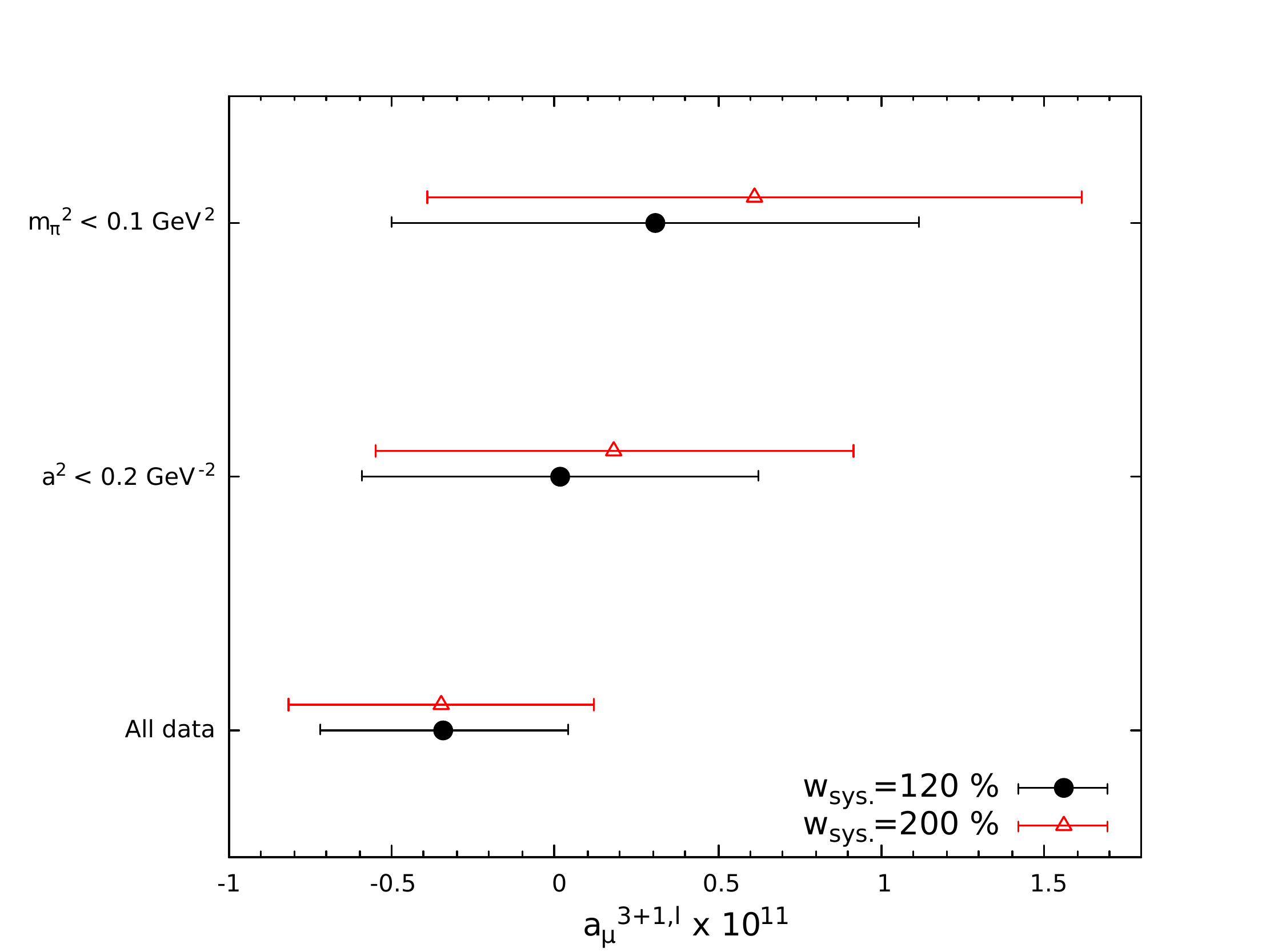}
\caption{Left: Extrapolation of the $(3+1)_{\rm{light}}$ contribution to $\ahlbl$, determined using $w_{\rm{sys.}}=120\%$. The points show the results from each ensemble and the vertical green line indicates physical meson masses. The orange dashed lines show the extrapolation to the physical point with Eq.~\eqref{eq:3p1_extrap}, excluding the coarsest ensemble A654. Right: Fit results of Eq.~\eqref{eq:3p1_extrap} after applying cuts to the data, with two choices of $w_{\rm{sys.}}$.}\label{fig:3p1_summary}
\end{figure}

Table~\ref{ho:tab:3p1_ycut} shows our choice of $|y|_{\rm{cut}}$ and the value of $a_\mu^{(3+1)}$ computed on each ensemble, with $w_{\rm{sys.}}=120\%$; these results are also plotted in Fig.~\ref{fig:3p1_summary} (left).
It is already clear from this plot that there is no distinguishable $O(a)$-dependence in the data at our level of precision.
The same is true of volume effects.
This leads us to parameterise our data in a very simple form, namely
\begin{equation}\label{eq:3p1_extrap}
a_\mu^{\text{(3+1)-}l} = A (m_K^2 - m_\pi^2).
\end{equation}
Such a mass-dependence is motivated by the fact that this contribution must vanish at the $\text{SU}(3)_f$-symmetric point. This fit describes our data well and we investigate the stability of our final fit result through applying several cuts in our data, as is shown in Fig.~\ref{fig:3p1_summary} (right).
Also shown in the same figure are the results obtained by applying the same procedure with a different choice of the weight, $w_{\rm{sys.}}=200\%$, for the estimate of the systematic error of the omitted tail contribution.
Note that a bigger value of $w_{\rm{sys.}}$ implies that one cuts the lattice data at larger $|y|$.
As the lattice data become noisier with increasing $|y|$, the fluctuations of the central value determined from the lattice data also become larger, especially for our ensembles with lighter pion mass.   
However, from the consistency between fits with different cuts in the data, it seems that our choice of $w_{\rm{sys.}}=120 \%$ is reasonable enough without being too conservative. 
For our final determination, including our fit-systematic, we choose to quote the determination excluding the coarsest ensemble A654 ($a^2<0.2\text{ GeV}^{-2}$):
\begin{equation}
a_\mu^{(3+1)\text{-}l} = 0.0(0.6)\times10^{-11}.
\end{equation}

\begin{figure}[h!]
\includegraphics[scale=0.35]{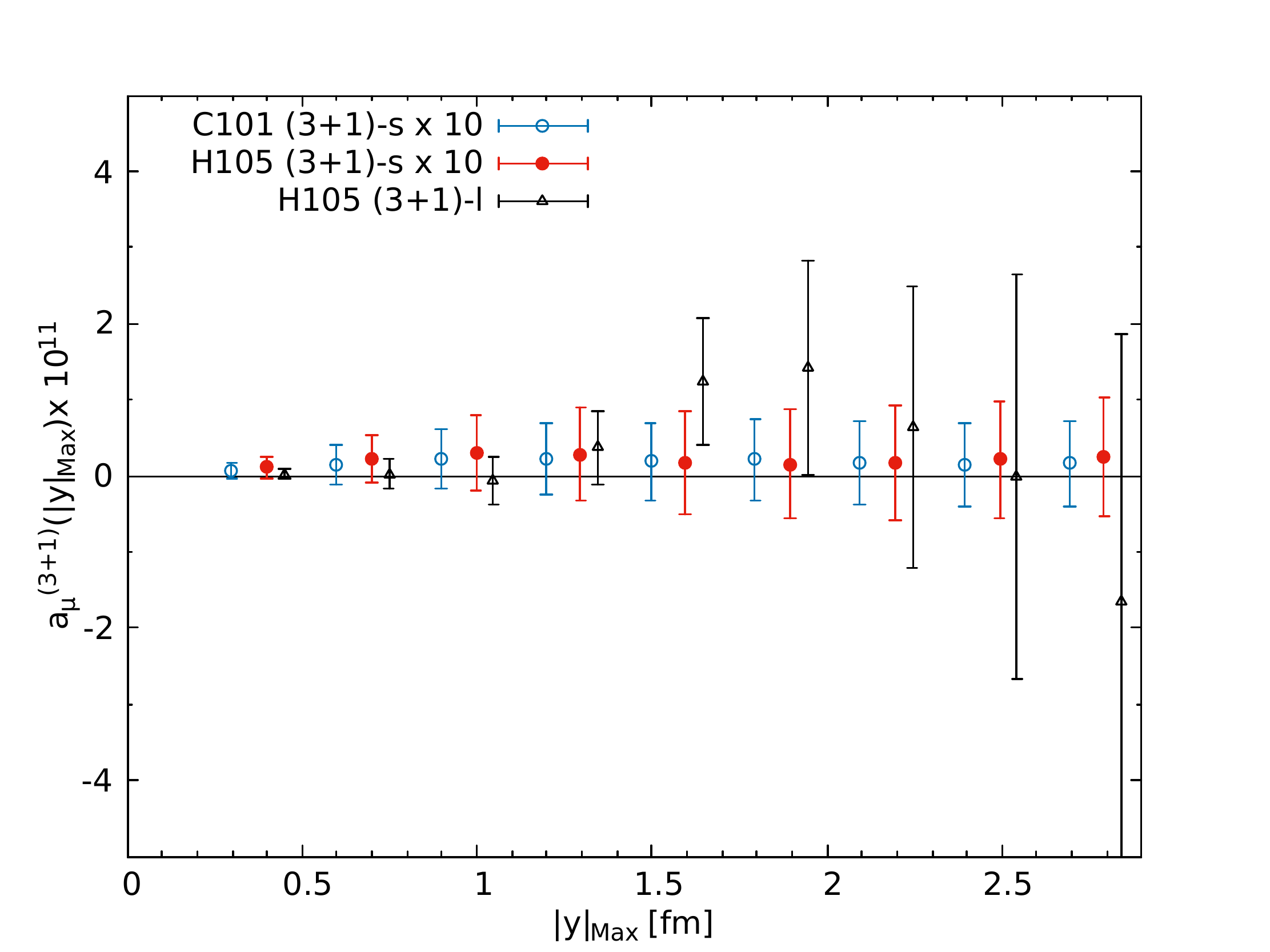}
\caption{Partially-integrated $(3+1)_{\rm{strange}}$ contribution to $\ahlbl$ for ensembles C101 ($m_\pi\approx 220$ MeV, $m_K\approx 470$ MeV) and H105 ($m_\pi\approx 280$ MeV, $m_K\approx 460$ MeV), compared to the $(3+1)_{\rm{light}}$ for the ensemble H105. The $(3+1)_{\rm{strange}}$ data are multiplied by 10 for visibility. As for the statistics for the $(3+1)_{\rm{strange}}$, it is $50\%$ compared to the $(3+1)_{\rm{light}}$ for C101 and about $15\%$ for H105.}\label{strange3p1:fig:comp_3p1_ls_C101_H105}
\end{figure}  

\subsubsection{The \texorpdfstring{$(3+1)_{\rm{strange}}$}{(3+1)-strange} contribution}\label{ho:sect:strange3p1}

We have computed the $(3+1)_{\rm{strange}}$ contribution using two ensembles: C101 and H105. For both ensembles, the partially-integrated $a_\mu$ is shown in Fig.~\ref{strange3p1:fig:comp_3p1_ls_C101_H105}, and this is compared with the $(3+1)_{\rm{light}}$ for ensemble H105. 
It is clear from the lattice data that $(3+1)_{\rm{strange}}$ is at least ten times smaller than our bound on $(3+1)_{\rm{light}}$ and can be entirely neglected for our target precision compared to the leading contributions.
As it involves the strange-quark triangle, we expect that this quantity depends only on hadronic states which are at least as heavy as the kaon. From this point of view, because the kaon masses on the used ensembles are somewhat lighter than the physical one, we find it exceedingly unlikely that it would grow significantly as the quark masses approach their physical values. 

\subsection{The \texorpdfstring{$(2+1+1)$}{(2+1+1)} and the \texorpdfstring{$(1+1+1+1)$}{(1+1+1+1)} results}

Due to the much-higher computational cost of the lattice-wide object (Eq.~\eqref{formalism:eq:pi2}) used in our computational strategy, we only determined the light-quark, $(2+1+1)$ contribution for the ensembles N203 and C101. Also, we only computed the $(1+1+1+1)$ contribution for the ensemble C101 because of our expectation for its insignificance to the final error. The results for the partially integrated $a_\mu(|y|)$ for both of these ensembles are shown in Fig.~\ref{ho:fig:comp_ho}. 

A computation from PQChPT (see Appendix~\ref{matching_nf2p1:sect}) shows that at leading order, these two topologies receive neither contributions from the neutral pseudoscalar-meson poles, nor from charged pseudoscalar-meson loops.
It is therefore hard to decide at which value of $|y|$ one can cut the lattice data and apply a model prediction afterwards. However, one can see from Fig.~\ref{ho:fig:comp_ho} that the $(2+1+1)$-contribution for both ensembles is smaller than the $(3+1)$ at small $|y|$.

\begin{figure}[h!]
  \begin{minipage}{0.495\textwidth}
    N203\vspace{-12pt}
  \includegraphics[scale=0.35]{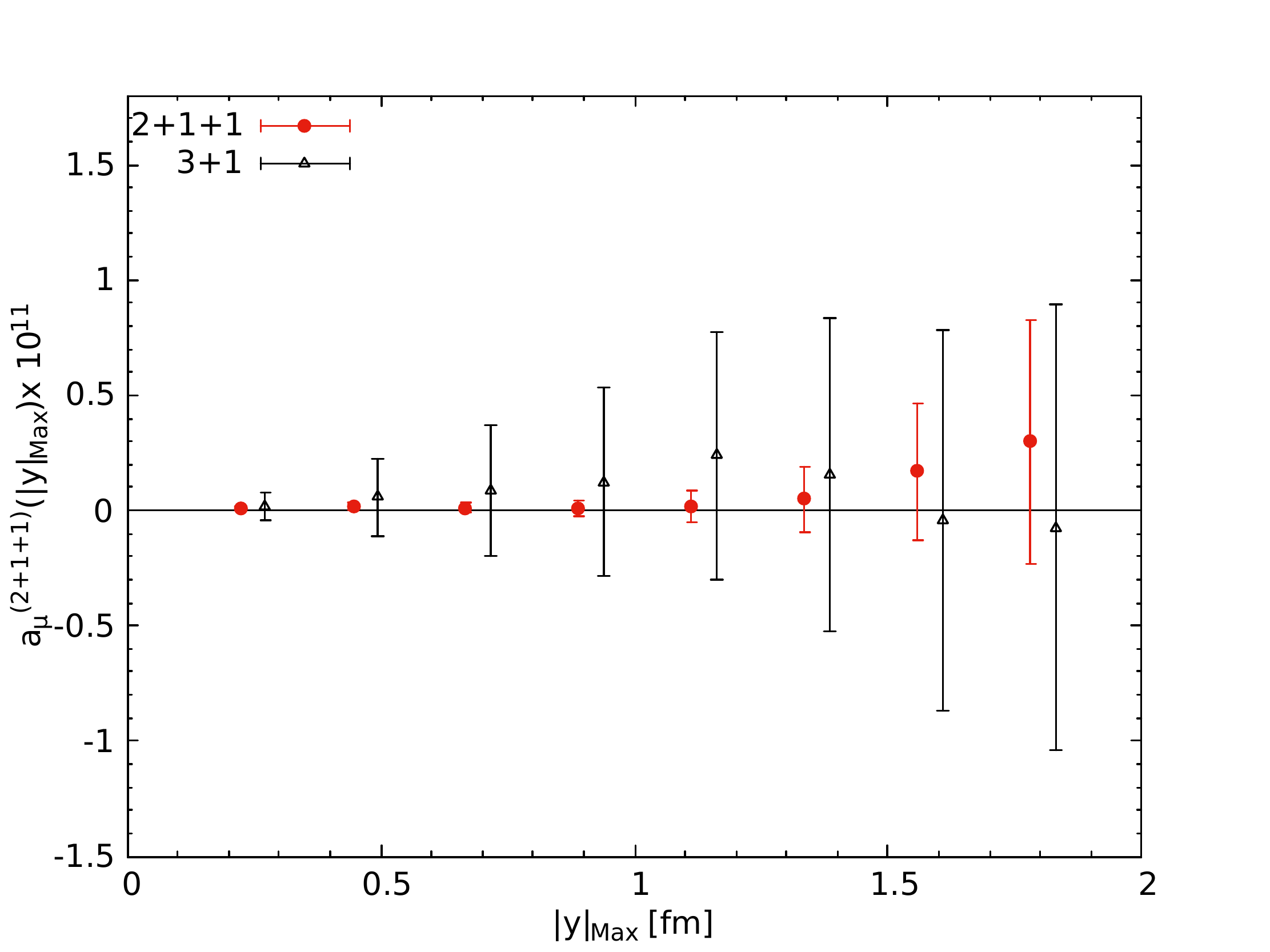}\label{ho:fig:comp_ho_N203}
  \end{minipage}
  \hspace{-8pt}
  \begin{minipage}{0.495\textwidth}
    C101\vspace{-12pt}
  \includegraphics[scale=0.35]{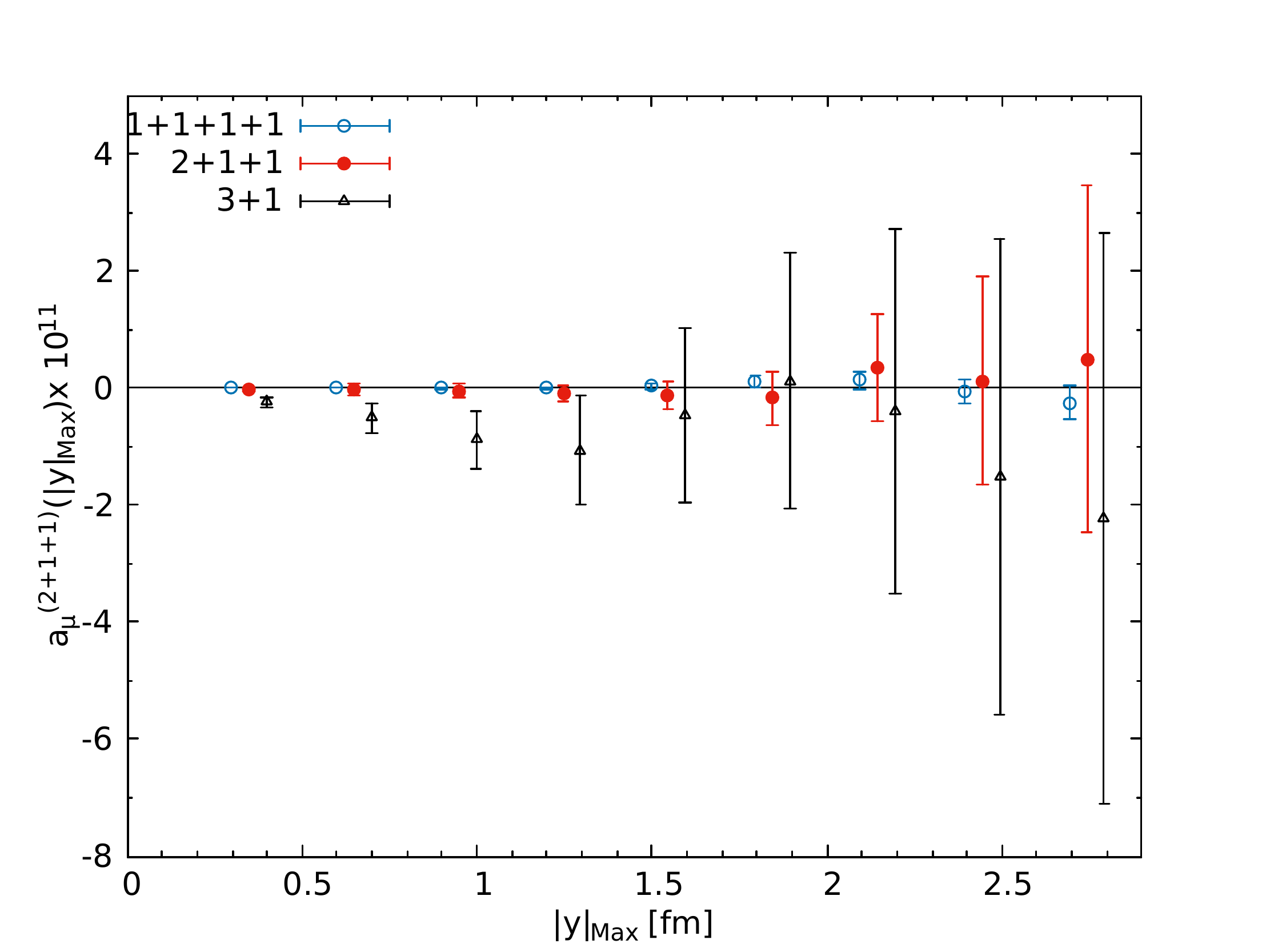}\label{ho:fig:comp_ho_C101}
  \end{minipage}
  \caption{Partially-integrated higher-order contributions to $\ahlbl$, in comparison with $(3+1)_{\rm{light}}$. Left: Ensemble N203 ($m_\pi\approx 340$ MeV). Right: Ensemble C101 ($m_\pi\approx 220$ MeV).}\label{ho:fig:comp_ho}
\end{figure}   

Although the rapid degradation of the signal of the $(2+1+1)$-contribution is expected, our strategy of averaging over possible ways of constructing the vector $y$ for a given $|y|$ appears to work well at suppressing the statistical noise of this quantity at short distances. In the end, we conservatively estimate this quantity to be zero with half the error of the $(3+1)$ contribution. 
From the smallness of the light-quark contribution of the $(2+1+1)$ topology, we deem it legitimate to assign the value of zero to the strange contribution. This comes with no contribution to the error budget, as this will be irrelevant compared to our overall level of precision for $\ahlbl$.
Note that the mere charge factor suppresses the strange $(2+1+1)$ contribution relative to the light $(2+1+1)$ by a factor of five.
As for the $(1+1+1+1)$ contribution, its observed smallness on the right panel of Fig.\ \ref{ho:fig:comp_ho} does not come as a surprise, in particular since its charge factor weights it five times less than the already-small $(2+1+1)$.
Any improvement to either of these quantities would have a completely negligible effect on the final result for $\ahlbl$, at our current level of precision.

%% file: combined_result.tex
\section{The total \texorpdfstring{$\ahlbl$}{a-mu HLBL}}\label{sec:final_res}

In this section we investigate two approaches to determining the contribution of the two leading light-quark contributions to $\ahlbl$: the first consists of fitting the sum of the two contributions and the second consists of adding the results of individual fits to the fully-connected and $(2+2)$ contributions using various ans\"atze. We investigate possible systematics in our approach by comparing the results with terms in $a$ or $a^2$ and by performing cuts in $m_\pi L$, $a^2$, and $m_\pi^2$.

\subsection{Sum and fit}

\begin{figure}[h!]
\includegraphics[scale=0.35]{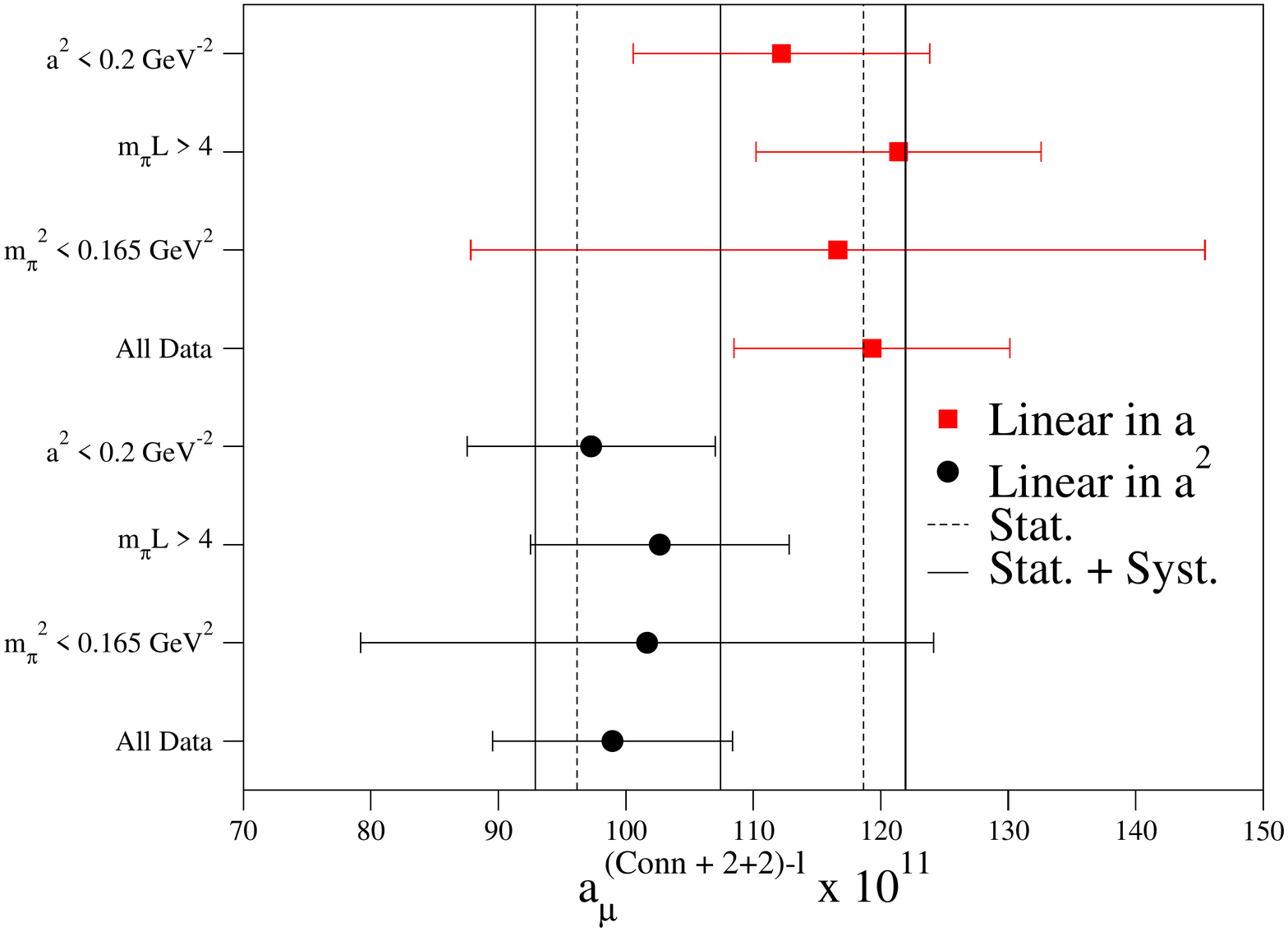}
\caption{Chiral-continuum-infinite-volume fits to the sum of the light-quark fully-connected and $(2+2)$ disconnected contributions. The vertical lines represent the result given in Eq.\ (\ref{eq:finalres}) with its statistical (dashed lines) and full uncertainty (solid lines).}\label{fig:syst_comb}
\end{figure}

We find that the fit Ansatz of Eq.~\eqref{eq:fvol2} describes our data well. At the same time, a term linear in $a$ instead of $a^2$ also gives a good fit ($\chi^2/\text{dof} < 1$ for all fits in this section).
The results for these fits can be found in Fig.~\ref{fig:syst_comb} and are listed in Tab.~\ref{tab:sumfit_results} in Appendix~\ref{app:data_tables}. There is a systematic difference between the fits in $a$ and $a^2$, with the former pulling the final value up a little. 
Applying the various cuts has little impact on the central value, and only the cut on the pion mass removing the $\text{SU}(3)_f$-symmetric-point data leads to a significant increase of the statistical error. This is not surprising, as the larger the volume and the closer the pion mass to its physical value, the larger the cancellation between the fully-connected and $(2+2)$ contributions becomes, and therefore the relative error on their sum.

It is thus clear from Fig.~\ref{fig:syst_comb} that our main systematic in this approach comes from the continuum extrapolation, as was the case in our previous work \cite{Chao:2020kwq}. For the final result from this analysis, we treat the two ans\"atze for the parametrization of cutoff effects on an even footing and perform a fit to a constant to all the possibilities. As an estimate of the systematic error, we compute the root-mean-squared deviation of the fit results $y_i$ compared to the average result $\bar{y}$, i.e.\ $(\sum_{i=1}^N (y_i-\bar{y})^2/N)^{1/2}$. We finally end up with a value of
\begin{equation}\label{eq:finalres}
a_\mu^{(\text{Conn.}+(2+2))\text{-}l} = 107.4(11.3)(9.2) \times 10^{-11},
\end{equation}
with the first error being statistical and the second systematic.

\subsection{Individual fits}

Considering Fig.~\ref{fig:connplot}, it does appear that our data exhibits some curvature in $m_\pi^2$ going towards the physical pion mass, however the underlying functional form is unclear. We have identified several ans\"atze to describe this non-analytic term in Eq.~\eqref{eq:singterm}, all of which provide acceptable descriptions ($\chi^2/\text{dof}\approx 1$) of our data. A plot summarising the values obtained for $a_\mu$ at the physical point is shown in Fig.~\ref{fig:syst_conndisc} and these values can be found in Tab.~\ref{tab:fitsum_results} in Appendix~\ref{app:data_tables}. A fit without some kind of curvature term poorly describes the connected data ($\chi^2/\text{dof}\approx 2.5$), but for the disconnected a good fit is still possible without such a term due to the relatively low statistical precision of the data.

\begin{figure}[h!]
\includegraphics[scale=0.35]{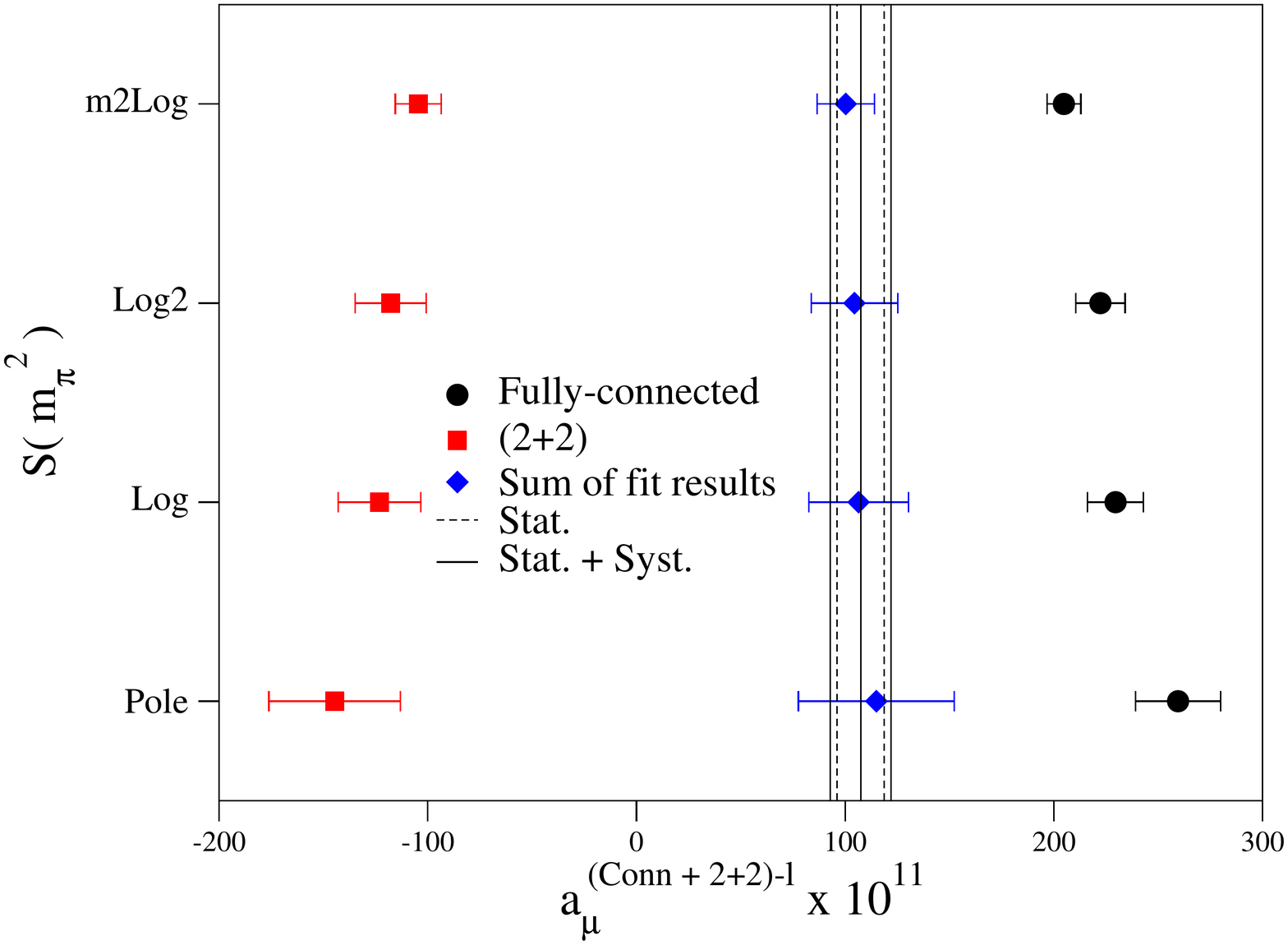}
\caption{Individual fit results to Eq.~(\ref{eq:fvol3}) for different choices of the curvature function $S(m_\pi^2)$, also shown is the result from the previous section.}\label{fig:syst_conndisc}
\end{figure}

It is clear from Fig.~\ref{fig:syst_conndisc} that there is considerable ambiguity on the resulting individual contributions from choosing the functional form of this curvature term, although this somewhat ``washes out" when the sum is taken. Due to the difficulty of resolving this term precisely using a global fit, we view combining fits to the individual contributions as a suboptimal procedure, especially without a result very close to the physical pion mass to help constrain this possible curvature. It is, however, reassuring that the two approaches have good agreement. In conclusion, we choose to quote the fit to the sum of the two contributions for our final result.

%% file: conclu.tex
\section{Conclusions\la{sec:concl}}

Our final estimate of the light and strange quark contributions to $\ahlbl$ is
\begin{equation}
\ahlbl = 106.8(14.7)\times 10^{-11}.
\end{equation}
This result includes all systematics (added in quadrature) as well as previously-unmeasured $(2+1+1)$ and $(1+1+1+1)$ higher-order contributions. The overall precision is about $14\%$. A breakdown of the individual contributions to this result can be found in Tab.~\ref{tab:final_res}.

%% table of results
\begin{table}[h!]
\begin{tabular}{c|c}
\toprule
Contribution & Value$\times 10^{11}$ \\
\hline
Light-quark fully-connected and $(2+2)$ & 107.4(11.3)(9.2) \\
Strange-quark fully-connected and $(2+2)$ & $-0.6(2.0)$ \\
$(3+1)$ & 0.0(0.6) \\
$(2+1+1)$ & 0.0(0.3) \\
$(1+1+1+1)$ & 0.0(0.1) \\
\hline
Total & 106.8(14.7) \\
\botrule
\end{tabular}
\caption{A breakdown of our result for $\ahlbl$.}\label{tab:final_res}
\end{table}

We find that, as we approach the physical pion mass, the two leading contributions to the total $\ahlbl$, the light-quark fully-connected and $(2+2)$ disconnected, yield significant cancellations. This makes a precise measurement at low pion mass and large-$m_\pi L$ extremely challenging. In fact, without the data from the $\text{SU}(3)_f$-symmetric-point data our determination would be considerably less precise (see Fig.~\ref{fig:syst_comb}). It is also clear that the only quantities really needed in the determination of $\ahlbl$ are the fully-connected and $(2+2)$ light-quark contributions.
We find that all of the sub-leading contributions are consistent with zero within the desired precision. For the first time, we have performed a direct calculation of the $(2+1+1)$ and $(1+1+1+1)$ contributions and again find these contributions to be consistent with zero and smaller than the $(3+1)$, which is expected to be the case from large-$N_c$ arguments, and naively from the magnitude of their charge factors.

\begin{figure}
\includegraphics[scale=0.30]{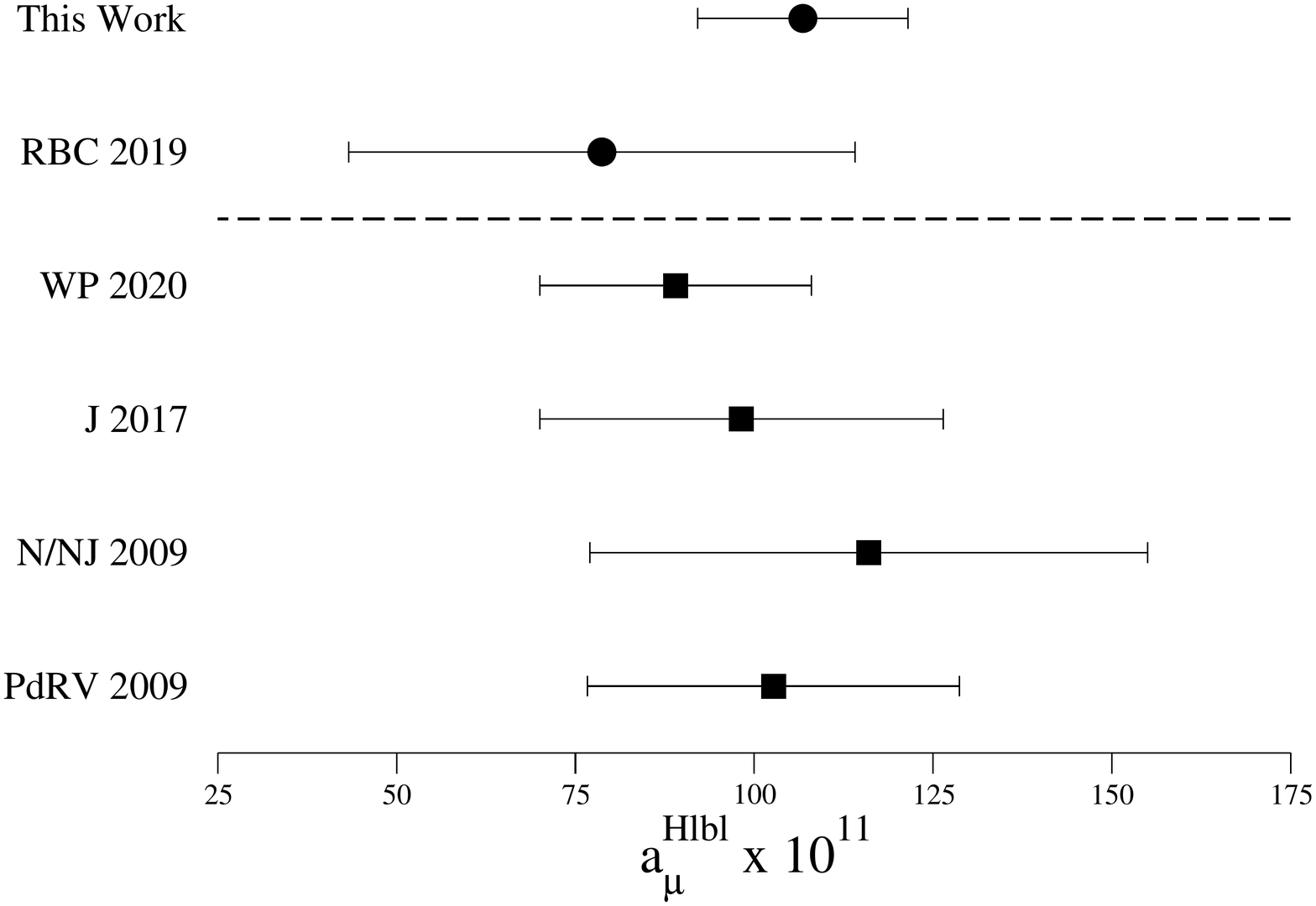}
\caption{A comparison of our result for the $u$, $d$, and $s$ contributions to $\ahlbl$ with the literature. The results in circles are the two available lattice determinations (this work and \cite{Blum:2019ugy}, above the horizontal dashed line). The results in squares are phenomenological predictions from \cite{Aoyama:2020ynm},\cite{Jegerlehner:2017gek},
\cite{Nyffeler:2009tw,Jegerlehner:2009ry}, and \cite{Prades:2009tw}. 
All errors have been added in quadrature.}\label{fig:summary_plot}
\end{figure}

As suggested in previous lattice determinations \cite{Blum:2016lnc,Blum:2019ugy} and several phenomenological predictions (e.g. \cite{Prades:2009tw,Nyffeler:2009tw,Jegerlehner:2009ry,Jegerlehner:2017gek} 
and discussion/references in \cite{Aoyama:2020ynm}), the hadronic light-by-light contribution is in no way large enough to bridge the current gap between theory and experiment for the overall $(g-2)_\mu$. In Fig.~\ref{fig:summary_plot}, we illustrate that there is excellent agreement between our determination and the literature. An uncorrelated fit to a constant of the upper three values of Fig.~\ref{fig:summary_plot} yields $\ahlbl=98.9(11.1)\times 10^{-11}$. We remind the reader that we consistently omit the contribution of the charm quark, which in \cite{Aoyama:2020ynm} is estimated to be $3(1)\times 10^{-11}$.
Whether one performs an average of different $\ahlbl$ determinations or not, with the level of precision and consistency achieved,
the highest priority in improving the overall $(g-2)_\mu$ theory prediction is now to sharpen the HVP determination.
% the dominant theoretical uncertainty to the overall $(g-2)_\mu$ now stems from the HVP contribution. 

Still, further improvements in the lattice determination of $\ahlbl$ are clearly possible with the formalism we have employed.
It is worth reiterating once more that a lattice determination of $\ahlbl$ needs only to focus on the light fully-connected and $(2+2)$ contributions as, at the required accuracy to make an impact on the theory prediction of $(g-2)_\mu$, these are the only parts that matter.

%% file: matching_nf2p1.tex
\newcommand{\str}{\mathrm{str}}
\newcommand{\dd}{\mathrm{d}}
\newcommand{\diag}{\mathrm{diag}}
\section{Diagram matching using Partially-Quenched Chiral Perturbation Theory}\label{matching_nf2p1:sect}

Partially-Quenched Chiral Perturbation Theory (PQChPT), an effective field theory (EFT) of Partially-Quenched Quantum Chromodynamics (PQQCD), is a commonly-used tool for matching different QCD Wick-contractions to Feynman diagrams in an EFT (see e.g. \cite{DellaMorte:2010aq}).

In our previous study at the SU$(3)_f$-symmetric point~\cite{Chao:2020kwq}, we have shown how one applies PQChPT to obtain the corresponding contributions for the finite size effect correction. 
For the purpose of this paper, we shall consider PQChPT with SU$(3)_f$-breaking because our calculations are performed much closer to the physical-quark-mass point. 
In particular, we will focus on two hadronic contributions from the EFT which are considered to be dominant at large distances for the light-by-light scattering: the neutral pseudoscalar meson exchange and the charged pseudoscalar meson loop.

In all of the cases discussed below, we will consider $N_f=2+1$ QCD as the underlying theory. We will study the matching for the pseudoscalar meson loop using SU$(N|M)$-theories because only the charged mesons are expected to contribute. 
On the other hand, we will also consider U$(N|M)$-theories for the neutral pseudoscalar meson exchange, because it allows one to investigate how the flavour-singlet meson, $\eta'$, contributes to each individual QCD Wick-contraction, as well as how $\eta/\eta'$-mixing might arise in diagram matching. 

(S)U($N|M$)-PQQCD is a theory with $N-M$ sea quarks, $M$ (quenched) valence quarks and $M$ ghost quarks. 
The (S)U$(N|M)$ flavour symmetry is explicitly broken due to the non-degenerate quark masses. 
Starting from PQQCD, one constructs the corresponding PQChPT in almost the same way as one obtains ChPT from QCD, but for the Goldstone bosons (the pseudoscalar mesons) being graded Lie-group (S)U$(N|M)$-valued fields.  The terms in the Lagrangian are the same as in ChPT up to replacements of the traces by \textit{super}-traces, which guarantees the invariance of the Lagrangian under symmetry transformations. 
Since the purpose of this study is to understand how different Feynman diagrams are matched between theories, we will only consider the lowest order terms for both contributions and renormalisation will not be taken into account.

At lowest order in perturbation theory, the flavour-breaking effect is due to the mass term.  This manifests itself in the mixing of the propagators if we consider the Goldstone boson fields as living in the adjoint representation of the symmetry group;
the interaction vertices are not affected at this level.
Denoting $m_l$ the light quark mass and $m_s$ the strange quark mass, and starting from $N_f=2+1$ QCD, we will need two different PQChPTs to single-out different Wick contractions, those in which quenched quarks/ghosts either have mass values of $m_l$ or $m_s$.
We enumerate these as: 
\begin{itemize}
\item[(\textit{i})] (S)U$(5|2)$ with light quenched quarks/ghosts 
\item[(\textit{ii})] (S)U$(6|3)$ with strange quenched quarks/ghosts. 
\end{itemize}

In the following, we will first give the expressions for the propagators, and the relevant interaction vertices will then be given in dedicated sub-sections for the neutral pseudoscalar exchange and the charged pseudoscalar loop.
Then, we apply the usual routine as described in~\cite{Chao:2020kwq} to build different four-point functions which give exactly the desired diagrams from an adequate choice of PQChPT.
The expressions will be given in momentum space in Euclidean spacetime. 

Note that we give the matching relations between different contraction topologies in QCD and PQChPT:
one has to include the correct charge factors in order to recover the full light-by-light scattering result.

\subsection{Propagators in \texorpdfstring{$N_f = 2+1$}{Nf=2+1} QCD}

\subsubsection{\texorpdfstring{SU$(N|M)$}{SU(N|M)}}

Under the framework of SU$(N|M)$, we define the \textit{super-trace} of an $(N+M)\times(N+M)$ matrix $A$ as
\begin{equation}
\str(A) = \sum_{i=1}^{N} A_{ii} - \sum_{i=N+1}^{M+N}A_{ii}\,.
\end{equation}

The graded group SU$(N|M)$ is generated by super-traceless matrices.
A convenient choice of generator basis $\{T^a\}$ is a set of super-traceless Hermitian matrices, such that 
\begin{equation}\label{matching_nf2p1:eq:othonorm}
\str(T^a T^b) = \frac{1}{2}g_{ab},
\end{equation}
where
\begin{equation}\label{matching_nf2p1:eq:metric}
g = \begin{pmatrix}
\mathbb{I}_{(N^2-1)\times (N^2-1) } && 0 && 0  \\ 
0  && \mathbb{I}_{MN\times MN } \otimes (-\sigma_2) && 0 \\
0  && 0 && - \mathbb{I}_{M^2\times M^2}
\end{pmatrix}
\,,\quad
\sigma_2 = \begin{pmatrix}
0 && -i \\ 
i && 0 
\end{pmatrix}.
\end{equation}
The kinetic part of the PQChPT Lagrangian can then be arranged as
\begin{equation}
\begin{split}
\mathcal{L}_{\mathrm{kin}}
=& \frac{1}{2}g_{ab}\Big(
\partial_\mu \phi^a \partial_\mu \phi^b 
+ M_\pi^2 \phi^a\phi^b 
\Big)
+ \frac{1}{2}\Delta M^2 K_{ab}\phi^a\phi^b,
\end{split}
\end{equation}
where the $\phi_a$s are Goldstone bosons/fermions and
\begin{equation}\label{matching_nf2p1:eq:kab}
\Delta M^2 = M_K^2 -  M_\pi^2, \quad
K_{ab} = 4 \,\str[\mathcal{S}T^a T^b].
\end{equation}
Here $\mathcal{S}$ is an $(N+M)\times (N+M)$ diagonal matrix which is related to the quark species of the underlying PQQCD, where the quark fields live in the fundamental representation of SU$(N|M)$.
For our purpose, we require the underlying PQQCD to be an extension of $N_f=2+1$ QCD by setting $\mathcal{S}=\diag(0,0,1,\cdots)$.
Under this convention, the first two positions in the fundamental representation correspond to the $u$- and $d$-quark, the third to the $s$-quark and the rests are quenched quarks and their ghost counterparts.
The remaining elements of $\mathcal{S}$ are set in the following way: $\mathcal{S}_{ii} = 0$ if the $i$-th quark in the underlying PQQCD is of mass $m_l$ and $\mathcal{S}_{ii} = 1$ if it is of mass $m_s$.
We shall specify the generator basis before proceeding further, as this affects the definition of $K_{ab}$ and the form of the propagators that we give later. 
For SU$(N|M)$, we can write down a basis for the generators with $N+M-1$ that are diagonal, and $(N+M)(N+M-1)$ generators with 0's in the diagonal.
We will call the first category \textit{neutral} and the second \textit{charged}. 

Analogously to the SU$(3)$ Gell-Mann matrices, the set of charged generators can be written as a union of two sets of matrices $I\cup J$, where
\begin{equation}
\begin{split}
& I = \Big\{C^{ij} | 1\leq i < j \leq N+M,\quad (C^{ij})_{ab} =  \frac{1}{2}(\delta_{ai}\delta_{bj} + \delta_{aj}\delta_{bi})\Big\}\,,
\\
& J = \Big\{D^{ij} | 1\leq i < j \leq N+M,\quad (D^{ij})_{ab} =  \frac{i}{2}(\delta_{ai}\delta_{bj} - \delta_{aj}\delta_{bi})\Big\}\,.
\end{split}
\end{equation}

As for the neutral generators, the $j$-th of them, $A^{j}$, is defined by
\begin{equation}\label{matching_nf2p1:eq:neutral_gen}
A^j = \frac{1}{\sqrt{2|\str[(B^j)^2]}|}B^j
,\quad
B^j = \diag(\underbrace{1,\cdots,1}_{j},a,\underbrace{0,\cdots,0}_{N+M-1-j})
,\quad
a = 
\begin{cases}
-j  & \textrm{if $j\leq N-1$}, \\
2N-j  & \textrm{otherwise}.
\end{cases}
\end{equation} 
It is straightforward to see that, upon re-enumeration, the basis of generators constructed in this way satisfies Eq.~\eqref{matching_nf2p1:eq:othonorm}.
Note that the neutral generator defined by $j=1$ in Eq.~\eqref{matching_nf2p1:eq:neutral_gen} has a special r\^ole  because it does not mix with other neutral generators under flavour-breaking. 
We will name it as \textit{neutral pion} ($\pi^0$). 

For both PQChPT theories (\textit{i}) and (\textit{ii}) in which we are interested, the propagator takes the form
\begin{equation}\label{nf2p1prop}
S_{ab}(x) = 
 g_{ab} G_1(x,M_{ab}) - \Delta M^2 \tilde{H}_{ab}G_2(x,M_{ab}, \tilde{M}_{ab})\,,
\end{equation}
where
\begin{equation}
G_1(x,m) = \int\dd^4 p \,\frac{e^{ipx}}{p^2+m^2}, \quad 
G_2(x,m_1,m_2) = \int\dd^4 p \,\frac{e^{ipx}}{(p^2 +m_1^2)(p^2 + m_2^2)}.
\end{equation}
The mass parameters, $M_{ab}$ and $\tilde{M}_{ab}$, and the matrix $\tilde{H}$ depend on the PQChPT and only the neutral non-pion sector is affected by the propagator mixing induced by the second term on the right-hand side of Eq.~\eqref{nf2p1prop},  due to flavour-symmetry breaking.

In the charged meson sector, for both (\textit{i}) and (\textit{ii}), we have $H_{ab}=0$ and
\begin{equation}\label{matching_nf2p1:eq:mab}
M_{ab}^2 = 
\begin{cases}
M_\pi^2 \quad\textrm{if the underlying PQQCD content is light-light},\\
M_K^2 \quad\textrm{if the underlying PQQCD content is light-strange}, \\
M^2_{\bar{s}s} \equiv M_\pi^2 + 2\Delta M^2 \quad\textrm{if the underlying PQQCD content is strange-strange}.
\end{cases}
\end{equation}

For the neutral sector, the results for (\textit{i}) and (\textit{ii}) are as follows:
\begin{itemize}
\item[(\textit{i})]
\begin{equation}
M_{ab}^2 = M_\pi^2 \,,\quad 
\tilde{M}_{ab}^2 = M_\pi^2 + \frac{4}{3}\Delta M^2 \,,\quad
\tilde{H}_{ab} = (g K g)_{ab},
\end{equation}
\item[(\textit{ii})]
\begin{equation}
\begin{split}
& M_{ab}^2 = 
\begin{cases}
M_\pi^2 \quad\textrm{for $a=b=\pi^0$}, \\
M_{\bar{s}s}^2 \quad\textrm{else},
\end{cases}
\\
& \tilde{K}_{ab} = 
\begin{cases}
K_{ab} \quad\textrm{for $a=b=\pi^0$}, \\
K_{ab} - 2g_{ab} \quad\textrm{else},
\end{cases}
\\
& \tilde{M}_{ab}^2 =
M_{ab}^2 - \frac{2}{3} \Delta M^2
\,,\quad
\tilde{H}_{ab} = (g\tilde{K}g)_{ab}.
\end{split}
\end{equation}
\end{itemize}
\subsubsection{\texorpdfstring{U$(N|M)$}{U(N|M)}}

To make connection with the physical parameters in U(3) ChPT, we define
\begin{equation}
\begin{split}
& \delta\bar{m}^2 = M_{\eta'}^2 - M_\eta^2\,,
\quad 
\Delta\mathring{M}_{\eta'}^2 = M_\eta^2 + M_{\eta'}^2 - 2M_K^2\,,\quad 
\Lambda = \frac{\frac{3}{2}\Delta \mathring{M}^2_{\eta'} - \Delta M^2}{2\sqrt{2}\Delta M^2}\tan(2\delta)\,,
\end{split}
\end{equation}
where $\delta$ is the $\eta/\eta'$-mixing angle. 

One can build a partially-quenched U$(N|M)$ theory from a U$(N-M)$ theory in the same fashion as illustrated in the previous section; 
the only difference is the presence of an extra generator with a non-vanishing super-trace. This has already been done in the literature but in a different generator basis~\cite{Bernard:1993sv}. 
For our purpose, we stick to our already-built generator basis for SU$(N|M)$ plus the diagonal generator $\Lambda^{-1} \mathbb{I}$ for our generator basis.
We extend the definition of the metric defined for SU$(N|M)$ in Eq.~\eqref{matching_nf2p1:eq:metric} to
\begin{equation}
\tilde{g} = 
\begin{pmatrix}
\Lambda^{-2} & \mathbf{0} \\
\mathbf{0} & g
\end{pmatrix}\,,
\end{equation}
where the top-left corner of the matrix on the right-hand side represents the flavour-singlet sector. 
We will use the number $0$ to indicate this sector.
Instead of introducing new notation, we extend naturally the definition of the matrix $K$ defined in Eq.~\eqref{matching_nf2p1:eq:kab} to include the flavour-singlet sector.

After matching to the U$(N-M)$ theory, one obtains for the leading order of the chiral Lagrangian:
\begin{equation}\label{matching_nf2p1:eq:unm_lagrangian}
\mathcal{L}_{\rm{kin}} = \frac{1}{2}\tilde{g}_{ab}\Big( \partial_\mu\tilde{\phi}^a \partial_\mu\tilde{\phi}^b + M_\pi^2 \tilde{\phi}^a\tilde{\phi}^b \Big) + \frac{1}{2}\Delta M^2 K_{ab}\tilde{\phi}^a\tilde{\phi}^b +\frac{1}{2}\Delta M_{\eta'}^2 (\tilde{\phi}^0)^2,
\end{equation}
where $\tilde{\phi}^a$ is the same field as $\phi^a$ in the SU$(N|M)$-theory if $a \neq 0$ and $\tilde{\phi}^0 = \eta'$, and 
\begin{equation}
\Delta M_{\eta'}^2 = \Lambda^{-2}\Delta \mathring{M}_{\eta'}^2 + (\Lambda^{-2}-1)K_{00}\Delta M^2\,.
\end{equation}

The derivation of the propagators for the Lagragian Eq.~\eqref{matching_nf2p1:eq:unm_lagrangian} can be performed following a similar procedure as in~\cite{Golterman:2009kw}.
We introduce here a matrix $\tilde{H}$, two constants, $\lambda$ and $D_{00}$, and mass parameters, $M_{ab}$ and $\tilde{M}_{ab}$,  which will be specified later according to the partially-quenched theory considered.

We define
\begin{equation}
\Sigma_{ab} = (\tilde{g}^{-1})_{a0}(\tilde{g}^{-1})_{0b}
\,,\quad
\Theta_{ab} = -\Big( (\tilde{g}^{-1})_{a0} H_{0b} + H_{a0}(\tilde{g}^{-1})_{0b} \Big)
\,,\quad
\Xi_{ab} = H_{a0}H_{0b},
\end{equation}
\begin{equation}
G_3(x,a,b,c)=\int_p \frac{e^{ipx}}{(p^2+a^2)(p^2+b^2)(p^2+c^2)}\,,
\end{equation}
\begin{equation}
G_4(x,a,b,c,d)=\int_p\frac{e^{ipx}}{(p^2+a^2)(p^2+b^2)(p^2+c^2)(p^2+d^2)}\,.
\end{equation}
The U$(N|M)$-propagator is then given by
\begin{equation}\label{matching_nf2p1:eq:prop_unm}
\begin{split}
S_{ab}(x) = &
\quad
(\tilde{g}^{-1})_{ab}G_1(x,M_{ab})
\\&
-\Delta M^2 H_{ab}G_2(x,M_{ab},\tilde{M}_{ab}) 
\\
& -\Delta \bar{M}_{\eta'}^2\Sigma_{ab} G_2(x,M_{ab},\tilde{M}_{ab})
\\ &
-\Delta M^2 \Delta \bar{M}_{\eta'}^2 (\Theta_{ab} + \lambda \Sigma_{ab})G_3(x,M_{ab},M_\eta,M_{\eta'})
\\
&-\Delta \bar{M}_{\eta'}^2(\Delta M^2)^2 \Xi_{ab} G_4(x,M_{ab},\tilde{M}_{ab}, M_{\eta}, M_{\eta'})\,,
\end{split}
\end{equation}
where
\begin{equation}
\Delta\bar{M}^{2}_{\eta'} = \Delta M^2_{\eta'} + (1-\Lambda^2)D_{00}\Delta M^2\,.
\end{equation}

For the charged sector, only the first term on the right-hand side of Eq.~\eqref{matching_nf2p1:eq:prop_unm} does not vanish and the mass $M_{ab}$ parameter is the same as defined in Eq.~\eqref{matching_nf2p1:eq:mab}; for the neutral sector, we have for the cases ($i$) and ($ii$):
\begin{itemize}
\item[(\textit{i})]
\begin{equation}
\begin{split}
& M_{ab}^2 = M_\pi^2 \,,\quad 
\tilde{M}_{ab}^2 = M_\pi^2 + \lambda\Delta M^2 
\,,\quad
\lambda = \frac{2}{3}(2+\Lambda^2) \,,
\\
& 
D_{00} = 0\,,\quad
H_{ab} = (\tilde{g}^{-1}K \tilde{g}^{-1})_{ab}\,.
\end{split}
\end{equation}
\item[(\textit{ii})]
\begin{equation}
\begin{split}
& M_{ab}^2 = 
\begin{cases}
M_\pi^2 \quad\textrm{for $a=b=\pi^0$}, \\
M_{\bar{s}s}^2 \quad\textrm{else},
\end{cases}
\\
& \tilde{K}_{ab} = 
\begin{cases}
K_{ab} - 2\delta_{ab} \quad\textrm{($a \neq \pi^0$ and has no ghost constituent) or ($a = \eta'$)}, \\
K_{ab} + 2\delta_{ab} \quad\textrm{$a \neq \pi^0, \eta'$ and has ghost contituents}, \\
K_{ab} \quad\textrm{else},
\end{cases}
\\
&
\tilde{M}_{ab}^2 =
M_{ab}^2 +\lambda \Delta M^2
\,,\quad
\lambda = -\frac{2}{3}(1+2\Lambda^2)
\,,\quad
D_{00} = 2 \,,\quad
H_{ab} = (\tilde{g}^{-1}\tilde{K}\tilde{g}^{-1})_{ab}\,.
\end{split}
\end{equation}
\end{itemize}

\subsection{Charged pseudoscalar meson loop}

Consider here an SU$(N|M)$ theory.
Up to $O(p^2)$, the interaction Lagrangian is~\cite{DellaMorte:2010aq}
\begin{equation}
\mathcal{L}^{(2)} = 
 -\sum_{k} C_{k}^{bc}g^{ak}\partial_\mu \phi^a v_\mu^b\phi^c
+\frac{1}{2}\sum_{k,l}C_k^{ab}C_{l}^{cd}g^{kl}v_\mu^a\phi^b\phi^c v_\mu^d\,,
\end{equation} 
where 
\begin{equation}
C_a^{bc} 
=
-2i\sum_k \str\{[T^b, T^c]T^k\}g^{ka}\,,
\end{equation}
and $[\cdot,\cdot]$ acts as commutator or anti-commutator according to the generators that it applies to (for definition cf.~\cite{DellaMorte:2010aq}).
The summed indices run over the whole basis of generators.

The 3-point and 4-point vertices in Euclidean space-time are then given by
\\
\begin{minipage}{.3\linewidth}
\centering
\includegraphics[scale=0.4]{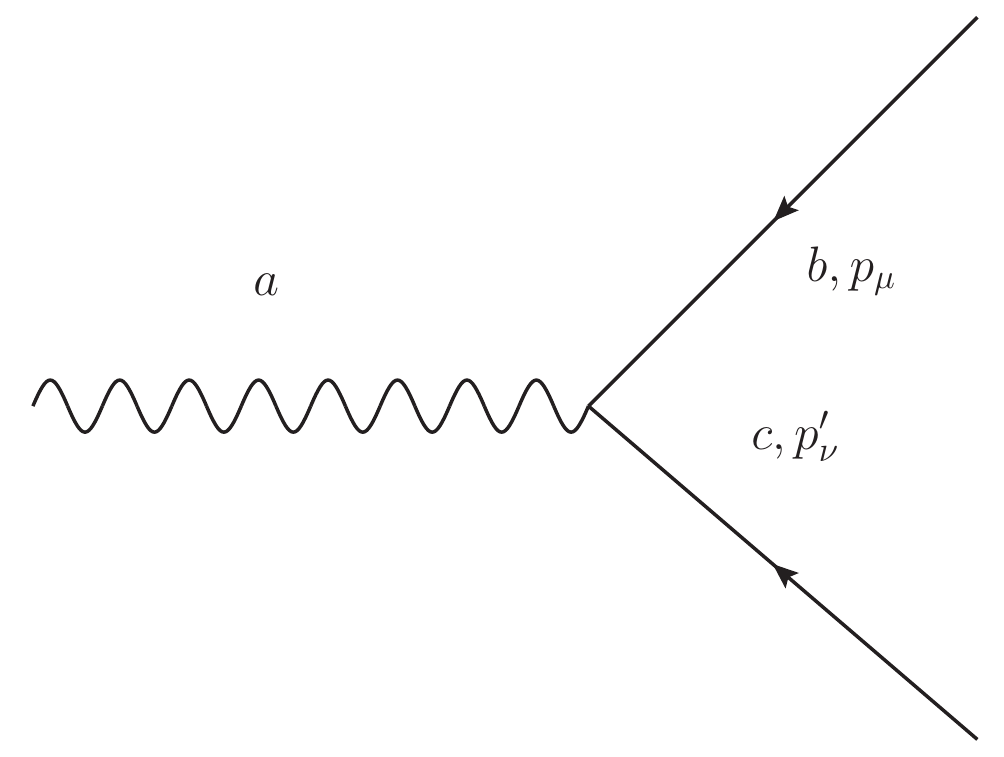}
\end{minipage}
\begin{minipage}{.7\linewidth}
\begin{equation}
\begin{split}
& V^{cb ;a}_{3;\mu} = -\Big(\tilde{C}^{ab}_c p'_\mu + \tilde{C}^{ac}_b p_\mu\Big)\,, \\
& \tilde{C}^{bc}_a \equiv 2 \str([T^b, T^c]T^a) = i C^{bc}_i g^{ai}\,,
\end{split}
\end{equation}
\end{minipage}
\begin{minipage}{.3\linewidth}
\centering
\includegraphics[scale=0.5]{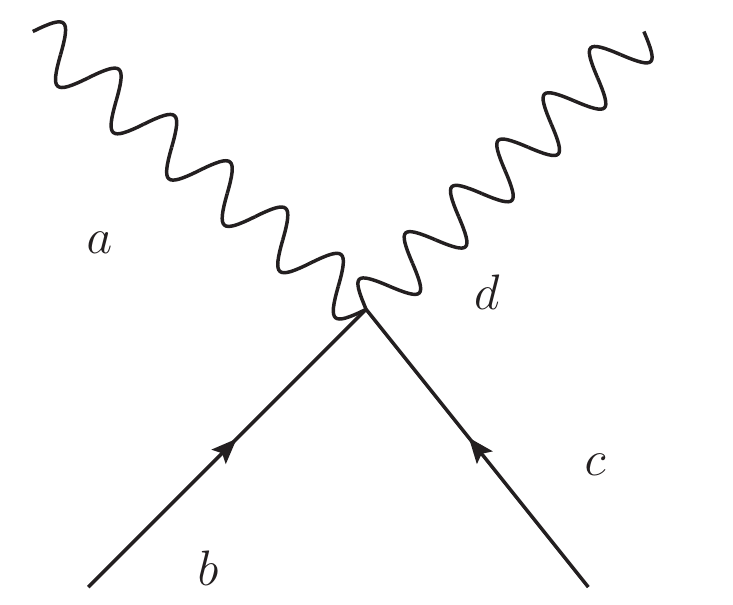}
\end{minipage}
\begin{minipage}{.7\linewidth}
\begin{equation}
\begin{split}
V^{a d ; b c}_{4;\mu\nu} 
= & - \delta_{\mu\nu}g^{mn}(C^{ab}_m C^{cd}_n + C^{ac}_m C^{bd}_n) \\
= & 
2 \delta_{\mu\nu}\Big( \str([T^a, T^b][T^c, T^d]) + \str([T^a, T^c][T^b, T^d]) \Big) .
\end{split}
\end{equation}
\end{minipage}

\begin{figure}[h]
  \centering
  \includegraphics[scale=0.5]{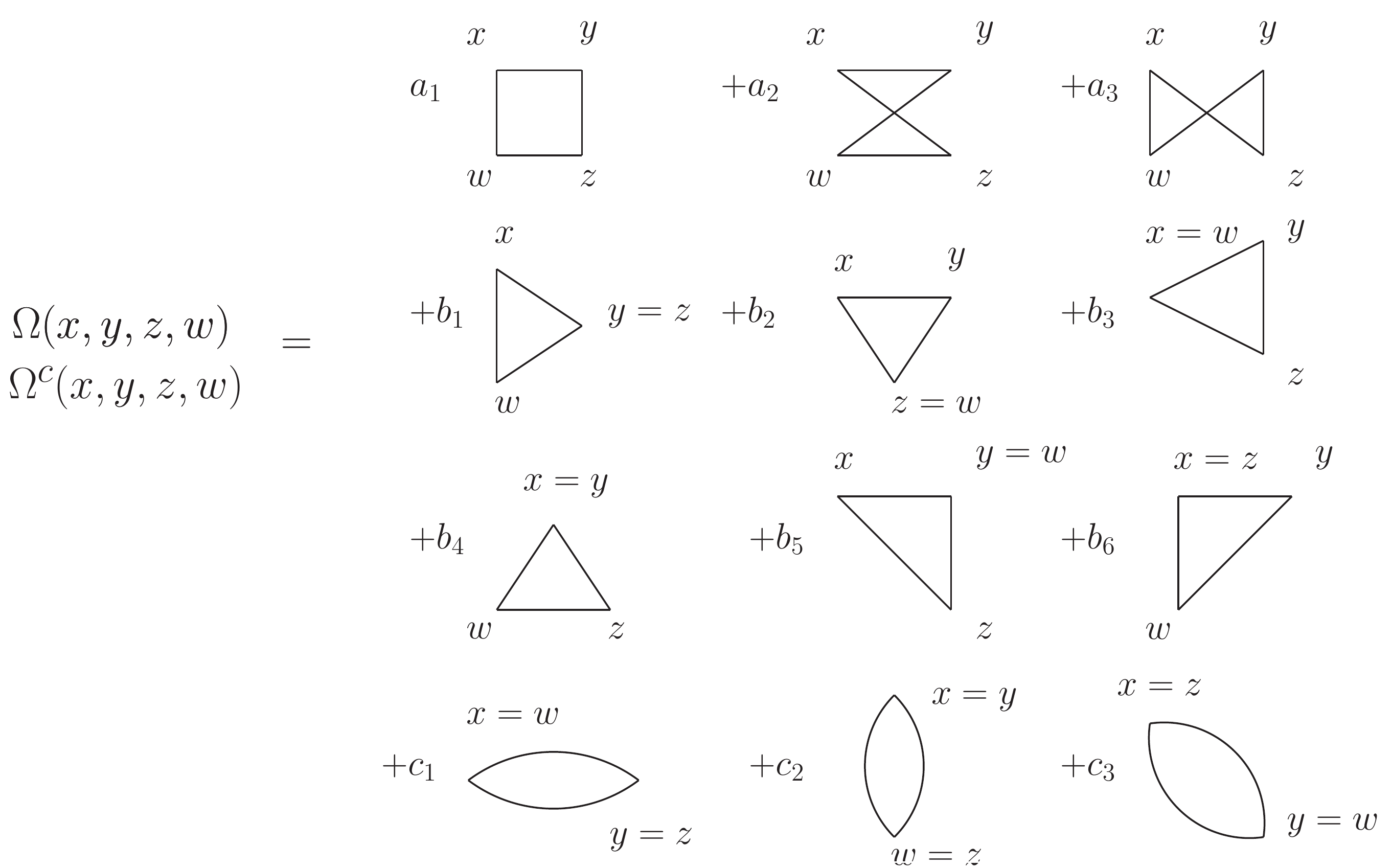}
  \caption{Diagrammatic definition of $\Omega$ and $\Omega^c$}\label{matching_nf2p1:fig:omega_def}
\end{figure}

\begin{figure}
 \centering
 \includegraphics[scale=0.5]{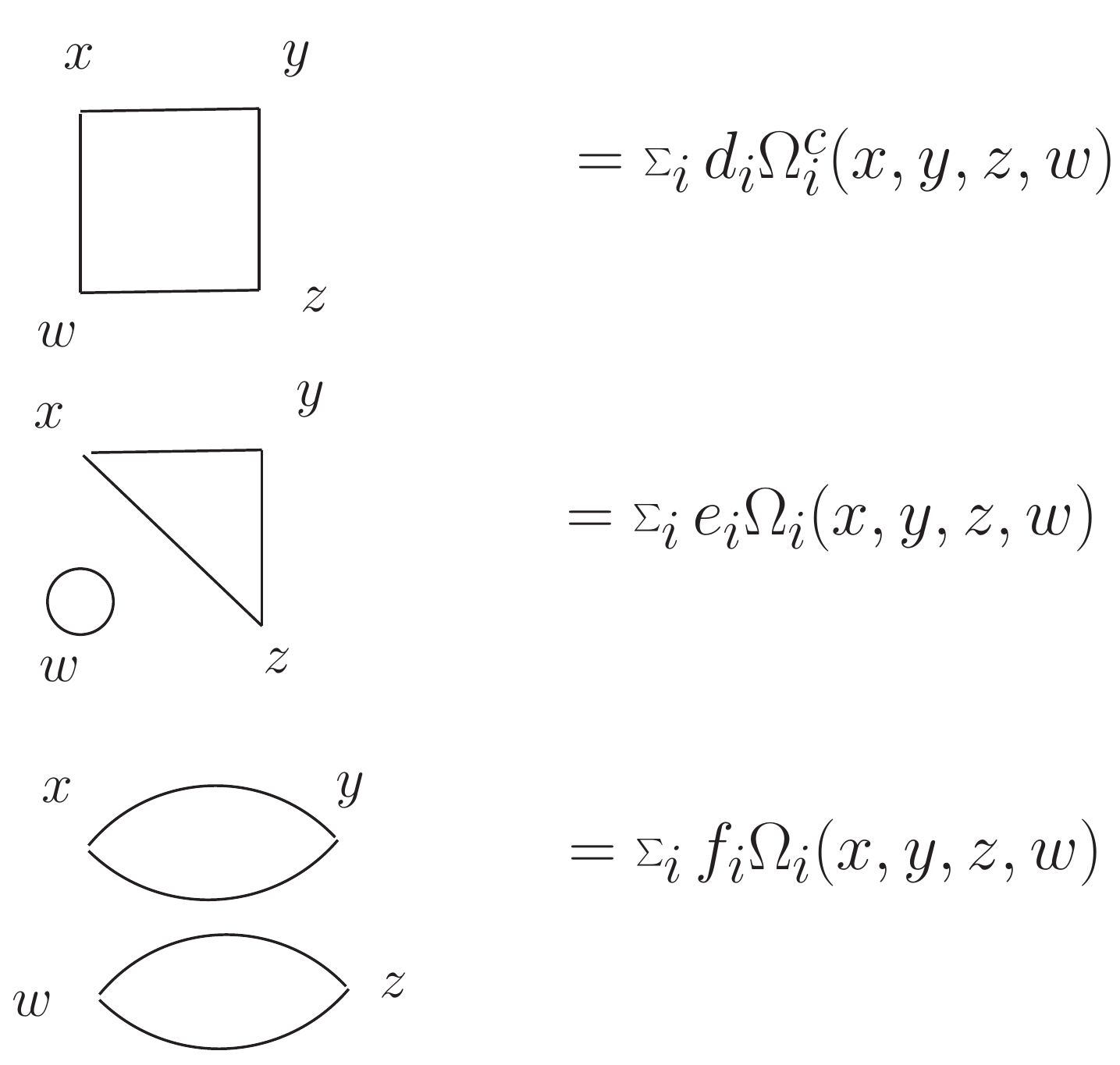}
 \caption{Matching of different QCD-Wick contraction to charged pseudoscalar loop contributions.}\label{matching_nf2p1:fig:loop}
\end{figure}

In Fig.~\ref{matching_nf2p1:fig:omega_def}, we define two functions: $\Omega$ and $\Omega^c$. These are linear combinations of different diagrammatic classes appearing in the charged pseudoscalar meson loop computation (cf. Table~\ref{matching_nf2p1:tab:omega_coef}), and 
$\Omega$ and $\Omega^c$ are computed with a given pseudoscalar meson mass.
Different QCD Wick-contractions are matched to the charged pseudoscalar loop in the way described in Fig.~\ref{matching_nf2p1:fig:loop} with coefficients given in Table~\ref{matching_nf2p1:tab:loop_coef}.
The first column of Table~\ref{matching_nf2p1:tab:loop_coef} indicates the pseudoscalar meson mass that $\Omega$ or $\Omega^c$ should take, and only the non-vanishing contributions are listed.\footnote{All $(2+1+1)$ and $(1+1+1+1)$ diagrams computed with light minus strange disconnected loops vanish.}
One can check that, with the correct charge factors, the total $\ahlbl$ receives contributions from one charged pion-loop and one charged kaon-loop.

\begin{table}[h]
 \begin{tabular}{|c|c|c|c|c|c|c|c|c|c|c|c|c|}
 \toprule
  & $a_1$ & $a_2$ & $a_3$ & $b_1$ & $b_2$ & $b_3$ & $b_4$ & $b_5$ & $b_6$ & $c_1$ & $c_2$ & $c_3$ \\
  \hline
 $\Omega$ & 1 & 1 & 1 & 1 & 1 & 1 & 1 & 1 & 1 & 1 & 1 & 1\\
 $\Omega^c$ & 1 & 0 & 0 & $\frac{1}{2}$ & $\frac{1}{2}$ & $\frac{1}{2}$ & $\frac{1}{2}$ & 0 & 0 & $\frac{1}{2}$ & $\frac{1}{2}$ & 0 \\
 \botrule
 \end{tabular}
 \caption{Coefficients defined in Fig.~\ref{matching_nf2p1:fig:omega_def}.}\label{matching_nf2p1:tab:omega_coef}
\end{table}

\begin{table}
 \begin{tabular}{|c|c|c|c|c|c|c|c|}
 \toprule
 $i$ 		&  (4)-$l$ 	& (4)-$s$ 	& (2+2)-$ll$ 	& (2+2)-$ls$ 	& (2+2)-$ss$ 	& (3+1)-$l$ 	& (3+1)-$s$ \\ 
 \hline
 $\pi$ 		& 2 			& 0 			& 1 			& 0 			& 0 			& -1 			& 0 \\
 $K$ 		& 1 			& 2 			& 0			 	& 2 			& 0 			& 1 			& -1 \\
 $\bar{s}s$ & 0 			& 1 			& 0 			& 0 			& 1 			& 0 			&1\\
 \botrule
 \end{tabular}
 \caption{Coefficients $d_i$ (for fully-connected), $e_i$ (for (3+1)) and $f_i$ (for (2+2)) for the contribution of different charged pseudoscalar mesons defined in Fig.~\ref{matching_nf2p1:fig:loop}. Here, the same convention as for our lattice computation described in the main text is used. In particular, ``(4),$l$/$s$" refers to the fully-connected light/strange contribution, ``(2+2)-$ls$" refers to the sum of the 2+2 diagrams with one light and one strange quark, and ``(3+1)-$l$/$s$" refers to the light/strange quark triangle correlated with light minus strange disconnected loop.}\label{matching_nf2p1:tab:loop_coef} 
\end{table}

\subsection{Neutral pseudoscalar meson exchange}

The neutral pseudoscalar meson exchange is possible due to the chiral anomaly. 
To make the analysis simpler, we only consider the coupling of a pseudoscalar meson with two photons, $P\gamma\gamma$, via the Wess-Zumino-Witten term~\cite{Wess:1971yu, Witten:1983tw}, whose Feynman rule up to an irrelevant factor for our analysis is given by:
\\
\begin{minipage}{.5\linewidth}
\centering
\includegraphics[scale=0.5]{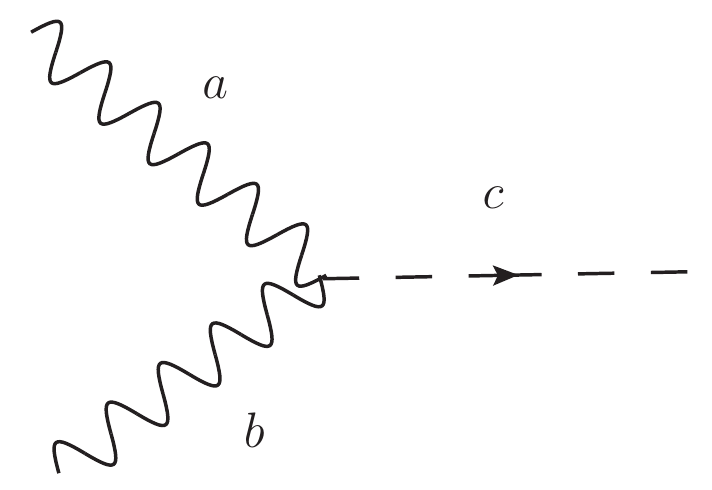}
\end{minipage}
\begin{minipage}{.5\linewidth}
\begin{equation}\label{matching_nf2p1:eq:vwzw}
V_{\rm{WZW}}^{ab;c} \propto \str[\{T^a,T^b\}T^c]\,.
\end{equation}
\end{minipage}
 
We compute the matching coefficients for the cases with and without the flavour-singlet pseudoscalar meson $\eta'$.
\textit{A priori}, the $P\gamma\gamma$ vertex receives contributions from flavour symmetry breaking effects already at tree-level.
In addition, if one is to include the flavour-singlet pseudoscalar meson, additional terms that violate the OZI rule also have an impact at tree-level~\cite{Kaiser:2000gs}. As our goal is to have a qualitative idea of the diagram matching and for a more quantitative analysis, a more realistic transition form factor is required anyway, these effects are not included in our analysis.

\begin{figure}[h!]
\centering
\includegraphics[scale=0.5]{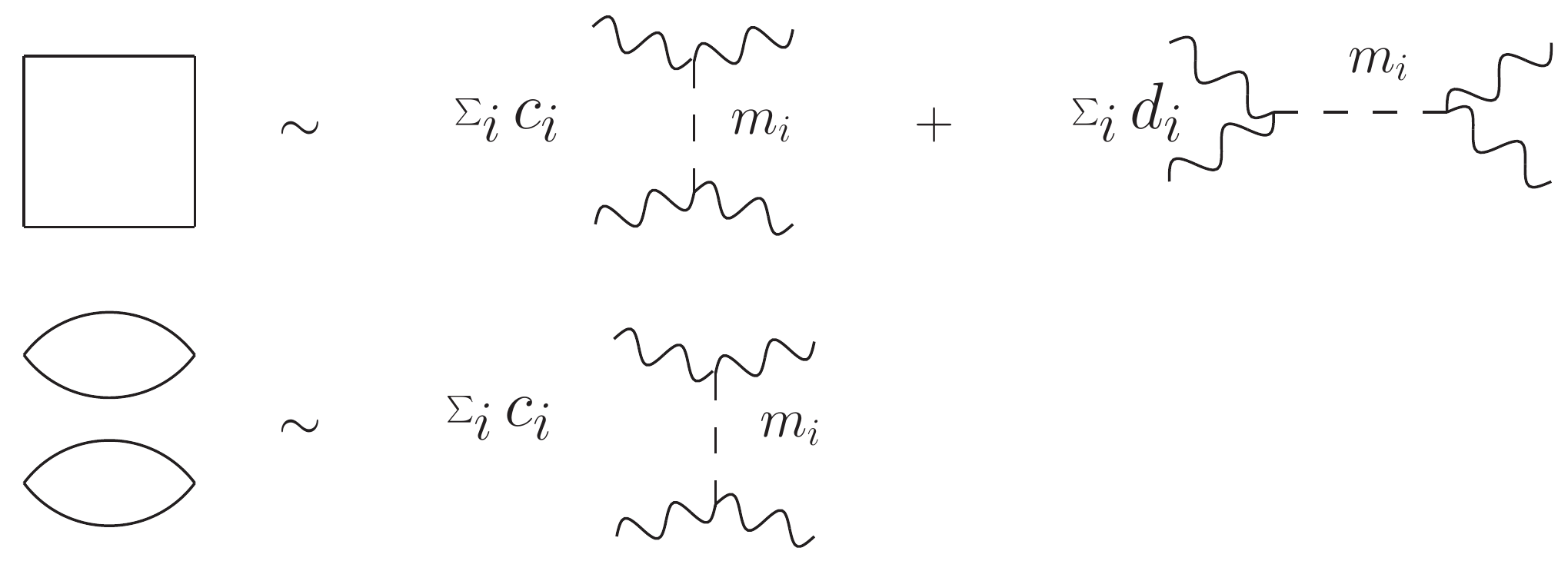}
\caption{Matching pattern between QCD Wick-contractions and pseudoscalar pole exchange channels.}\label{matching_nf2p1:fig:pionpole_def}
\end{figure}

With either the incusion of the $\eta'$ or not, we found that only the fully-connected and the $(2+2)$-disconnected receive contributions from the neutral pseudoscalar exchange.\footnote{All diagrams from the other 3 topologies computed with light minus strange disconnected loops vanish.}
The matching patterns are given in Fig.~\ref{matching_nf2p1:fig:pionpole_def}, where QCD Wick-contractions are placed on the left hand side and the contributing pseudoscalar exchange channels are displayed on the right hand side. 
The coefficients $c_i$ and $d_i$ are given in each of the following sub-sections.
Note that these are only given up to a common factor.
The normalisation convention that we choose here is to make $c_{\pi}$ for (4)-$l$ equal to 1. 

\subsubsection{Without the flavour-singlet meson}

We first study the matching without the flavour-singlet meson using the SU$(N|M)$ theory, which gives rather simple results and are tabulated (Tab.~\ref{tab:simple_results}) below.
\begin{table}[h!]
 \begin{tabular}{|c|c|c|c|c|c|}
 \toprule
  				& (4)-$l$ & (4)-$s$ & (2+2)-$ll$ 	& (2+2)-$ls$ 	& (2+2)-$ss$ \\
 \hline
 $c_\pi$ 		& 1 		& 0 		& $-1$ 			& 0 			& 0 \\
 $c_{\eta}$ 	& 0 		& 0 		& $\frac{1}{3}$ & $-\frac{4}{3}$& $\frac{4}{3}$ \\
 $c_{\bar{s}s}$ & 0 		& 1 		& 0 			& 0 			& $-2$ \\
 \hline
 $d_\pi$ 		& 1 		& 0 \\
 $d_{\bar{s}s}$ & 0 		& 1 \\
 \cline{1-3}
 \end{tabular}
 \caption{Relative weights for different exchange channels for each contributing QCD Wick-contraction according to the definition of Fig.~\ref{matching_nf2p1:fig:pionpole_def} }\label{tab:simple_results}
\end{table}
\subsubsection{With the flavour-singlet meson}

The inclusion of the flavour-singlet meson does not change the matching for the fully-connected diagrams. However, for the $(2+2)$-disconnected, we have:
\begin{itemize}
\item \textbf{(2+2), light-light}
\begin{equation}
c_{\pi^0} = -1,
\end{equation}
\begin{equation}
\begin{split}
c_{\eta} = & \frac{2}{27}\frac{1}{\delta \bar{m}^2(-2\Delta M^2 - \Delta \mathring{M}_{\eta'}^2 + \delta \bar{m}^2)}
\\ &\times
\Big(
-6\Delta M^2 \Delta \mathring{M}_{\eta'}^2(1+2\Lambda^2) 
+4\Delta M^4(1-\Lambda^2)(1-4\Lambda^2)
\\&+6\Delta M^2(-1 + \Lambda^2)\delta \bar{m}^2
+9\Delta \mathring{M}_{\eta'}^2\Lambda^2(\Delta\mathring{M}_{\eta'}^2 - \delta\bar{m}^2)
\Big),
\end{split}
\end{equation}
\begin{equation}
\begin{split}
c_{\eta'} = & \frac{2}{27}\frac{1}{\delta \bar{m}^2(2\Delta M^2 + \Delta \mathring{M}_{\eta'}^2 + \delta \bar{m}^2)}
\\ &\times
\Big(
-6\Delta M^2 \Delta \mathring{M}_{\eta'}^2(1+2\Lambda^2) 
+4\Delta M^4(1-\Lambda^2)(1-4\Lambda^2)
\\& -6\Delta M^2(-1 + \Lambda^2)\delta \bar{m}^2
+9\Delta \mathring{M}_{\eta'}^2\Lambda^2(\Delta\mathring{M}_{\eta'}^2 + \delta\bar{m}^2)
\Big).
\end{split}
\end{equation}
\item \textbf{(2+2), light-strange}
\begin{equation}
\begin{split}
c_\eta = &
\frac{2}{9\delta\bar{m}^2}\Big(
-2\Delta M^2(-1+\Lambda^2) - 3\Delta \mathring{M}_{\eta'}^2(1+\Lambda^2) + 3(-1+\Lambda^2)\delta\bar{m}^2
\Big),
\end{split}
\end{equation}
\begin{equation}
\begin{split}
c_{\eta'} = &
\frac{2}{9\delta\bar{m}^2}\Big(
2\Delta M^2(-1+\Lambda^2) + 3\Delta \mathring{M}_{\eta'}^2(1+\Lambda^2) + 3(-1+\Lambda^2)\delta\bar{m}^2
\Big).
\end{split}
\end{equation}
\item \textbf{(2+2), strange-strange}
\begin{equation}
c_{\bar{s}s} = -2,
\end{equation}
\begin{equation}
\begin{split}
c_\eta = &-\frac{4}{27}\frac{\Lambda^2[3\Delta \mathring{M}^2_{\eta'}+4\Delta M^2(-1+\Lambda^2)]}{\delta \bar{m}^2(2\Delta M^2-\Delta \mathring{M}_{\eta'}^2+\delta\bar{m}^2)[-3\Delta\mathring{M}^2_{\eta'}+2\Delta M^2(1-4\Lambda^2)+3\delta\bar{m}^2]}
\\ &\times\Big(
4\Delta M^4(5+4\Lambda^2)-9\Delta \mathring{M}_{\eta'}^2(-\Delta \mathring{M}_{\eta'}^2 + \delta \bar{m}^2) - 6\Delta M^2(-2\Delta \mathring{M}_{\eta'}^2 +3\delta\bar{m}^2)
\Big),
\end{split}
\end{equation}
\begin{equation}
\begin{split}
c_{\eta'} = &\frac{4}{27}\frac{\Lambda^2[3\Delta \mathring{M}^2_{\eta'}+4\Delta M^2(-1+\Lambda^2)]}{\delta \bar{m}^2(-2\Delta M^2+\Delta \mathring{M}_{\eta'}^2+\delta\bar{m}^2)[3\Delta\mathring{M}^2_{\eta'}-2\Delta M^2(1-4\Lambda^2)+3\delta\bar{m}^2]}
\\ &\times\Big(
4\Delta M^4(5+4\Lambda^2)+9\Delta \mathring{M}_{\eta'}^2(\Delta \mathring{M}_{\eta'}^2 + \delta \bar{m}^2) + 6\Delta M^2(2\Delta \mathring{M}_{\eta'}^2 +3\delta\bar{m}^2)
\Big).
\end{split}
\end{equation}
\end{itemize}
It is worth noting that, when the $\eta'$ is integrated out, which amounts to setting $\Lambda=1$ and $\Delta \mathring{M}_{\eta'}^2=\infty$, we recover the conclusion obtained for the case of SU$(N|M)$. 
Following the $N_c$-counting rules outlined in~\cite{Gasser:1984gg} one can also verify that the total $(2+2)$ contribution vanishes in the large-$N_c$ limit.

%% file: data_tables.tex
\section{Tables of data}\label{app:data_tables}

\begin{table}[h!]
  \begin{tabular}{c|c|c}
    \toprule
    Cut & $O(a)$ Ansatz & Result$\times 10^{11}$ \\
    \hline
    $a^2 < 0.2 \text{ GeV}^{-2}$ & $a^2$ & 97.3(9.7) \\
    $m_\pi L > 4$ & $a^2$ & 102.7(10.1) \\
    $m_\pi^2<0.165 \text{ GeV}^2 $ & $a^2$ & 101.7(22.5) \\
    All Data & $a^2$ & 99.0(9.4) \\
    \hline
    $a^2 < 0.2 \text{ GeV}^{-2}$ & $a$ & 112.2(11.6) \\
    $m_\pi L > 4$ & $a$ & 121.4(11.2) \\
    $m_\pi^2<0.165 \text{ GeV}^2 $ & $a$ & 116.6(28.8) \\
    All Data & $a$ & 119.3(10.8) \\
    \botrule
  \end{tabular}
  \caption{Results of the fit to the sum of the fully-connected and $(2+2)$ light-quark contributions in Fig.~\ref{fig:syst_comb}.}\label{tab:sumfit_results}
\end{table}

\begin{table}[h!]
  \begin{tabular}{c|cc|cc}
    \toprule
    $S\left(m_\pi^2 \right)$ & Fully-connected$\times 10^{11}$ & $\chi^2/\text{dof}$ & $(2+2)$ Disconnected$\times 10^{11}$ & $\chi^2/\text{dof}$ \\
    \hline
    m2Log2 & 204.8(8.1) & 1.1 & $-104.5(11.0)$ & 0.9 \\
    Log2 & 229.5(13.4) & 1.1 & $-123.1(19.8)$ & 0.9 \\
    Log & 222.2(11.8) & 1.1 & $-117.8(17.0)$ & 0.9 \\
    Pole & 259.5(20.4) & 1.1 & $-144.6(31.5)$ & 0.9 \\
    \botrule
  \end{tabular}
  \caption{Results of the individual fits to the fully-connected and $(2+2)$ light-quark contributions in Fig.~\ref{fig:syst_conndisc}.}
  \label{tab:fitsum_results}
\end{table}